\documentclass[]{elsarticle}

\usepackage[margin=3cm]{geometry}
\usepackage{amsmath,accents,siunitx,mathtools,amssymb,bm}
\usepackage{algorithm, algorithmic}
\usepackage{graphicx}
\usepackage[font=small,labelfont=bf]{caption}
\usepackage{subcaption}
\usepackage{enumitem}
\usepackage{scalerel}
\usepackage{booktabs}
\usepackage[multiple]{footmisc}
\usepackage{hyperref}
\hypersetup{
	linkcolor  = black,
	citecolor  = red,
	urlcolor   = red,
	colorlinks = true,
}
\usepackage[capitalize]{cleveref}

\def\appendixname{}

\newlength\figH
\newlength\figW
\usepackage{tikz}
\usepackage{pgfplots} 
\pgfplotsset{compat=newest} 
\pgfplotsset{plot coordinates/math parser=false}

\makeatletter
\def\ps@pprintTitle{%
	\let\@oddhead\@empty
	\let\@evenhead\@empty
	\def\@oddfoot{\centerline{\thepage}}%
	\let\@evenfoot\@oddfoot}
\makeatother
\begin{document}

\begin{frontmatter}

\title{Efficient multi-partition topology optimization}


\author[mymainaddress]{Koppen, S.\corref{mycorrespondingauthor}}
\ead{s.koppen@tudelft.nl}

\author[mymainaddress]{Langelaar, M.}
\author[mymainaddress]{van Keulen, F.}

\address[mymainaddress]{Department of Precision and Microsystems Engineering, Delf University of Technology, Mekelweg 2, 2628 CD Delft, The Netherlands}

\begin{abstract}
In topology optimization, the state of structures is typically obtained by numerically evaluating a discretized PDE-based model. 
The degrees of freedom of such a model can be partitioned in free and prescribed sets to define the boundary conditions. 
A multi-partition problem involves multiple partitions of the same discretization, typically corresponding to different loading scenarios.
As a result, solving multi-partition problems involves multiple factorization/preconditionings of the system matrix, requiring a high computational effort.
In this paper, a novel method is proposed to efficiently calculate the responses and accompanying design sensitivities in such multi-partition problems using static condensation for use in gradient-based topology optimization.
A main problem class that benefits from the proposed method is the topology optimization of small-displacement multi-input-multi-output compliant mechanisms. 
However, the method is applicable to any linear problem.
We present its formulation and an algorithmic complexity analysis to estimate computational advantages for both direct and iterative solution methods to solve the system of equations, verified by numerical experiments. 
It is demonstrated that substantial gains are achievable for large-scale multi-partition problems. This is especially true for problems with both a small set of number of degrees of freedom that fully describes the performance of the structure and with large similarities between the different partitions.
A major contribution to the gain is the lack of large adjoint analyses required to obtain the sensitivities of the performance measure. 
\end{abstract}


\end{frontmatter}


\newcommand*{\tran}{^{\mkern-1.5mu\mathsf{T}}} 
\newcommand*{\traninv}{^{\mkern-1.5mu-\mathsf{T}}} 
\renewcommand{\dm}{\ensuremath{\mathrm{d}}} 

\newcommand{\eg}{\textit{e.g.}\xspace}
\newcommand{\ie}{\textit{i.e.}\xspace}

\newenvironment{comment}{\color{gray}}{\newline}

\newcommand{\sm}{%
	\mathbin{{\vee}\mkern-6mu{\wedge}}%
}
\renewcommand{\ms}{%
	\mathbin{{\wedge}\mkern-6mu{\vee}}%
}
\renewcommand{\mm}{%
	\mathbin{{\wedge}\mkern-6mu{\wedge}}%
}
\renewcommand{\ss}{%
	\mathbin{{\vee}\mkern-6mu{\vee}}%
}

\newcommand{\abf}[5][]{\accentset{#1}{#2}_{\text{#3}{#4}}^{#5}}

\renewcommand{\i}[1][]{\scaleto{(#1)}{5pt}}
\newcommand{\ii}{^{\i[i]}}

\renewcommand{\m}{\wedge}
\renewcommand{\s}{\vee}

\renewcommand{\u}{\mathbf{u}}
\newcommand{\x}{\mathbf{x}}
\newcommand{\U}{\mathbf{U}}
\renewcommand{\K}{\mathbf{K}}
\newcommand{\f}{\mathbf{f}}
\renewcommand{\F}{\mathbf{F}}
\newcommand{\nrhs}{l}
\renewcommand{\dim}{d}

\newcommand{\setsdof}{\mathbb{D}}
\newcommand{\setmdof}{\mathbb{M}}
\newcommand{\setddof}{\accentset{\vee}{\mathbb{P}}}
\newcommand{\setcdof}{\accentset{\vee}{\mathbb{F}}}
\newcommand{\setqdof}{\mathbb{P}}
\newcommand{\sethdof}{\mathbb{F}}
\newcommand{\setxdof}{\mathbb{A}}
\newcommand{\setpdof}{\accentset{\wedge}{\mathbb{P}}}
\newcommand{\setfdof}{\accentset{\wedge}{\mathbb{F}}}

\newcommand{\nsdof}{n}
\newcommand{\nmdof}{m}
\newcommand{\nhdof}{f}
\newcommand{\nqdof}{p}
\newcommand{\ncdof}{\abf[\s]{f}{}{}{}}
\newcommand{\nddof}{\abf[\s]{p}{}{}{}}
\newcommand{\nfdof}{\abf[\m]{f}{}{}{}}
\newcommand{\npdof}{\abf[\m]{p}{}{}{}}

\newcommand{\Zi}{\abf[\sm]{\mathbf{X}}{f}{}{}}
\newcommand{\Zit}{\abf[\ms]{\mathbf{X}}{f}{}{}}
\newcommand{\Zc}{\abf[\s]{\mathbf{V}}{f}{}{}}
\newcommand{\ses}{\mathbf{A}}
\newcommand{\bes}{\mathbf{B}}
\newcommand{\lambdaa}{\abf[\sm]{\bm{\Lambda}}{f}{}{}}
\newcommand{\lambdas}{\abf[\m]{\bm{\Lambda}}{f}{}{}}
\newcommand{\lambdak}{\abf{\tilde{\bm{\Lambda}}}{f}{}{}}
\newcommand{\lag}{\mathcal{L}}

\newcommand{\mydot}{\protect\scalebox{1.5}{$\bullet$}}
\newcommand{\mytimes}{\protect\scalebox{1}{\color{red}{$\times$}}}
\newcommand{\mytimesb}{\protect\scalebox{1}{$\times$}}
\newcommand{\myotimes}{\protect\scalebox{1.5}{$\otimes$}}
\newcommand{\myocircle}{\protect\scalebox{1.5}{$\circ$}}
\newcommand{\myrotbox}{\protect\scalebox{1.5}{\color{red}{$\diamond$}}}
\newcommand{\mybox}{\protect\scalebox{1}{$\Box$}}

\newcommand{\pder}[2][]{\ensuremath{\frac{\partial#1}{\partial#2}}}
\newcommand{\der}[2][]{\ensuremath{\frac{\text{d}#1}{\text{d}#2}}}

\section{Introduction}
\label{sec:introduction}
Finding a solution to large scale gradient-based topology optimization problems in a nested analysis and design approach requires a significant amount of computational effort due to the need to perform one or multiple expensive simulations per design iteration \citep{Amir2009}.
These simulations are often based on solving discretized governing partial differential equations with many Degrees of Freedom (DOFs).
A solution to such a linear system of equations is an assignment of values to the unknown variables at each DOF for given Boundary Conditions (BCs), such that all the equations are satisfied simultaneously.
In this setting, we call this set of variables the state, which may represent nodal values such as, \textit{e.g.}, temperature, displacement, pressure or velocity.

One may opt for a direct or iterative solution method to solve the involved system of linear equations; both methods consist of two subsequent steps. Direct methods require the (generally expensive) construction of factorization, and subsequently find the exact solution via comparatively inexpensive back-substitutions. In contrast, iterative methods require the construction of a preconditioner, and subsequently generate a sequence of improving approximate solutions until convergence. Relative cost of preconditioner construction and the iterative solution process depends on many factors, such as the type of preconditioner and condition number. For conciseness, we will hereafter denote the construction of factorization/preconditioner simply by \emph{preprocessing} and the back-substitution/iterative method simply by the \emph{solve}.

A priori to finding the solution to the governing equations, the DOFs are generally partitioned in two unique sets based on the applied BCs.
One may either specify (i) the magnitude of the solution at a DOF, that is the state is prescribed, or (ii) the magnitude of the applied discretized load, and thus the state is free.
These types are complementary---all DOFs for which the state is free will have an applied load (although possibly with zero magnitude), and vice versa.

\begin{figure}
	\centering
	\begin{subfigure}[t]{0.22\textwidth}
		\includegraphics[width=\textwidth]{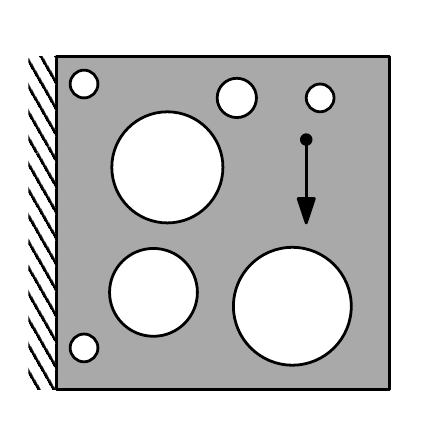}
		\caption{First scenario, providing state $\abf[]{\u}{}{1}{\i[1]}$. The state is prescribed on the left side and the body is subjected to a load $\abf[]{\f}{}{1}{\i[1]}$.}
		\label{fig:lc11a}
	\end{subfigure}
	~
	\begin{subfigure}[t]{0.22\textwidth}
		\includegraphics[width=\textwidth]{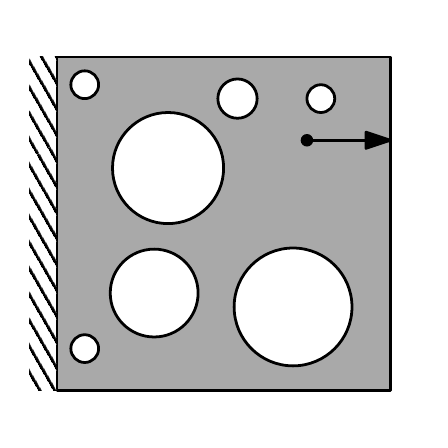}
		\caption{Second scenario, providing state $\abf[]{\u}{}{2}{\i[1]}$. The state is prescribed on the left side and the body is subjected to a load $\abf[]{\f}{}{2}{\i[1]}$.}
		\label{fig:lc12a}
	\end{subfigure}
	~
	\begin{subfigure}[t]{0.22\textwidth}
		\includegraphics[width=\textwidth]{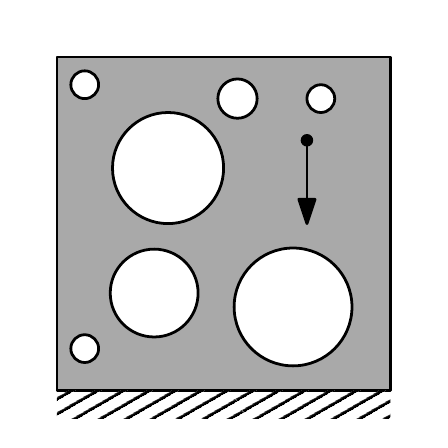}
		\caption{Third scenario, providing state $\abf[]{\u}{}{1}{\i[2]}$. The state is prescribed on the bottom and the body is, in this case a similar to \cref{fig:lc11a}, point load $\abf[]{\f}{}{1}{\i[2]}$.}
		\label{fig:lc21a}
	\end{subfigure}
	~
	\begin{subfigure}[t]{0.22\textwidth}
		\includegraphics[width=\textwidth]{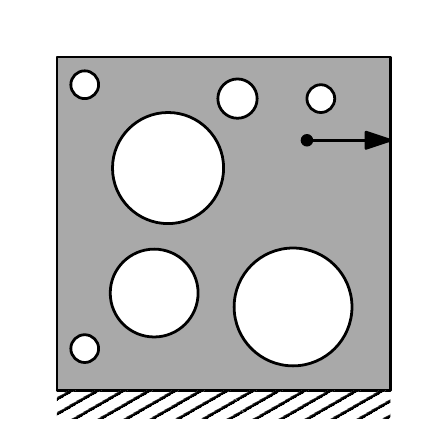}
		\caption{Fourth scenario, providing state $\abf[]{\u}{}{2}{\i[2]}$. The state is prescribed on the bottom and the body is, in this case a similar to \cref{fig:lc12a}, point load $\abf[]{\f}{}{2}{\i[2]}$.}
		\label{fig:lc22a}
	\end{subfigure}
	\caption{A structural topology to be optimized for four different scenarios, as shown in \cref{fig:lc11a,fig:lc12a,fig:lc21a,fig:lc22a}. As a result, this Multi-Partition Optimization Problem (MPOP) consists of two unique analysis sets, both comprising two analyses.}
	\label{fig:multilc}
\end{figure}

As \citet{Rozvany1993} pointed out almost three decades ago, real-world optimization problems generally involve multiple scenarios, see \textit{e.g.} \citet{Cavazzuti2010}.
We define a scenario as an engineering description of a specific loading condition, see for example the four scenarios depicted in \cref{fig:multilc}. Each scenario is translated to BCs, \textit{i.e.} prescribed state(s) and applied load(s), that represent the scenario as good as possible.
In principle, each scenario comprise a unique partitioning of DOFs.
However, scenarios of which only the magnitude of the applied BCs vary, but the type remains the same, share equal partitioning.

In this work, we define an analysis as the process of finding a solution to a properly constrained set of discretized governing equations---that is finding the state $\u \in \mathbb{R}^n$, with $n \in \mathbb{N}$ the number of nodal DOFs.
Each analysis relates to a single scenario, and thereto also comprises a unique partitioning of DOFs.
A combination of analyses with equal partitioning is hereafter called an analysis set.
A single analysis set thus requires finding the states to multiple loads (possibly with varying magnitude) with the same partitioning of DOFs.
These different states can generally be efficiently found, since only a single preprocessing process is required \citep{Diaz1992}.\footnote{For very large number of loads one can opt for efficient sampling schemes, see \textit{e.g.} \citet{Zhang2017,Zhang2020}.}

Topology optimization problem formulations involving responses (objective and constraints) that depend on the states of multiple analyses sets are called Multi-Partition Optimization Problems (MPOPs).
In contrast to problems involving a single partition, in MPOPs the preprocessing cannot be reused. 
This is evident in the design of, for example, multi-input-multi-output compliant mechanisms (see \textit{e.g.} \citet{Liu2009, Jin2017}).  

For clarification, consider a response, herafter denoted by $g \in \mathbb{R}$, that depends on the state(s) of $a \in \mathbb{N}$ analysis sets. Thus, $g = g\left[\abf[]{\U}{}{}{\i[1]},...,\abf[]{\U}{}{}{\i[a]}\right]$, with $\abf[]{\U}{}{}{\i[i]} \in \mathbb{R}^{\nsdof \times l\ii}$ and $l\ii \in \mathbb{N}$ a positive natural number of analyses contained in analysis set $i$, thus $\abf[]{\U}{}{}{\i[i]}:= \left[\abf[]{\u}{}{1}{\i[i]},...,\abf[]{\u}{}{l\ii}{\i[i]}\right]$ with each state $\mathbf{u}\ii \in \mathbb{R}^n$.
The states of an analysis set $\abf[]{\U}{}{}{\i[i]}$ implicitly depend on the design variables $\mathbf{x} \in \mathbb{R}^N$ via the discretized governing equations, that is
\begin{equation}
	\label{eq:kufshort}
	\abf[]{\K}{}{}{}\left[\mathbf{x}\right]\abf[]{\U}{}{}{\i[i]} = \abf[]{\F}{}{}{\i[i]}, \quad \forall ~ i \in ~ \mathbb{A}
\end{equation}
with the index set of all analyses sets $\mathbb{A} := \left\{i \in \mathbb{N} ~ | ~ i \leq a\right\}$, the design dependent system matrix
$\K\left[\mathbf{x}\right] \in \mathbb{R}^{n \times n}$ and loads $\abf[]{\F}{}{}{\i[i]} \in \mathbb{R}^{\nsdof \times l\ii}$.
Hereafter we assume the system matrix is symmetric and, without loss of generality, assume the applied loads are design independent. Note that, in \cref{eq:kufshort}, the states and loads of analysis set $i$ contain all DOFs; both free and prescribed states as well as applied and reaction loads.

A simple example of an MPOP is the optimization of a structure considering four scenarios, schematically shown in \cref{fig:multilc}. 
Note that between the four scenarios the set of DOFs for which the state is prescribed differs. 
Therefore, in this example, two unique analysis sets exist, comprising of two analyses each.
The response function $g$ of this artificial problem depends on the states of these two unique analysis sets, \textit{i.e.}  $g\left[\abf[]{\U}{}{}{\i[1]}, \abf[]{\U}{}{}{\i[2]}\right]$, where for this problem $\abf[]{\U}{}{}{\i[i]} =\left[ \abf[]{\u}{}{1}{\i[i]}, \abf[]{\u}{}{2}{\i[i]}\right]$ are the states (in this case the displacement) of analysis set $i$.


In spite of the fact that MPOPs are relevant for many applications, the computational efficiency of both the forward and sensitivity analyses involved in such problems has, to the best knowledge of the authors, not been investigated yet. 
With an increasing number of analysis sets, the computational effort to solve MPOPs increases quickly.
Since the preprocessing used in the analysis occurring in MPOPs is partition-specific, in a straightforward approach the computational effort of solving these problems scales with the number of analysis sets considered. On the contrary, within reasonable limits, the number of analyses per set is of less importance, given the fact that a solve requires negligible computational effort as compared to the preprocessing.

The computational efficiency of an analysis in the context of structural optimization can be increased by efficient iterative solution strategies (see \textit{e.g.} \citet{Evgrafov2008, Amir2014, Aage2015, Wu2016}), or approximated reanalysis, see \textit{e.g.} \citet{Kirsch1993, Amir2009, Amir2015}. Alternatively, one can opt for model-order reduction techniques, which reduce the computational effort of both analysis and optimization, see \textit{e.g.} \citet{Yoon2010a, Choi2019}. 
Typically, however, model-order reduction techniques introduce a loss of accuracy. 
One special reduction method is what is called static condensation amongst engineers, for which the reduction is \textit{exact} \citep{Guyan1965, Irons1965}. 
As will be shown hereafter, in particular in MPOPs the use of static condensation is attractive.
Moreover, as static condensation, similar to an analysis as previously defined, relies on finding the solution to a set of linear system of equations, it can also be combined with efficient iterative solution techniques and/or approximated reanalysis.

Static condensation was originally developed by \citet{Guyan1965, Irons1965}, hence the common nomenclature Guyan-Irons reduction scheme. 
The term `static' was added to differentiate from its application in dynamic analysis, for which the reduction scheme is no longer exact. 
Static condensation is regularly applied in the process of substructuring, also known as the Schur complement domain decomposition method \citep{Przemieniecki1963, Schmit1976, Noor1978}. 
This technique is based on the decomposition of the original structure into one or multiple sub-structures, often referred to as super-elements. 
The dimensionality reduction is achieved by a procedure that is closely related to Kron's method of tearing and Gaussian elimination \citep{Kron1955, Wilson1974}.

\begin{figure*}
	\begin{center}
		\begin{subfigure}[t]{0.22\textwidth}
			\includegraphics[width=\textwidth]{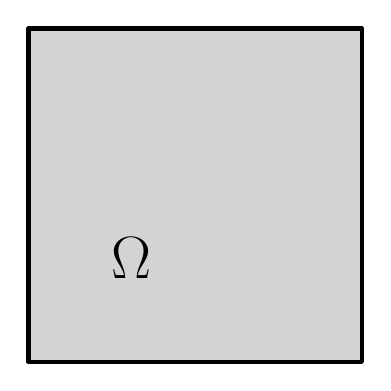}
			\caption{Bounded domain $\Omega$ occupied by a material with relevant physical properties, modeled as a continuum.}
			\label{fig:continuum}
		\end{subfigure}
		~
		\begin{subfigure}[t]{0.22\textwidth}
			\includegraphics[width=\textwidth]{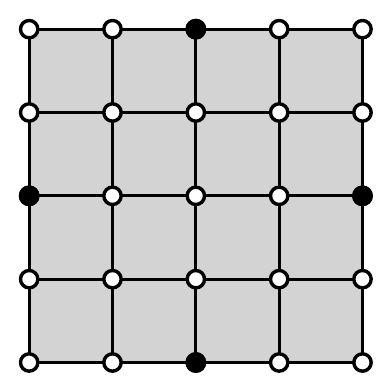}
			\caption{Domain from \cref{fig:continuum} discretized into 4 $\times$ 4 finite elements. The resulting nodes are split into primary (\mydot) and secondary (\myocircle) nodes. }
			\label{fig:discretization}
		\end{subfigure}
		~
		\begin{subfigure}[t]{0.22\textwidth}
			\includegraphics[width=\textwidth]{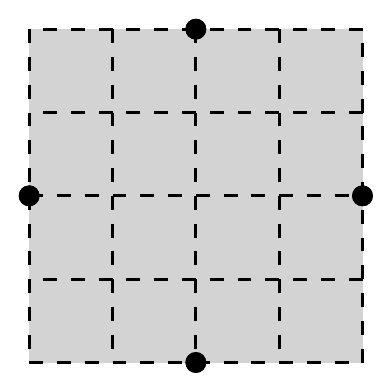}
			\caption{Superelement as obtained after static condensation. The information of the secondary DOFs is condensed into the primary DOFs.}
			\label{fig:cond}
		\end{subfigure}
		\caption{Step-wise illustration of static condensation. For simplicity, and without loss of generality, the domain is chosen to be a 2D square and discretized using a structured mesh of 4 $\times$ 4 quadrilateral finite elements. For simplicity of visualization, we consider all DOFs of a single node to be either primary or secondary DOFs. Primary nodes are indicated by a disk (\mydot), whereas secondary nodes are indicated by a circle (\myocircle).}
		\label{fig:condd}
	\end{center}
\end{figure*}

Consider the continuum from \cref{fig:continuum}.
After discretization of the governing equations in $n$ DOFs, as shown in \cref{fig:discretization}, the behaviour of the system can be described by the set of linear equations given by
\begin{equation}
	\label{eq:kuf}
	\K\left[\x\right] \u = \f,
\end{equation}
where, for simplicity, we consider a single analysis set consisting of a single analysis and thus $\u \in \mathbb{R}^n$ and $\mathbf{f} \in \mathbb{R}^n$.

The DOF index set $\mathbb{D} := \left\{i \in \mathbb{N} ~ | ~ i \leq n\right\}$ can be split in subset $\mathbb{M} \subset \mathbb{D}$ containing all indices of the primary DOFs, and its complement $\mathbb{S} := \mathbb{D} \setminus \mathbb{M}$ containing all indices of the secondary DOFs. The primary DOFs are those DOFs contained in the condensed model, whereas the secondary DOFs are eliminated.
After partitioning accordingly, one may rewrite the governing equations from \cref{eq:kuf} using oversets ($\wedge$) and ($\vee$) to represent the properties related to primary and secondary DOFs, respectively, as
\begin{equation}
	\label{eq:ms}
	\begin{bmatrix}
		\abf[\mm]{\K}{}{}{} & \abf[\ms]{\K}{}{}{}\\
		\abf[\sm]{\K}{}{}{} & \abf[\ss]{\K}{}{}{}
	\end{bmatrix}
	\begin{bmatrix}
		\abf[\m]{\u}{}{}{}\\
		\abf[\s]{\u}{}{}{}
	\end{bmatrix}
	=
	\begin{bmatrix}
		\abf[\m]{\f}{}{}{}\\
		\abf[\s]{\f}{}{}{}
	\end{bmatrix},
\end{equation}
where the principal sub-matrix $\abf[\ss]{\K}{}{}{}$ is \textit{nonsingular}, such that $\abf[\ss]{\K}{}{}{}$ is invertible \citep{Benscoter1948}. 
This partitioning is completely independent of the partitioning in free (f) and prescribed (p) DOFs.
In case we are solely interested in the state of the primary DOFs, we can write the secondary state $\abf[\s]{\u}{}{}{}$ in terms of the primary state $\abf[\m]{\u}{}{}{}$ using the second row of \cref{eq:ms}, which yields
\begin{equation}
	\label{eq:usum}
	\abf[\s]{\u}{}{}{} = -\abf[\ss]{\K}{}{}{-1} \abf[\sm]{\K}{}{}{}\abf[\m]{\u}{}{}{},
\end{equation}
assuming $\abf[\s]{\f}{}{}{}=\mathbf{0}$ for simplicity (the general case will be treated in \cref{sec:method}).
Subsequently, $\abf[\s]{\u}{}{}{}$ can be eliminated by substitution of \cref{eq:usum} in the first row of \cref{eq:ms}, which gives
\begin{equation}
	\left(\abf[\mm]{\K}{}{}{} - \abf[\ms]{\K}{}{}{}\abf[\ss]{\K}{}{}{-1}\abf[\sm]{\K}{}{}{}\right) \abf[\m]{\u}{}{}{} = \abf[\m]{\f}{}{}{}.
\end{equation}
The result of this condensation is a reduced system matrix $\tilde{\K} \in \mathbb{R}^{\nmdof \times \nmdof}$, with the number of primary DOFs $m := |\mathbb{M}|$, that exactly describes the behaviour of the full-order model.
This reduced system matrix follows as
\begin{equation}
	\label{eq:schur}
	\tilde{\K} := \abf[\mm]{\K}{}{}{} - \abf[\ms]{\K}{}{}{}\abf[\ss]{\K}{}{}{-1}\abf[\sm]{\K}{}{}{},
\end{equation}
and was originally proposed as the ``Schur complement of $\abf[\ss]{\K}{}{}{}$ in $\K$", first introduced in this form and notation by \citet{Haynsworth1968}. 
The eponymous adjective ``Schur" was chosen by Haynsworth because of the first use of the complement in the``Schur determinant lemma" published by \citet{Schur1917}. 

\begin{figure*}
	\centering
	\begin{subfigure}[t]{0.22\textwidth}
		\includegraphics[width=\textwidth]{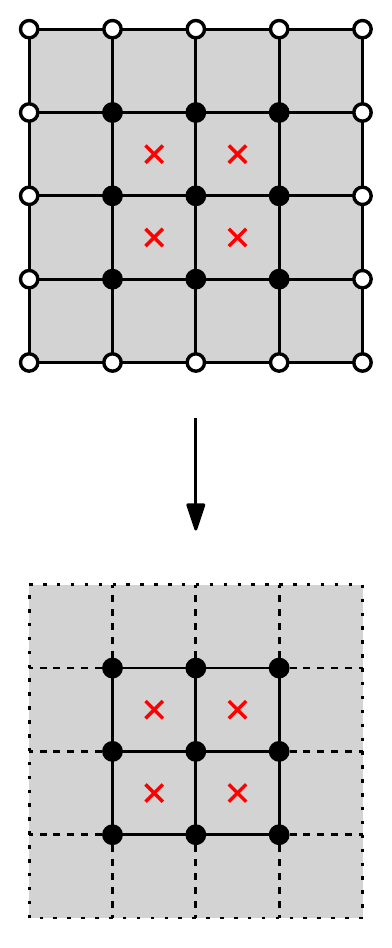}
		\caption{Condensation of the DOFs in non-design domain(s), noticeable by de lack of influence by design variables in those regions, reducing the dimensionality of the resulting design dependent system matrix.}
		\label{fig:nondes}
	\end{subfigure}
	~
	\begin{subfigure}[t]{0.22\textwidth}
		\includegraphics[width=\textwidth]{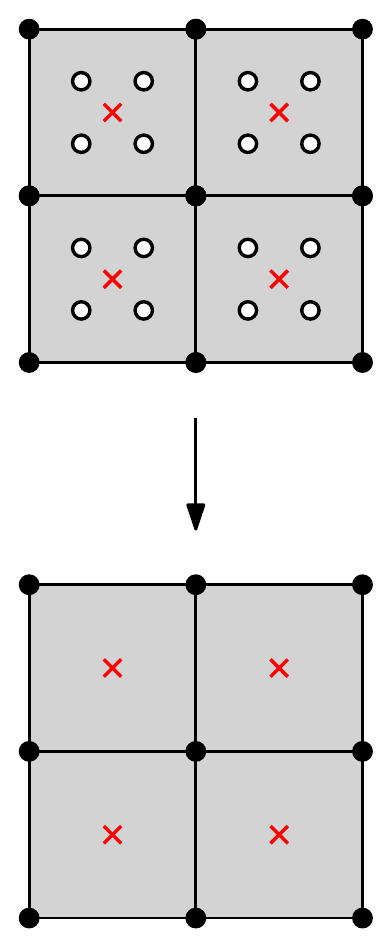}
		\caption{Condensation of four internal DOFs of four structural or finite elements, effectively reducing the dimensionality of each element.}
		\label{fig:element}
	\end{subfigure}
	~
	\begin{subfigure}[t]{0.22\textwidth}
		\includegraphics[width=\textwidth]{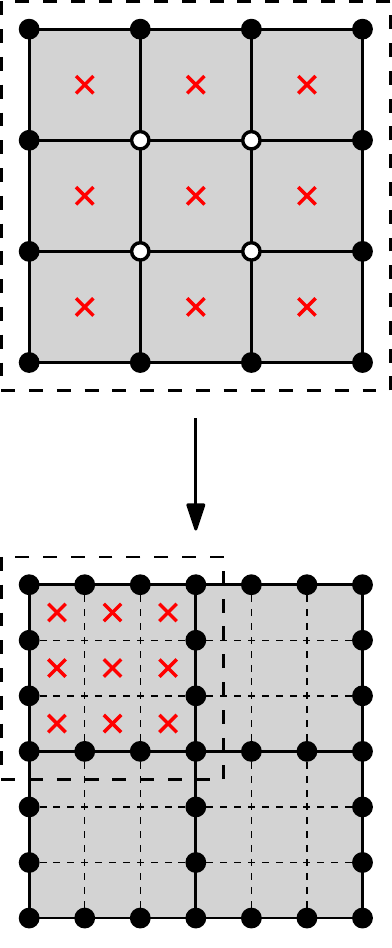}
		\caption{Condensation of the internal DOFs of a periodic unit cell. The resulting equivalent superelements of lower dimensionality are subsequently assembled in a 2 $\times$ 2 grid.}
		\label{fig:unitcell}
	\end{subfigure}
	~
	\begin{subfigure}[t]{0.22\textwidth}
		\includegraphics[width=\textwidth]{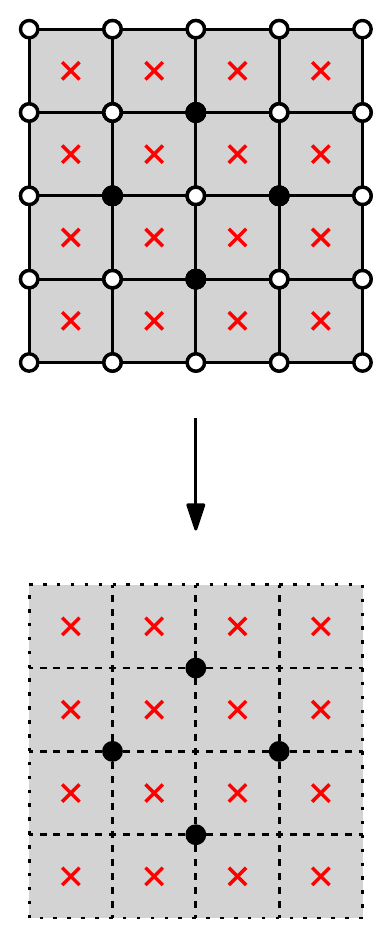}
		\caption{Condensation of all DOFs of a structure that do not \textit{directly} influence any of the responses, resulting in a reduced system matrix consisting solely of the primary DOFs.}
		\label{fig:res}
	\end{subfigure}
	\caption{Four different settings in which static condensation can be used in structural optimization. Top and bottom figures illustrate discretized domains before and after the condensation process, respectively. Similar to \cref{fig:condd}, disks (\mydot) and circles (\myocircle) denote primary and secondary nodes, respectively. Red crosses (\mytimes) indicate the location of design variables, which only locally influence the design.}
	\label{fig:staticc}
\end{figure*}

Static condensation has been applied in the field of gradient-based structural optimization in a variety of settings. 
Contributions include formulations that condense the
\begin{itemize}
	\item DOFs within non-design domain(s) \citep{Botkin1989,Gangadharan1990,Yang1996},
	\item internal DOFs of (complex) finite elements or components \citep{Yoon2007,Xia2012,Groen2017,Amir2019}, 
	\item internal DOFs of repetitive unit cell(s) \citep{Zhang2006,Wu2019}, and/or
	\item DOFs of a structure that do not occur in the responses, \textit{e.g} all DOFs except input and output in a compliant mechanism \citep{Wang2009a,Hasse2009}.
\end{itemize}
\Cref{fig:staticc} schematically illustrates the use of static condensation in each of these applications, which are discussed in more detail hereafter.

Early studies by \citet{Botkin1989, Gangadharan1990, Yang1996} aim to reduce the computational effort of both the Finite Element Analysis (FEA) and/or the sensitivity analysis of 3D structural optimization problems by static condensation of specific sub-domains. 
The topology in those domains does not depend on the design variables and hence remain unchanged during optimization, as shown in \cref{fig:nondes}. 
Since only a small part of the structure is subject to change, significant computational efficiency increase can be achieved if the dimensionality of the problem is reduced before the optimization, such that repetitive analysis of the non-design domain(s) is avoided. 
Since these early studies, this use of condensation in optimization has become commonplace and is also available in commercial software packages.

In the fields of geometrically non-linear, high-resolution and multi-component layout structural optimization, static condensation is effectively applied to eliminate internal DOFs from finite elements before global assembly, as schematically shown in \cref{fig:element}. 
This process reduces element complexity and decreases the dimensionality of the assembly system matrix, see \textit{e.g.} \citet{Yoon2007, Xia2012, Groen2017, Amir2019}.

In addition to the aforementioned, the computational effort can be further reduced in case the structure is periodic. 
Since all repeated cells share the same topology, only static condensation of a single cell is required, as shown in \cref{fig:unitcell}. 
The system matrix of the cell is condensed to reduce the dimensionality of the assembled system matrix that forms the periodic structure, see \textit{e.g.} \citet{Zhang2006, Wu2019}. 

A rather underexposed application of the static condensation procedure is its use in a unique set of optimization problem formulations of which the responses target a small set of DOFs, see \cref{fig:res}. 
The states itself will generally depend on the full set of design variables.
Representative examples are studies proposing problem formulations for the synthesis of compliant mechanisms with desired kinematics \citep{Wang2009a, Hasse2017, Li2019a}. 
The added value of using static condensation in these formulations is the natural way of describing the performance of the structure as a function of a limited set of primary DOFs. 
In these formulations, the scenarios generally consist of a large number of DOFs that share the same type of BCs.

Although the usefulness of static condensation in FEA has been common knowledge, the usefulness in sensitivity analysis and the impact on the computational efficiency of structural optimization problem formulations has been unexplored so far. \citet{Gangadharan1990} acknowledge and demonstrate the potential usefulness of static condensation to reduce the computational effort in the sensitivity analysis for problems in which the design variables only affect a small part of the domain.
However, the usage of static condensation shows high potential in terms of computational effort for solving MPOPs as well, as it could reduce the cost of both the forward and sensitivity analysis, as will be shown in \cref{sec:method}. 
The findings in the present work explore the potential in this field, ready to be applied to real-world problems.

This paper presents a novel method to efficiently solve MPOPs of which the responses involve a relatively small set of DOFs, see \cref{fig:res}. 
The method exploits the concept of static condensation in a way not reported before. 
We analyse and discuss the method's characteristics and limitations, provide a comparison between the proposed method and the state-of-the-art, and demonstrate the computational efficiency by numerical examples.

This work is focused on gradient-based topology optimization problems, involving linear state equations, solved via nested analysis and design. 
In addition, we assume (quasi-)static design independent loading conditions and a square and symmetric system matrix.

Note that it is not per se true that analyses of the reduced-order model is cheaper than analyses of the full-order model under all conditions.
The reduced system matrix is not only of reduced dimensionality, but also non-sparse. 
This loss of sparsity may impair any efficiency increase due to dimension reduction. 
Hence, it is critical to understand under what conditions to apply static condensation in the context of gradient-based structural optimization.

A generalized mathematical description of the method and corresponding sensitivity analyses will be given in \cref{sec:method,sec:sensitivities}. 
Both direct and adjoint sensitivity analysis aspects will be discussed. 
The description forms the basis for the formulation of an computational effort estimate. This estimate is based on the algorithmic complexity as a function of both the number of required analyses and the reduction of number of DOFs.
The resulting estimation of computational efficiency with respect to a straightforward approach, \textit{i.e.} performing all analyses without condensation, is outlined in \cref{sec:compeffort}. 
The computational effort estimate is verified by numerical experiments. 
\Cref{sec:examples} describes a comparison of computational efficiency for a variety of relevant examples. 
For simplicity, the proposed method is demonstrated by, but by no means limited to, 2D single-physics static topology optimization problems. 
This is followed by a description of an example implementation and closed by discussions and conclusions.

\section{Method}\label{sec:method}

This section describes two different approaches to solve a generalized MPOP, namely (i) a elementary approach, and (ii) the novel work presented herein---the condensation approach.
We start the description of the elementary approach of solving MPOPs, which will act as a reference.

\subsection{elementary approach}
\label{sec:clapp}

In the elementary approach one first defines the set of prescribed DOFs for each scenario. 
For each of the resulting sets of prescribed DOFs one can partition the system of linear equations accordingly. 
Subsequently, as in standard FEA one solves for the free state(s) \citep{Cook2001}.

The index set of all DOFs $\mathbb{D}$ can be split in subset $\mathbb{P}^{\i[i]} \subset \mathbb{D}$ containing the indices of all prescribed, and its complement $\mathbb{F}^{\i[i]} := \mathbb{D} \setminus \mathbb{P}^{\i[i]}$ containing the indices of all free DOFs of analysis set $i$, hereafter denoted by subscripts `p' and `f', respectively.
After partitioning accordingly, one may write the governing equations as
\begin{equation}\label{eq:eqofmotion}
	\begin{bmatrix}
		\abf[]{\K}{ff}{}{\i[i]} & \abf[]{\K}{fp}{}{\i[i]}\\
		\abf[]{\K}{pf}{}{\i[i]} & \abf[]{\K}{pp}{}{\i[i]}
	\end{bmatrix}
	\begin{bmatrix}
		\abf[]{\U}{f}{}{\i[i]}\\
		\abf[]{\U}{p}{}{\i[i]}
	\end{bmatrix}
	=
	\begin{bmatrix}
		\abf[]{\F}{f}{}{\i[i]}\\
		\abf[]{\F}{p}{}{\i[i]}
	\end{bmatrix}, \quad \forall ~ i \in \mathbb{A},
\end{equation}
where $\abf[]{\U}{f}{}{\i[i]}$ are the free state(s), $\abf[]{\U}{p}{}{\i[i]}$ the prescribed state(s), $\abf[]{\F}{f}{}{\i[i]}$ the applied load(s) and $\abf[]{\F}{p}{}{\i[i]}$ the reaction load(s) of analysis set $i$. Note that $\abf[]{\U}{f}{}{\i[i]} \in \mathbb{R}^{f^{\i[i]} \times \nrhs^{\i[i]}}$, where $f^{\i[i]}:= |\mathbb{F}^{\i[i]}|$. Where subscript `f' indicates the subset of the state and italic `$f$' indicates the corresponding dimensionality of this set.
The solutions to \cref{eq:eqofmotion}, $\abf[]{\U}{f}{}{\i[i]}$, now can be obtained by solving the system of linear equations
\begin{equation}
	\label{eq:y2}
	\abf[]{\K}{ff}{}{\i[i]} \abf[]{\U}{f}{}{\i[i]} = \abf[]{\F}{f}{}{\i[i]} - \abf[]{\K}{fp}{}{\i[i]} \abf[]{\U}{p}{}{\i[i]}, \quad \forall ~ i \in \mathbb{A}.
\end{equation}
Subsequently, if required, the reaction loads are obtained via
\begin{equation}
	\label{eq:fpfull}
	\abf[]{\F}{p}{}{\i[i]} = \abf[]{\K}{pf}{}{\i[i]} \abf[]{\U}{f}{}{\i[i]} + \abf[]{\K}{pp}{}{\i[i]} \abf[]{\U}{p}{}{\i[i]}, \quad \forall ~ i \in \mathbb{A}.
\end{equation}
Note that because of the uniqueness of each analysis set, following our definition, $\mathbb{F}^{\i[i]} \not= \mathbb{F}^{\i[j]}$ and $\mathbb{P}^{\i[i]} \not= \mathbb{P}^{\i[j]}$ for all $i\not=j \in \mathbb{A}$. 
Therefore, the number of times one has to preprocess a large system of equations equals $a$.

\subsection{Condensation approach}
\label{sec:coapp}
We propose a novel method, aimed at solving MPOPs, consisting of two subsequent steps: condensation and analyses.
First, one defines a set of primary DOFs that can fully describe the response function(s) and one condenses the remaining secondary DOFs to obtain a reduced system matrix. With this step, one basically solves for the secondary DOFs applying the same BCs for all analysis sets.
Subsequently, a variety of analyses is performed based on this reduced system matrix with varying partitions, originating from the BCs of the scenarios.
The latter step is fully equivalent to the elementary approach, however, as a result of the condensation, involves a much smaller system of linear equations. As such, one basically circumvents repeatedly solving for the secondary DOFs.
Key is to define the set of primary DOFs in an adequate manner to omit repetitive analyses of large systems of equations, by restricting the size of the set of primary DOFs.

\subsubsection{Definition of primary DOFs}
\label{sec:sets}
\begin{figure*}
	\begin{center}
		\begin{subfigure}[t]{0.23\textwidth}
			\includegraphics[width=\textwidth]{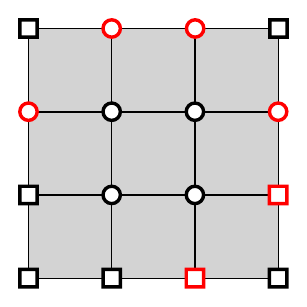}
			\caption{First example analysis. Note the unique subdivision of DOFs in free and prescribed, visualized by shape (\myocircle ~vs. \mybox) as well as the DOFs of interest, visualized by color (black vs. \textcolor{red}{red})}
			\label{fig:lc1}
		\end{subfigure}
		~
		\begin{subfigure}[t]{0.23\textwidth}
			\includegraphics[width=\textwidth]{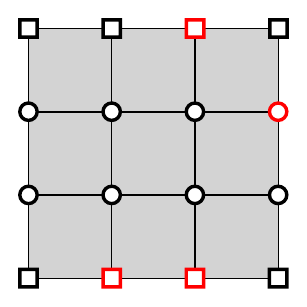}
			\caption{Second example analysis. Note the change in partitioning (free vs. prescribed) of DOFs according to the boundary conditions of this scenario as well as the change in DOFs of interest.}
			\label{fig:lc2}
		\end{subfigure}
		~
		\begin{subfigure}[t]{0.23\textwidth}
			\includegraphics[width=\textwidth]{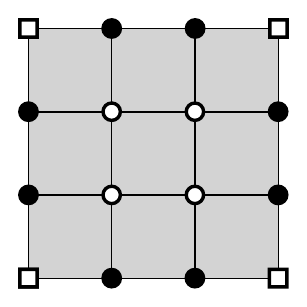}
			\caption{Subdivision of the set of DOFs $\mathbb{D}$ in disjoint sets $\setddof$ (\mybox), $\setcdof$ (\myocircle) and $\mathbb{M}$ (\mydot), according to the definitions in \cref{eq:setddof,eq:setmdof} and the partitioning as visualized in \cref{fig:lc1,fig:lc2}.}
			\label{fig:sets}
		\end{subfigure}
		\caption{Example of the discretized domain from \cref{fig:continuum} subjected to two distinct analyses; as shown in \cref{fig:lc1,fig:lc2}. The DOFs with prescribed states are denoted by a square (\mybox), the DOFs with free states by a circle (\myocircle) and primary DOFs by a black dot (\mydot) (analogously to \cref{fig:condd,fig:staticc}). DOFs of interest are coloured in red (\textcolor{red}{\myocircle}, \textcolor{red}{\mybox}). The combination of analyses leads to a subdivision in sets following \cref{eq:setddof,eq:setmdof}, as shown in \cref{fig:sets}.}
		\label{fig:buildsets}
	\end{center}
\end{figure*}

Consider again the structure as visualized in \cref{fig:continuum}. For the purpose of explanation, after discretization, we subject this arbitrary structure to two different scenarios as illustrated in \cref{fig:lc1,fig:lc2}. 
For each scenario one defines the DOFs with a prescribed state, indicated by a square in \cref{fig:buildsets}.
In addition, one defines per scenario which DOFs are of interest for the response function of the MPOP, in \cref{fig:buildsets} indicated in red. 
Note that both free and prescribed DOFs may be of interest. 
In what follows we explain how to find the smallest possible set of primary DOFs that can fully represent the behaviour of all scenarios \underline{and} contains all DOFs of interest for a given MPOP. 

For the condensation approach the index set of all DOFs $\setsdof$ is split in three disjoint parts, namely $\setsdof = \setddof \sqcup \setcdof \sqcup \setmdof$ as shown in \cref{fig:sets}. 
We define the proper index subset of prescribed secondary DOFs, $\setddof \subsetneq \setsdof$, to consist of all indices of DOFs for which the state is prescribed in all analyses excluding those that are a DOF of interest for any of the analyses.
Mathematically, this can be written as the intersection between all index sets of prescribed DOFs $\setqdof^{\i[i]}$ for which the DOFs are not a DOF of interest, \textit{i.e.}
\begin{equation}
	\label{eq:setddof}
	\setddof := \left\{x \in \left\{\cap_{i \in \setxdof} ~\setqdof^{\i[i]}\right\}~ |~ x \not\in \left\{\cup_{i\in\setxdof}~\setmdof^{\i[i]}\right\}\right\}.
\end{equation}
Here $\setmdof\ii$ consists of the indices of all DOFs of interest of analysis $i$.
Fully analogously, the index set of free secondary DOFs, $\setcdof \subsetneq \setsdof$, is defined as all indices of DOFs for which the state is free in all analyses, excluding those that are a DOF of interest for any of the analyses.
%
Based on these chosen definitions, the index set of primary DOFs, $\setmdof$, is simply defined as the union between all indices of DOFs that are of interest in any of the analysis sets and the indices of DOFs that change freedom between analysis sets, \textit{i.e.}
\begin{equation}
	\setmdof := \left(\cup_{i\in\setxdof}~\setmdof\ii\right) \cup \left(\left(\cup_{i\in\setxdof}~\setqdof\ii\right) \setminus \left(\cap_{i\in\setxdof}~\setqdof\ii\right)\right).
\end{equation}
Note this is also the complement of all secondary DOF indices, that is
\begin{equation}
	\label{eq:setmdof}
	\setmdof := \setsdof \setminus \left(\setddof \cup \setcdof\right).
\end{equation}
In many relevant applications, $1 < \nmdof \ll \nsdof$, meaning the number of primary DOFs is much lower than the total number of DOFs.

\subsubsection{Static condensation}
Considering a general MPOP, the response can now be rewritten as $g\left[ \abf[\m]{\U}{}{}{\i[1]}, \abf[\m]{\U}{}{}{\i[2]},...,\abf[\m]{\U}{}{}{\i[a]} \right]$, with $\abf[\m]{\U}{}{}{\i[i]}$ the state of the primary DOFs of analysis set $i$. 
From the system matrix $\K \in \mathbb{R}^{\nsdof \times \nsdof}$ one can obtain a reduced, but dense, system matrix $\tilde{\K} \in \mathbb{R}^{\nmdof \times \nmdof}$ through static condensation, which exactly represents the system behaviour in terms of the primary DOFs.
We can partition the system of linear equations in line with the subdivision in sets as proposed in \cref{sec:sets}, which gives
\begin{equation}\label{eq:systemofeq}
	\begin{bmatrix}
		\abf[\mm]{\K}{}{}{} & \abf[\ms]{\K}{f}{}{} & \abf[\ms]{\K}{p}{}{} \\
		\abf[\sm]{\K}{f}{}{} & \abf[\ss]{\K}{ff}{}{} & \abf[\ss]{\K}{fp}{}{} \\
		\abf[\sm]{\K}{p}{}{} & \abf[\ss]{\K}{pf}{}{} & \abf[\ss]{\K}{pp}{}{} 
	\end{bmatrix}
	\begin{bmatrix}
		\abf[\m]{\U}{}{}{}\\
		\abf[\s]{\U}{f}{}{}\\
		\abf[\s]{\U}{p}{}{}
	\end{bmatrix}
	=
	\begin{bmatrix}
		\abf[\m]{\F}{}{}{}\\
		\abf[\s]{\F}{f}{}{}\\
		\abf[\s]{\F}{p}{}{}
	\end{bmatrix},
\end{equation}
where $\abf[\s]{\F}{f}{}{} \in \mathbb{R}^{\abf[\s]{f}{}{}{} \times l}$ are the loads on the secondary DOFs and $\abf[\s]{\U}{p}{}{} \in \mathbb{R}^{\abf[\s]{p}{}{}{} \times \nrhs}$ the prescribed states of the secondary DOFs. Note that the primary states $\abf[\m]{\U}{}{}{}$ contain both free and prescribed DOFs in accordance with the scenarios imposed, as will become evident from \cref{sec:analysesonkbar}.
Here we have defined $\nrhs$ to be the sum of the loads in all analysis sets, \textit{i.e.}
\begin{equation}
	l:= \sum_{i \in \setxdof} \nrhs\ii.
\end{equation}
Hence, one can write for all $i \in \setxdof$
\begin{equation}
	\abf[\s]{\F}{f}{}{} = 
	\left[
	\abf[\s]{\F}{f}{}{\i[1]}  ,...,  \abf[\s]{\F}{f}{}{\i[a]}
	\right], \quad \text{with} \quad \abf[\s]{\F}{f}{}{\i[i]} = \left[\abf[\s]{\f}{f}{,1}{\i[i]}  ,...,  \abf[\s]{\f}{f}{,\nrhs\ii}{\i[i]}\right]\in \mathbb{R}^{\abf[\s]{f}{}{}{\i[i]} \times \nrhs\ii}.
\end{equation}
Analogously to \cref{eq:usum}, from the second row of \cref{eq:systemofeq}, we obtain
\begin{equation}\label{eq:ubc}
	\abf[\ss]{\K}{ff}{}{}\abf[\s]{\U}{f}{}{} = \abf[\s]{\F}{f}{}{} - \abf[\sm]{\K}{f}{}{} \abf[\m]{\U}{}{}{} - \abf[\ss]{\K}{fp}{}{}\abf[\s]{\U}{p}{}{}.
\end{equation}
Substitution of \cref{eq:ubc} into the first row of \cref{eq:systemofeq} gives
\begin{equation}
	\label{eq:redshort}
	\abf[]{\tilde{\K}}{}{}{}\abf[\m]{\U}{}{}{} = \abf[\m]{\F}{}{}{} + \abf[]{\tilde{\F}}{}{}{}.
\end{equation}
Here $\abf[]{\tilde{\K}}{}{}{}$ is the reduced system matrix defined as
\begin{equation}\label{eq:kbarfbar}
	\abf[]{\tilde{\K}}{}{}{} := \abf[\mm]{\K}{}{}{} - \abf[\ms]{\K}{f}{}{} \Zi, \quad \abf[]{\tilde{\K}}{}{}{} \in \mathbb{R}^{\nmdof \times \nmdof},
\end{equation}
and $\abf[]{\tilde{\F}}{}{}{}$ is the \textit{reduced load}, defined as
\begin{equation}
	\label{eq:fbar}
	\abf[]{\tilde{\F}}{}{}{} := \abf[\ms]{\K}{f}{}{}\Zc - \abf[\ms]{\K}{p}{}{}\abf[\s]{\U}{p}{}{}, \quad \abf[]{\tilde{\F}}{}{}{}  \in \mathbb{R}^{\nmdof \times l}.
\end{equation}
In order to obtain this reduced system of equations one has to solve the linear systems of equations
\begin{equation}
	\label{eq:solves}
	\abf[\ss]{\K}{ff}{}{}  \Zi = \abf[\sm]{\K}{f}{}{}, 
\end{equation}
and
\begin{equation}
	\label{eq:solvesq}
	\abf[\ss]{\K}{ff}{}{}  \Zc = \abf[\ss]{\K}{fp}{}{}\abf[\s]{\U}{p}{}{} - \abf[\s]{\F}{f}{}{},
\end{equation}
where we have introduced the `condensation states' $\Zi \in \mathbb{R}^{\ncdof \times \nmdof}$ and $\Zc \in \mathbb{R}^{\ncdof \times \nrhs}$. Note that the total number of analyses thus equals $\nmdof + \nrhs$.

The reduced system matrix $\abf[]{\tilde{\K}}{}{}{}$ consists of two parts: $\abf[\mm]{\K}{}{}{}$ containing the contributions of the primary DOFs, and $\abf[\ms]{\K}{f}{}{} \Zi$ containing the contributions of the secondary DOFs reflected onto the primary DOFs. 
Moreover, by definition of \cref{eq:schur}, $\abf[]{\tilde{\K}}{}{}{}$ is square and symmetric. 
One should add the appropriate reduced loads to the loads on the primary DOFs in all subsequent analyses based upon the reduced system matrix.
If the loads on the secondary DOFs $\abf[\s]{\F}{f}{}{} = \mathbf{0}$ and the prescribed state on the secondary DOFs $\abf[\s]{\U}{p}{}{} = \mathbf{0}$, then the reduced load vanishes. 

After solving \cref{eq:redshort}, as will be described in the next section, one can retrieve the secondary states via
\begin{equation}
	\label{eq:retubc}
	\abf[\s]{\U}{f}{}{} = - \left(\Zi \abf[\m]{\U}{}{}{} + \Zc\right),
\end{equation}
and subsequently the corresponding reaction loads on the secondary DOFs via
\begin{equation}
	\label{eq:retfbd}
	\abf[\s]{\F}{p}{}{} = \abf[\sm]{\K}{p}{}{} \abf[\m]{\U}{}{}{} + \abf[\ss]{\K}{pf}{}{} \abf[\s]{\U}{f}{}{} + \abf[\ss]{\K}{pp}{}{} \abf[\s]{\U}{p}{}{}.
\end{equation}

\subsubsection{Analyses based on the reduced system matrix}
\label{sec:analysesonkbar}
After condensation, the reduced system matrix $\abf{\tilde{\K}}{}{}{}$ can be used to repeatedly analyse the reduced governing equations, and obtain the required states and/or reaction loads for the response(s) of the considered MPOP. 

\begin{figure*}
	\begin{center}
		\begin{subfigure}[t]{0.23\textwidth}
			\includegraphics[width=\textwidth]{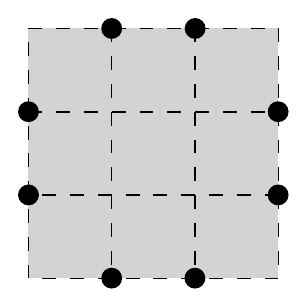}
			\caption{Superelement as obtained after condensation of the system in \cref{fig:sets}.}
			\label{fig:sets2}
		\end{subfigure}
		~
		\begin{subfigure}[t]{0.23\textwidth}
			\includegraphics[width=\textwidth]{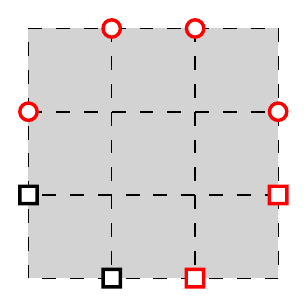}
			\caption{First partitioning of the primary DOFs. The partitioning is in line with the partitioning of the full system, see \cref{fig:lc1}.}
			\label{fig:lc12b}
		\end{subfigure}
		~
		\begin{subfigure}[t]{0.23\textwidth}
			\includegraphics[width=\textwidth]{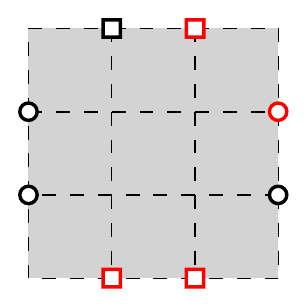}
			\caption{Second partitioning, in accordance with the scenario in \cref{fig:lc2}.}
			\label{fig:lc22b}
		\end{subfigure}
		\caption{Analyses on using the reduced system matrix, based on the example in \cref{fig:buildsets}. \cref{fig:sets2} shows the condensed domain of \cref{fig:sets}, resulting in a reduced system matrix. \cref{fig:lc12b,fig:lc22b} illustrate the analyses based upon the reduced system matrix, in accordance with the scenarios of \cref{fig:lc1,fig:lc2}.}
		\label{fig:buildsets2}
	\end{center}
\end{figure*}
Reconsider the system of equations given in \cref{eq:redshort}, \textit{i.e.}
\begin{equation}
	\label{eq:redshorti}
	\abf[]{\tilde{\K}}{}{}{} \abf[\m]{\U}{}{}{} = \abf[\m]{\F}{}{}{} + \abf[]{\tilde{\F}}{}{}{}.
\end{equation}
Note that the dimensionality of this system of equations is $\nmdof$.
Similarly to the elementary approach, \cref{sec:clapp}, one can split the indices in $\setmdof$ according to the type of boundary conditions, in the disjoint sets $\setmdof = \setfdof\ii \sqcup \setpdof\ii$ for each analysis set $i$. We define
\begin{align}
	\begin{aligned}
		\setfdof\ii & := \setmdof \cap \sethdof\ii,\\
		\setpdof\ii & := \setmdof \cap \setqdof\ii.
	\end{aligned}
\end{align}
Partitioning of the system of linear equations accordingly gives
\begin{equation}
	\label{eq:auf}
	\begin{bmatrix}
		\abf[]{\tilde{\K}}{ff}{}{\i[i]} & \abf[]{\tilde{\K}}{fp}{}{\i[i]}\\
		\abf[]{\tilde{\K}}{pf}{}{\i[i]} & \abf[]{\tilde{\K}}{pp}{}{\i[i]}
	\end{bmatrix}
	\begin{bmatrix}
		\abf[\m]{\U}{f}{}{\i[i]}\\
		\abf[\m]{\U}{p}{}{\i[i]}
	\end{bmatrix}
	=
	\begin{bmatrix}
		\abf[\m]{\F}{f}{}{\i[i]}\\
		\abf[\m]{\F}{p}{}{\i[i]}
	\end{bmatrix} +
	\begin{bmatrix}
		\abf[]{\tilde{\F}}{f}{}{\i[i]}\\
		\abf[]{\tilde{\F}}{p}{}{\i[i]}
	\end{bmatrix}, \quad \forall ~ i \in \setxdof.
\end{equation}
Note again, that this division of DOFs is unique for each set of analyses based upon $\abf[]{\tilde{\K}}{}{}{}$, since $\setfdof\ii \not= \setfdof^{\i[j]}$ for all $i\not=j \in \setxdof$. 
In addition to the partitioning, for each analysis set the appropriate loads are selected. 
Thus, whereas $\abf[]{\tilde{\F}}{}{}{} \in \mathbb{R}^{\nmdof \times l}$, $\abf[]{\tilde{\F}}{f}{}{\i[i]} \in \mathbb{R}^{\abf[\m]{f}{}{}{\i[i]} \times \nrhs\ii}$. 
The subdivision is schematically visualized in \cref{fig:buildsets2}, in line with the representation in \cref{fig:buildsets}.

The states of a variety of analyses sets can be obtained by solving the set of linear equations similar to the elementary approach, \textit{i.e.}
\begin{equation}
	\label{eq:y}
	\abf[]{\tilde{\K}}{ff}{}{\i[i]} \abf[\m]{\U}{f}{}{\i[i]} = \abf[\m]{\F}{f}{}{\i[i]} - \abf[]{\tilde{\K}}{fp}{}{\i[i]} \abf[\m]{\U}{p}{}{\i[i]} + \abf[]{\tilde{\F}}{f}{}{\i[i]}, \quad \forall~ i \in \setxdof.
\end{equation}
Note that system matrix $\abf[]{\tilde{\K}}{ff}{}{\i[i]}$ is square and non-singular for all $i$. 
If required, the reaction loads are obtained via
\begin{equation}
	\label{eq:fb}
	\abf[\m]{\F}{p}{}{\i[i]} = \abf[]{\tilde{\K}}{pf}{}{\i[i]}\abf[\m]{\U}{f}{}{\i[i]} + \abf[]{\tilde{\K}}{pp}{}{\i[i]} \abf[\m]{\U}{p}{}{\i[i]} - \abf[]{\tilde{\F}}{p}{}{\i[i]}, \quad \forall ~ i \in \setxdof.
\end{equation}

\subsection{Comparison of response evaluation}

Although the approaches obtain \emph{exactly} the same state and thereto response value(s), they differ in implementation, with critical consequences for the required computational effort.
In order to keep an overview of the large number of sets, subscripts and different operators presented, 
the most important operators, their dimensionality and reference to their first use can be found in \cref{tab:index}. 

\begin{table}
	\centering
	\caption{Commonly used operators, dimensionality and first appearance equation number.}
	\begin{tabular}{r|ll}
		\toprule
		Op & Dim ($\mathbb{R}$) & Eq.\\
		\midrule
		$\K$ & $\nsdof \times \nsdof$ & \ref{eq:kuf}\\
		$\abf[]{\K}{ff}{}{\i[i]}$ & $\abf[]{f}{}{}{\i[i]} \times \abf[]{f}{}{}{\i[i]}$ & \ref{eq:y2}\\
		$\abf[]{\U}{f}{}{\i[i]}$ & $\abf[]{f}{}{}{\i[i]} \times \nrhs\ii$ & \ref{eq:y2}\\
		\midrule
		$\abf[]{\tilde{\K}}{}{}{}$ & $\nmdof \times \nmdof$& \ref{eq:kbarfbar}\\
		$\abf[]{\tilde{\F}}{}{}{}$ & $\nmdof \times \nrhs$ & \ref{eq:fbar}\\
		$\abf[\ss]{\K}{ff}{}{}$ & $\abf[\s]{f}{}{}{} \times \abf[\s]{f}{}{}{}$ & \ref{eq:solves}\\
		$\Zi$ & $\abf[\s]{f}{}{}{} \times \nmdof$ & \ref{eq:solves}\\
		$\Zc$ & $\abf[\s]{f}{}{}{} \times \nrhs$ & \ref{eq:solves}\\
		\midrule
		$\abf[]{\tilde{\K}}{ff}{}{\i[i]}$ & $\abf[\m]{f}{}{}{\i[i]}  \times \abf[\m]{f}{}{}{\i[i]}$ & \ref{eq:y}\\
		$\abf[\m]{\U}{f}{}{\i[i]}$ & $\abf[\m]{f}{}{}{\i[i]} \times \nrhs\ii$ & \ref{eq:y}\\
		\bottomrule
	\end{tabular}
	\label{tab:index}
\end{table}

To highlight the differences, consider the pseudo codes in \cref{alg:clappforwards,alg:coappforwards,}, that describe the procedures to obtain the same response function using the elementary and condensation approach, respectively. 
Note the \textit{critical} difference between the solves in \cref{alg:clappforwards}: Line 2 and \cref{alg:coappforwards}: Line 10. To obtain the state(s) of the DOFs of interest, in the condensation approach, one solves for the state using the partitioned reduced system matrices $\abf[]{\tilde{\K}}{ff}{}{\i[i]} \in \mathbb{R}^{\abf[\m]{f}{}{}{\i[i]} \times \abf[\m]{f}{}{}{\i[i]}}$, where the number of free primary DOFs $\abf[\m]{f}{}{}{\i[i]}:= \nmdof - \abf[\m]{p}{}{}{\i[i]}$ and $\abf[\m]{p}{}{}{\i[i]} := |\setpdof\ii|$. In the elementary approach, one solves for the state using system matrices $\abf[]{\K}{ff}{}{\i[i]} \in \mathbb{R}^{\abf[]{f}{}{}{\i[i]}  \times \abf[]{f}{}{}{\i[i]} }$, where the number of free DOFs $\abf[]{f}{}{}{\i[i]} := \nsdof - \abf[]{p}{}{}{\i[i]} $ and $\abf[]{p}{}{}{\i[i]}  := |\setqdof\ii|$. Since $\abf[]{f}{}{}{\i[i]} \approx \nsdof$ and $\abf[\m]{f}{}{}{\i[i]} \approx m$, for sufficiently large systems, the dimensionality of the constrained system matrix $\abf[]{\K}{ff}{}{\i[i]}$ is much larger than the constrained reduced system matrix $\abf[]{\tilde{\K}}{ff}{}{\i[i]}$, that is $n \gg m$ for all analysis sets $i$.
The difference in computational effort of preprocessing a system matrix of dimensionality $n \times n$ as compared to $m \times m$ is large. 
The condensation approach only requires a single preprocessing of a large system matrix (\cref{alg:coappforwards}: Line 4), compared to $a$ preprocessing steps of a large system matrix for the elementary approach (\cref{alg:clappforwards}: Line 2). As a result, the computational effort to evaluate the response(s) can be highly reduced by using the condensation approach, when multiple scenarios are considered.

Upon close examination of the approaches, it becomes apparent the elementary approach contains the differences between the BCs of analysis sets in the large system of equations (\cref{alg:clappforwards}: Line 2), effectively solving \emph{repeatedly} for the large amount of secondary DOFs.
In contrast, the condensation approach basically uses some pre-processing (the static condensation in \cref{alg:coappforwards}: Line 4) to solve for the secondary DOFs. The different partitionings are `moved' to the system of equations of reduced dimensionality, see \cref{alg:coappforwards}: Line 10.

In terms of implementation, little is required to extend the elementary approach to the condensation approach. Apart from additional `administrative tasks' involving index sets, \textit{i.e.} \cref{alg:coappforwards}: Lines 1---3 and 8---9, the extension requires additional matrix-vector multiplications (Lines 5---6) and solves (Lines 4 and 10). Irrespective of the programming language of choice, those operations are also required for the elementary approach. The similarity between approaches in terms of implementation becomes all the more evident upon examination of the attached MATLAB code.

\begin{algorithm}
	\caption{elementary approach: response evaluation}
	\begin{algorithmic}[1]
		\FORALL{$i \in \mathbb{A}$}
		\STATE \textcolor{red}{solve} $\quad \abf[]{\K}{ff}{}{\i[i]}\abf[]{\U}{f}{}{\i[i]} = \abf[]{\F}{f}{}{\i[i]} - \abf[]{\K}{fp}{}{\i[i]}\abf[]{\U}{p}{}{\i[i]}, \quad \abf[]{\K}{ff}{}{\i[i]} \in \mathbb{R}^{\abf[]{f}{}{}{\i[i]} \times \abf[]{f}{}{}{\i[i]}} $
		\ENDFOR
		\RETURN $g \leftarrow g\left[\abf[]{\U}{}{}{\i[1]}, ..., \abf[]{\U}{}{}{\i[a]}\right]$
	\end{algorithmic}
	\label{alg:clappforwards}
\end{algorithm}

\begin{algorithm}
	\caption{Condensation approach: response evaluation}
	\begin{algorithmic}[1]
		\STATE $\setddof := \left\{x \in \left\{\cap_{i \in \setxdof} ~\setqdof^{\i[i]}\right\}~ |~ x \not\in \left\{\cup_{i\in\setxdof}~\setmdof^{\i[i]}\right\}\right\}$
		\STATE $\setcdof := \left\{x \in \left\{\cap_{i\in\setxdof} ~\sethdof\ii\right\}~ |~ x \not\in \left\{\cup_{i\in\setxdof}~\setmdof\ii\right\}\right\}$
		\STATE $\setmdof \leftarrow \setsdof \setminus \left(\setddof \cup \setcdof\right)$
		\STATE \textcolor{red}{solve} $\quad \abf[\ss]{\K}{ff}{}{} \begin{bmatrix}
			\Zi & \Zc\end{bmatrix} = \begin{bmatrix}
			\abf[\sm]{\K}{f}{}{} & 	\abf[\ss]{\K}{fp}{}{} \abf[\s]{\U}{p}{}{}  - 	\abf[\s]{\F}{f}{}{} 
		\end{bmatrix}, \quad  \abf[\ss]{\K}{ff}{}{} \in \mathbb{R}^{\abf[\s]{f}{}{}{}  \times \abf[\s]{f}{}{}{} }$
		\STATE $\abf[]{\tilde{\K}}{}{}{}  \leftarrow \abf[\mm]{\K}{}{}{} - \abf[\ms]{\K}{f}{}{}\Zi$
		\STATE $\abf[]{\tilde{\F}}{}{}{} \leftarrow \abf[\ms]{\K}{f}{}{}\Zc - \abf[\ms]{\K}{p}{}{}\abf[\s]{\U}{p}{}{}$
		\FORALL{$i \in \setxdof$}
		\STATE $\setfdof\ii \leftarrow \setmdof \cap \sethdof\ii$
		\STATE $\setpdof\ii \leftarrow \setmdof \cap \setqdof\ii$
		\STATE \textcolor{red}{solve} $\quad \abf[]{\tilde{\K}}{ff}{}{\i[i]} \abf[\m]{\U}{f}{}{\i[i]} = \abf[\m]{\F}{f}{}{\i[i]} - \abf[]{\tilde{\K}}{fp}{}{\i[i]} \abf[\m]{\U}{p}{}{\i[i]} + \abf[]{\tilde{\F}}{f}{}{\i[i]}, \quad \abf[]{\tilde{\K}}{ff}{}{\i[i]} \in \mathbb{R}^{\abf[\m]{f}{}{}{\i[i]}  \times \abf[\m]{f}{}{}{\i[i]} }$
		\ENDFOR
		\RETURN $g\leftarrow g\left[\abf[\m]{\U}{}{}{\i[1]},...,\abf[\m]{\U}{}{}{\i[a]}\right]$
	\end{algorithmic}
	\label{alg:coappforwards}
\end{algorithm}

\section{Sensitivity analysis}
\label{sec:sensitivities}
Design sensitivities are essential for gradient-based structural optimization schemes, as they provide the basis for the design update. 
This section describes the sensitivity analysis of all relevant entries used in the condensation approach, as described in \cref{sec:coapp}. 
Since the second part of the condensation approach, as described in \cref{sec:analysesonkbar}, also covers the elementary approach we omit a separate description of the sensitivities for the elementary approach.

Making use of the chain rule of differentiation, the full derivative of a response function $g\left[\abf[\m]{\U}{}{}{}\left[\tilde{\K}\left[\K\left[\mathbf{x}\right]\right], \tilde{\F}\left[\K\left[\mathbf{x}\right]\right]\right]\right]$ to design variable $x_k$ reads as
\begin{equation}
	\der[g]{x_k} = \der[g]{\abf[\m]{\U}{}{}{}} : \left(\pder[{\abf[\m]{\U}{}{}{}}]{\tilde{\K}}:\pder[\tilde{\K}]{\K} +\pder[{\abf[\m]{\U}{}{}{}}]{\tilde{\F}}:\pder[\tilde{\F}]{\K}  \right): \der[\K]{x_k}.
\end{equation}
Herein, we assume $\der[g]{\abf[\m]{\U}{}{}{}}$ and $\der[\K]{x_k}$ are known, since they are problem and response dependent. In addition, for the discussion in this section, we assume design independent loads.
This section is subdivided into the separate sensitivity analysis of 
\begin{enumerate}
	\item the reduced system matrix $\tilde{\K}$ to the system matrix $\K$, 
	\item the reduced load(s) $\tilde{\F}$ to the system matrix $\K$, and
	\item the primary state(s) $\abf[\m]{\U}{}{}{}$ to the reduced system matrix $\tilde{\K}$ and load(s) $\tilde{\F}$.
\end{enumerate}
The remaining sensitivity analyses, \textit{e.g} the sensitivities of the reaction loads $\abf[\m]{\F}{p}{}{}$ with respect to the reduced system matrix, are not considered critical for understanding of the method and subsequent conclusions. 
However, for completeness those and additional sensitivities can be found in \cref{sec:sensitivityfull}.

In the following, the sub-matrices of the system matrices in \cref{eq:systemofeq,eq:auf} are written using selection matrices. Some examples are:
\begin{align}\label{eq:submatrices}
	\begin{aligned}
		\abf[]{\K}{pf}{}{} &:=\abf[]{\mathbf{S}}{p}{}{\tran} \K\abf[]{\mathbf{S}}{f}{}{}\\ 
		\abf[\mm]{\K}{}{}{} &:=\abf[\m]{\mathbf{S}}{}{}{\tran} \K\abf[\m]{\mathbf{S}}{}{}{}\\ 
		\abf[\ms]{\K}{f}{}{} &:=\abf[\m]{\mathbf{S}}{}{}{\tran} \K\abf[\s]{\mathbf{S}}{f}{}{}\\ 
		\abf[]{\tilde{\K}}{pf}{}{} &:=\abf[]{\tilde{\mathbf{S}}}{p}{}{\tran} \tilde{\K}\abf[]{\tilde{\mathbf{S}}}{f}{}{}
	\end{aligned}
\end{align}
These highly sparse rectangular selection matrices $\accentset{(.)}{\mathbf{S}}_{(.)}$ can be used to select parts from the system matrix, which will prove useful during the sensitivity analysis.

\subsection{Sensitivities of the reduced system matrix}
\label{sec:sesensitivities}
Consider a response $g\big[\tilde{\K}\left[\K\right]\big]$, with
\begin{equation*}
	\abf[]{\tilde{\K}}{}{}{} := \abf[\mm]{\K}{}{}{} - \abf[\ms]{\K}{f}{}{} \Zi. \tag{\ref{eq:kbarfbar} revisited}
\end{equation*}
Here the system matrix is a function of design variables $\x$, \textit{i.e.} $\K\left[\mathbf{x}\right]$. 
However, for demonstration of the method we are solely interested in the sensitivities of the reduced system matrix with respect to the system matrix. 
Therefore, this dependency is omitted from here onward. 
For conciseness, we use the full differential.
The full derivative of the response function can be written as
\begin{equation}
	\label{eq:dA2}
	\text{d}g\big[\tilde{\K}\left[\K\right]\big] =  \der[g]{\tilde{\K}} : \der[\tilde{\K}]{\K}:\text{d}\K,
\end{equation}
where $\der[g]{\tilde{\K}}$ is the sensitivity of the response with respect to $\tilde{\K}$ and $\text{d}\K$ the sensitivity of the system matrix. 
In the following we assume both are known.
The aim is to find $\der[\tilde{\K}]{\K}$, \textit{i.e.} the sensitivities of the reduced system matrix with respect to the system matrix.

The total derivative of $\tilde{\K}$ is given by
\begin{equation}
	\label{eq:dAk}
	\text{d}\tilde{\K}\left[\K\right] = \text{d}\abf[\mm]{\K}{}{}{} - \text{d}\abf[\ms]{\K}{f}{}{} \Zi + \Zit\text{d}\abf[\ss]{\K}{ff}{}{} \Zi - \Zit\text{d} \abf[\sm]{\K}{f}{}{}.
\end{equation}
Here we have made use of
\begin{equation}
	\label{eq:dZ}
	\text{d}\Zi = -\abf[\ss]{\K}{ff}{}{-1} \text{d}\abf[\ss]{\K}{ff}{}{}  \Zi + \abf[\ss]{\K}{ff}{}{-1} \text{d}\abf[\sm]{\K}{f}{}{}.
\end{equation}
Substitution of \cref{eq:submatrices} into \cref{eq:dAk} yields
\begin{equation}
	\label{eq:kkbar}
	\text{d}\tilde{\K}\left[\K\right] = \left(\abf[\m]{\mathbf{S}}{}{}{} - \abf[\s]{\mathbf{S}}{f}{}{}\Zi \right)\tran\text{d}\K \left(\abf[\m]{\mathbf{S}}{}{}{} - \abf[\s]{\mathbf{S}}{f}{}{}\Zi \right).
\end{equation}
By rewriting\footnote{Here we introduce the dyadic product, also known and equivalent to the outer or tensor product. The dyadic product of two vectors $\mathbf{a} \in \mathbb{R}^n$ and $\mathbf{b} \in \mathbb{R}^m$ can be represented as $\mathbf{a} \otimes \mathbf{b} \equiv\mathbf{a}\mathbf{b}\tran \equiv  a_i b_j$. For matrices $\mathbf{A} \in \mathbb{R}^{n \times c}$ and $\mathbf{B} \in \mathbb{R}^{m \times c}$, the dyadic product is defined as $\mathbf{A} \otimes \mathbf{B} \equiv \mathbf{A}\mathbf{B}\tran \equiv A_{ij} B_{kj}$.}\textsuperscript{,}\footnote{\label{foot:atbc}$\mathbf{A}\tran \mathbf{B} \mathbf{C} \equiv A_{ij} B_{jk} C_{ki} = A_{ji} C_{ik} B_{jk} = \left(\mathbf{A}\mathbf{C}\tran\right): \mathbf{B} \equiv \left(\mathbf{A} \otimes \mathbf{C} \right) : \mathbf{B}$} \cref{eq:kkbar} we obtain the full derivative 
\begin{equation}
	\label{eq:xdfdkx}
	\text{d}\tilde{\K}\left[\K\right]= \left(\ses \otimes \ses \right) : \text{d}\K,
\end{equation}
with
\begin{equation}
	\label{eq:xghz}
	\ses := \abf[\m]{\mathbf{S}}{}{}{} - \abf[\s]{\mathbf{S}}{f}{}{}\Zi.
\end{equation}
Thus
\begin{equation}
	\label{eq:xx}
	\der[\tilde{\K}]{\K} = \ses \otimes \ses,
\end{equation}
which consists only of terms that have been previously computed in the condensation. 
The following section further explores this property of the reduced system matrix.
The full step-by-step derivation, both for the direct and adjoint approach, can be found in \cref{sec:senskbarkbdirect,sec:senskbarkbadjoint}.

\subsubsection{On the self-adjointness of the reduced system matrix}
\label{sec:adjoint}

Following the adjoint sensitivity approach \citep{Arora1979, Belegundu1986} we define
\begin{equation}
	\label{eq:initadj}
	\lag\big[\tilde{\K}\left[\K\right], \abf[\sm]{\bm{\Lambda}}{f}{}{}\big] = g \big[\tilde{\K}\left[\K\right]\big] + \abf[\sm]{\bm{\Lambda}}{f}{}{} : \left(\abf[\ss]{\K}{ff}{}{} \Zi - \abf[\sm]{\K}{f}{}{} \right),
\end{equation}
where $\abf[\sm]{\bm{\Lambda}}{f}{}{}$ are the Lagrange multipliers.
The sensitivities of the Lagrangian (extensive derivation can be found in \cref{sec:senskbarkbadjoint}) are obtained via
\begin{align}
	\label{eq:dLkbar}
	\begin{aligned}
		\text{d}\lag =\der[g]{\tilde{\K}} : \left(\abf[\m]{\mathbf{S}}{}{}{\tran}\text{d}\K\ses\right)+ \abf[\sm]{\bm{\Lambda}}{f}{}{} : \left(\abf[\s]{\mathbf{S}}{f}{}{\tran}\text{d}\K\ses\right).
	\end{aligned}
\end{align}
To find $\abf[\sm]{\bm{\Lambda}}{f}{}{}$, one has to solve the following system of equations:
\begin{equation}
	\label{eq:Lambda}
	\abf[\ss]{\K}{ff}{}{}\abf[\sm]{\bm{\Lambda}}{f}{}{} = \abf[\sm]{\K}{f}{}{} \der[g]{\tilde{\K}}.
\end{equation}
By suitable choice of calculation order \citep{Vanderplaats1980}, this can be rewritten as
\begin{equation}
	\label{eq:lambdaaa}
	\abf[\sm]{\bm{\Lambda}}{f}{}{} = \abf[\ss]{\K}{ff}{}{-1} \left(\abf[\sm]{\K}{f}{}{} \der[g]{\tilde{\K}}\right) =  \Zi \der[g]{\tilde{\K}}.
\end{equation}
In contrast to the general adjoint approach, where the calculation order is determined by the number of responses versus the number of design variables, the most suitable choice in this case is the same under all conditions, since $\Zi$ has already been calculated.
Thus, the sensitivities of the reduced system matrix can be calculated without solving an additional system of linear equations.
Since both the state and adjoint fields are exactly identical we can conclude that---for a symmetric system matrix---an optimization problem with response $g\left[\tilde{\K}\right]$ has a \textit{self-adjoint} operator \citep{Belegundu1986, Rozvany1993} \underline{independent} of the self-adjointness of the operator(s) $\tilde{\K}$ is acting upon.
As a result, no extra preprocessing is required to compute the sensitivities of $\tilde{\K}$. 
This has critical consequences with respect to the computational efficiency of the condensation approach. 

\subsection{Sensitivities of the reduced load}
\label{sec:serl}
Since the required steps to obtain the sensitivities of the reduced load are very similar to those in the previous section we solely present the result here.  
The full derivation---both for the direct and adjoint approach---can be found in \cref{sec:sensfbarkbdirect,sec:sensfbarkbadjoint}. 

Recall the definition of the reduced load, namely
\begin{equation*}
	\label{eq:fbart}
	\tilde{\F}\left[\K, \abf[\s]{\F}{f}{}{}, \abf[\s]{\U}{p}{}{}\right] := \abf[\ms]{\K}{f}{}{} \Zc - \abf[\ms]{\K}{p}{}{}\abf[\s]{\U}{p}{}{}. \tag{\ref{eq:fbar} revisited}
\end{equation*}
Let us consider a response function $g\big[\tilde{\F}\left[\K\right]\big]$, hence omitting the dependency on the free secondary load(s) $\abf[\s]{\F}{f}{}{}$ and prescribed secondary state(s) $\abf[\s]{\U}{p}{}{}$ here. 
Substitution of \cref{eq:submatrices} into \cref{eq:fbar} and application of either direct or adjoint sensitivity analysis yields
\begin{equation}
	\label{eq:dfdkb}
	\pder[\tilde{\F}]{\K} = \ses \otimes \bes
\end{equation}
with
\begin{equation}
	\label{eq:Y}
	\bes:= \abf[\s]{\mathbf{S}}{f}{}{} \Zc - \abf[\s]{\mathbf{S}}{p}{}{} \abf[\s]{\U}{p}{}{}.
\end{equation}
Note that \cref{eq:dfdkb} solely involves quantities that have been calculated in the condensation process and requires therefore negligible additional computational effort.

\subsection{Sensitivities of the state}
In an MPOP the response function, often $g\left[\abf[\m]{\U}{f}{}{}\left[\abf[]{\tilde{\K}}{}{}{},\abf[]{\tilde{\F}}{}{}{}, \abf[\m]{\F}{f}{}{}{}, \abf[\m]{\U}{p}{}{}{}, \right]\right]$, generally depends on multiple states, \textit{i.e.} $ g \left[\abf[\m]{\U}{f}{}{\i[1]}, ,..., \abf[\m]{\U}{f}{}{\i[a]}\right]$ is obtained by solving
\begin{equation*}
	\abf[]{\tilde{\K}}{ff}{}{\i[i]} \abf[\m]{\U}{f}{}{\i[i]} = \abf[\m]{\F}{f}{}{\i[i]} - \abf[]{\tilde{\K}}{fp}{}{\i[i]} \abf[\m]{\U}{p}{}{\i[i]} + \abf[]{\tilde{\F}}{f}{}{\i[i]}, \quad \forall~ i \in \setxdof. \tag{\ref{eq:y} revisited}
\end{equation*}
Therefore, one is required to determine the total derivative $\text{d}g\left[\abf[\m]{\U}{f}{}{\i[i]}\left[\abf[]{\tilde{\K}}{}{}{},\abf[]{\tilde{\F}}{}{}{\i[i]} ,\abf[\m]{\F}{f}{}{\i[i]}, \abf[\m]{\U}{p}{}{\i[i]}\right]\right]$ for all $i \in \setxdof$. 
For clarity, the superscript $i$ is omitted from here on. 
Moreover, for conciseness we omit the dependency on the free loads $\abf[\m]{\F}{f}{}{}{}$ and prescribed states $\abf[\m]{\U}{p}{}{}{}$ in this section. 
Although this section focuses on the condensation approach, the following is similar for the elementary approach, where $g\left[\abf[]{\U}{f}{}{}{}\left[\K\right]\right]$.  Complete and step-by-step derivations can be found in \cref{sec:sensubf}.

The full derivative of the response function is given by
\begin{equation}
	\label{eq:dguf}
	\text{d}g\left[\abf[\m]{\U}{f}{}{}\left[\tilde{\K}, \tilde{\F}\right]\right] = \der[g]{\abf[\m]{\U}{f}{}{} } :  \left( \pder[{\abf[\m]{\U}{f}{}{}}]{\tilde{\K}}:\text{d}\tilde{\K} + \pder[{\abf[\m]{\U}{f}{}{}}]{\tilde{\F}}:\text{d}\tilde{\F}\right).
\end{equation}
We aim to find $\pder[{\abf[\m]{\U}{f}{}{}}]{\tilde{\K}}$ and $\der[{\abf[\m]{\U}{f}{}{}}]{\tilde{\F}}$, while assuming $\text{d}\tilde{\K}$, $\text{d}\tilde{\F}$ and $\der[g]{\abf[\m]{\U}{f}{}{}}$ to be known.
The full derivative $\text{d} \abf[\m]{\U}{f}{}{}$ can be obtained by differentiation of \cref{eq:y}, which gives
\begin{equation}
	\label{eq:duf}
	\text{d}{\abf[\m]{\U}{f}{}{}}\left[\tilde{\K},\tilde{\F}\right] = \abf[]{\tilde{\K}}{ff}{}{-1} \left(\text{d}\abf[]{\tilde{\F}}{f}{}{} - \text{d}\abf[]{\tilde{\K}}{ff}{}{} {\abf[\m]{\U}{f}{}{}} - \text{d}\abf[]{\tilde{\K}}{fp}{}{} \abf[\m]{\U}{p}{}{}\right).
\end{equation}
Substitution of \cref{eq:submatrices} and simplification yields
\begin{equation}
	\pder[g]{\tilde{\K}} = -\abf[]{\tilde{\mathbf{S}}}{f}{}{} \lambdas \otimes \abf[\m]{\U}{}{}{}
\end{equation}
and
\begin{equation}
	\pder[g]{\tilde{\F}} = \abf[]{\tilde{\mathbf{S}}}{f}{}{} \lambdas,
\end{equation}
where
\begin{equation}
	\label{eq:yjadjoint2}
	\abf[]{\tilde{\K}}{ff}{}{}\lambdas = \der[g]{{\abf[\m]{\U}{f}{}{}}}
\end{equation}
is the adjoint equation and $\abf[\m]{\bm{\Lambda}}{}{}{\i[i]} \in \mathbb{R}^{m \times l\ii}$ is the state of the adjoint problem of analysis set $i$. 
Thus, to obtain $\text{d} {\abf[\m]{\U}{f}{}{}}\left[\tilde{\K}, \tilde{\F}\right]$ one requires one additional system of linear equations to be solved \underline{per analysis set}.

The total number of adjoint loads to solve an MPOP theoretically equals the number of responses times the number of states, that is the total number of adjoint loads
\begin{equation}
	b = hl = h \sum_{i=1}^{a} l\ii,
\end{equation}
with $h$ the number of responses of the MPOP.
However, if the right-hand-side is linear dependent on the applied load, the problem is self-adjoint and solving the corresponding adjoint equation can be omitted \citep{Rozvany1993}. This is for example the case in classical compliance minimization \citep{MartinPhilipBendsoe1988}.
What is more, in practice, the majority of responses depend on a single state or multiple states of a single analysis set, that is many adjoint loads $\der[g_k]{{\abf[\m]{\u}{}{j}{\i[i]}}}$ are zero. Thereto, generally the total number of adjoint loads satisfies
\begin{equation}
	b = \sum_{j=1}^h\sum_{i=1}^{a} b_{j}\ii,
\end{equation}
with $b_{j}\ii$ the number of non-zero adjoint loads of response function $j$ to the states of analysis set $i$, and thus $b\ii \leq l\ii$.

\subsection{Comparison of sensitivity analysis}
To compare the sensitivity analyses, consider the pseudo-codes \Cref{alg:clappsenss,alg:coappsenss}, respectively for the elementary and condensation approach. Herein, it is assumed both applied loads and prescribed states are design independent. 
Note the difference in dimensionality of the systems of equations to be solved in the second step of the algorithms, that is \cref{alg:clappsenss}: Line 2 and \cref{alg:coappsenss}: Line 2. It may be assumed here no additional preprocessing is required, as the preprocessing of the response evaluation can be reused. As a result, in the elementary approach, for each response $g_j$, $b_j$ adjoint states of substantial dimensionality are to be solved for. In contrast, assuming $m \ll n$, the computational effort of the sensitivity analysis in the condensation approach is negligible. Most notably is the fact that \emph{zero} solves of substantial dimensionality are required, see \cref{sec:adjoint}.

Apart from some additional matrix-vector operations, see \cref{alg:coappsenss}: Lines 4---5, implementation of the sensitivity analysis of the condensation approach is straightforward. Also note the similarity between the structure of \cref{alg:clappsenss}: Line 4 and \cref{alg:coappsenss}: Line 6.

Full step-by-step sensitivity analysis of $g\left[\abf[\m]{\U}{f}{}{}\left[\K, {\abf[\s]{\F}{f}{}{}},{\abf[\s]{\U}{p}{}{}}, {\abf[\m]{\F}{f}{}{}},{\abf[\m]{\U}{p}{}{}}\right]\right]$ and $g\left[\abf[\m]{\F}{p}{}{}{}\left[\K, {\abf[\s]{\F}{f}{}{}},{\abf[\s]{\U}{p}{}{}}, {\abf[\m]{\F}{f}{}{}},{\abf[\m]{\U}{p}{}{}}\right]\right]$, without those assumptions, can be found in \cref{sec:sensubf,sec:sensfbp}. In addition, the derivation of sensitivities for responses involving the state or reaction load of secondary DOFs, that is $g\left[{\abf[\s]{\U}{f}{}{}}\left[\K, {\abf[\s]{\F}{f}{}{}},{\abf[\s]{\U}{p}{}{}}, {\abf[\m]{\F}{f}{}{}},{\abf[\m]{\U}{p}{}{}}\right]\right]$ and $g\left[\abf[\s]{\F}{p}{}{}\left[\K, {\abf[\s]{\F}{f}{}{}},{\abf[\s]{\U}{p}{}{}}, {\abf[\m]{\F}{f}{}{}},{\abf[\m]{\U}{p}{}{}}\right]\right]$, can be found in \cref{sec:sensubc,sec:sensfbd}. All resulting sensitivities are summarized in \cref{tab:sensitivities} of \cref{sec:senssum}.

\begin{algorithm}
	\caption{elementary approach: sensitivity analysis}
	\begin{algorithmic}[1]
		\FORALL{$i \in \setxdof$}
		\STATE \textcolor{red}{solve} $\quad {\abf[]{\K}{ff}{}{\i[i]}} {\abf[]{\bm{\Lambda}}{f}{}{\i[i]}} = \pder[g]{{\abf[]{\U}{f}{}{\i[i]}}}, \quad  \abf[]{\K}{ff}{}{\i[i]} \in \mathbb{R}^{\abf[]{f}{}{}{\i[i]}  \times \abf[]{f}{}{}{\i[i]} }$
		\ENDFOR
		\RETURN $\pder[g]{x_k} \leftarrow -\sum_{i \in \setxdof} \sum_{j=1}^{l\ii} \abf[]{\bm{\lambda}}{}{j}{\i[i]}\cdot \pder[\K]{x_k} {\abf[]{\u}{}{j}{\i[i]}}$
	\end{algorithmic}
	\label{alg:clappsenss}
\end{algorithm}

\begin{algorithm}
	\caption{Condensation approach: sensitivity analysis}
	\begin{algorithmic}[1]
		\FORALL{$i \in \setxdof$}
		\STATE \textcolor{red}{solve} $\quad \abf[]{\tilde{\K}}{ff}{}{\i[i]} \abf[\m]{\bm{\Lambda}}{f}{}{\i[i]} = \pder[g]{{\abf[\m]{\U}{f}{}{\i[i]}}}, \quad \abf[]{\tilde{\K}}{ff}{}{\i[i]} \in \mathbb{R}^{\abf[\m]{f}{}{}{\i[i]}  \times \abf[\m]{f}{}{}{\i[i]} }$
		\ENDFOR
		\STATE $\ses \leftarrow \abf[\m]{\mathbf{S}}{}{}{} - \abf[\s]{\mathbf{S}}{f}{}{}\Zi$
		\STATE $\bes \leftarrow \abf[\s]{\mathbf{S}}{f}{}{} \Zc - \abf[\s]{\mathbf{S}}{p}{}{} \abf[\s]{\U}{p}{}{}$
		\RETURN $\pder[g]{x_k} \leftarrow \sum_{i \in \setxdof}\sum_{j=1}^{l\ii}  \left(\ses\abf[\m]{\bm{\lambda}}{}{j}{\i[i]}\right) \cdot \pder[\K]{x_k} \left({\abf[]{\mathbf{b}}{}{j}{\i[i]}} - \ses{\abf[\m]{\u}{}{j}{\i[i]}}\right)$
	\end{algorithmic}
	\label{alg:coappsenss}
\end{algorithm}

\section{Computational effort}
\label{sec:compeffort}
This section describes the computational effort involved in using the condensation approach compared to the elementary approach. 
First, predominant factors that influence the computational effort are discussed. 
Subsequently, the computational effort of both approaches is estimated using algorithmic complexity analysis, which yields an estimate of the computational efficiency of using the condensation approach with respect to the elementary approach as a function of the predominant factors.

\subsection{Predominant factors}
A critical factor influencing the computational effort of the condensation approach is the number of free secondary DOFs versus number of primary DOFs.
The number of free secondary DOFs $\ncdof$ depends on the system size $\nsdof$, the number of prescribed secondary DOFs $\nddof$ and the number of primary DOFs $\nmdof$ via $\ncdof = \nsdof - \nddof - \nmdof$. The number of free secondary DOFs $\ncdof$ is the primary contributor to the total effort of calculating the Schur complement, as it directly influences the cost of the preprocessing of $\abf[\ss]{\K}{ff}{}{}$.
The number of primary DOFs $\nmdof$ determines the number of right-hand-sides of the system of linear equations solved in the static condensation process. If the number of primary DOFs increases, the number of secondary DOFs decreases with equal amount.
In addition, the number of primary DOFs $\nmdof$ influences the computational effort of the subsequent analysis sets based upon $\tilde{\K}$.
The computational effort of subsequent analysis sets increases rapidly as $\nmdof$ increases, since $\tilde{\K}$ is a dense system matrix as compared to the generally sparse system matrix $\K$.

The similarity between analysis sets or, equivalently, the amount of overlap of sets $\sethdof\ii$ for all analysis sets is crucial. 
The higher the overlap of those sets, the smaller $\setmdof$ can be, and hence, the lower becomes the computational effort of both the static condensation and subsequent analysis sets.
The presence of non-zero prescribed values on the secondary DOFs, \textit{i.e.} $\abf[\s]{\U}{p}{}{} \not=\mathbf{0}$ and/or non-zero applied loads on the secondary DOFs $\abf[\s]{\F}{f}{}{} \not = \mathbf{0}$ introduces $\nrhs$ extra right-hand-sides for the static condensation and introduces a reduced load term.
In that case, the number of prescribed secondary DOFs $\nddof$ influences the effort of matrix-vector product $\abf[\ss]{\K}{fp}{}{}\abf[\s]{\U}{p}{}{}$ and $\abf[\ms]{\K}{p}{}{}\abf[\s]{\U}{p}{}{}$ if $\abf[\s]{\U}{p}{}{} \not=\mathbf{0}$, and influences $\ncdof$ via $\ncdof = \nsdof - \nmdof -\nddof$.
However, for the majority of large scale problems $\nddof \ll \nsdof$ and the additional effort becomes negligible.

Another clearly important factor influencing the computational effort is the number of analysis sets $a$.
The computational effort of the elementary approach increases linearly with $a$. 
In contrast, in the condensation approach $a$ has little influence on the computational effort if $\nmdof \ll \nsdof$, as analysis sets based upon $\tilde{\K}$ are---in that case---of negligible effort. 
In contrast, when $\nmdof \rightarrow \nsdof$ both the static condensation and all analysis sets based upon $\tilde{\K}$ become expensive, due to the high number of right-hand-sides in the static condensation and the large dense matrix $\tilde{\K}$.
For the elementary approach, most important is the number of free DOFs for each analysis set $\abf[]{f}{}{}{\i[i]}$, which determines the effort of the preprocessing of $\abf[]{\K}{ff}{}{\i[i]}$. 
Note that generally $\abf[]{f}{}{}{\i[i]} \approx \abf[]{f}{}{}{\i[j]}$ for all $i,j \in \setxdof$. 
The number of loads $\nrhs\ii$ for each analysis set will have a positive influence on the efficiency of the condensation approach with respect to the elementary approach for $\nmdof \ll \nsdof$, but \textit{vice versa} if $\nmdof \rightarrow \nsdof$.

Lastly, implementation-related factors play a role, \textit{e.g.} type of solver, compiled vs. interpreted code, vectorization and parallelization. 
Also the sparsity, bandedness and band structure of $\K$ influence the type of preprocessing and, hence, to a lesser extent impact the computational effort of both the condensation and/or elementary approach.
Note that some of the dominant parameters mostly affect the computational effort of the condensation approach (such as the number of analyses $a$ or dimensionality of $\tilde{\K}$), whereas others will have a comparable influence on both approaches, \textit{e.g.} the type of solver or specific implementation. 

\subsection{Prognosis}
Based on the previous discussion we expect the computational efficiency of the condensation approach with respect to the elementary approach to be high when $\nmdof \ll \nsdof$ and to increase linearly with $a$. Assuming $\nmdof \ll \nsdof$ and $a \geq 1$, we expect the source of computational efficiency of the condensation approach with respect to the elementary approach to be twofold:
\begin{enumerate}
	\item The condensation approach requires a single preprocessing of the system matrix, as opposed to $a$ for the elementary approach.\footnote{Note that in the elementary approach, if using a direct solution method, one also has to store $a$ system matrices of notable size in memory, as opposed to a single matrix in the condensation approach.} All subsequent analysis sets based on the reduced system matrix have relatively negligible computational effort. Thus, the more analysis sets are based upon $\tilde{\K}$, the higher will be the computational efficiency with respect to the elementary approach.
	\item As described in \cref{sec:adjoint}, the sensitivities of the reduced system matrix are self-adjoint, independent of the response. All information required to compute those sensitivities has been calculated during the condensation process. What remains is the computational effort of the adjoint problems based on the reduced system matrix. The problem may still consist of multiple non-self-adjoint responses, but when $\nmdof \ll \nsdof$, the associated adjoint problems are negligibly small compared to the condensation process required for the response evaluation. The sensitivity analysis of any response depending solely on the reduced system matrix is therefore of negligible computational effort.
\end{enumerate}

For a fixed number of analysis sets $a$, an increase in $\nmdof$ causes an increase in computational efficiency due to the low cost of repetitive analysis sets based upon the reduced system matrix in contrast to the high cost of each extra analysis set in the elementary approach. However, when $\nmdof$ becomes large, the static condensation becomes computationally more demanding. Of even more importance is the increase in dimensionality of the reduced system matrix, which is a dense matrix. Thus, it is expected that there exists a globally optimal ratio of $\nmdof$ to $\nsdof$ as a function of $a$ for any combination of other parameters that maximizes the computational efficiency of the condensation approach with respect to the elementary approach.

\subsection{Algorithmic complexity analysis}
To fairly compare the computational efficiency of the condensation approach to the elementary approach, we investigate the computational gain $\Xi$, defined as the relative decrease in computational effort through the use of the condensation approach compared to the elementary approach, as a function of the predominant parameters. We define the theoretical computational effort as the number of floating point operations (FLOPs) $\beta$ required and assume that the relative FLOP count of the two algorithms is an approximation that gives sufficient accuracy to estimate the resulting relative Central Processing Unit (CPU) time $t$. 
In what follows it will be shown that this assumption is valid under the chosen conditions. 
The computational gain can be based upon the theoretical estimated FLOP count, denoted by $\Xi_{\tilde{\beta}}$ or, alternatively, based on experimentally measured run time, hereafter denoted by $\Xi_\text{t}$.
First, this section presents an algorithmic complexity analysis, followed by a numerical study in \cref{sec:examples}.

Note that computational gain is measured in relative FLOPs or relative runtime, whereas---in some cases---storage considerations might be critical. 
Similar to runtime, the memory usage is expected to be high for large systems with low to moderate number of primary DOFs, and degrade with increasing number of primary DOFs.
A thorough investigation of this aspect is considered out of scope of the present work.

\subsubsection{Assumptions on FLOP count of predominant operations}
\label{ssec:assfc}

The FLOP count for the operations involved in the condensation and elementary approach are based upon the following assumptions:
\begin{itemize}
	\item Arithmetic operations with individual elements have computational complexity $\mathcal{O}(1)$, as is the case with fixed-precision floating-point arithmetic. CPU time is equal for all individual operations (addition, multiplication etc.).
	\item Linear systems of equations are either solved using direct (factorization plus back-substitution) or iterative (preconditioning plus iterative solve) solution methods.
	\item The computational effort is dominated by the preprocessing and solve. Thus, all other operations are of negligible cost. In other words, we assume problems to be sufficiently large, that is $n \gg 1$.
	\item The properties (sparsity, symmetry, band density, definiteness, band structure etc.) of all system matrices are known. Based on this the solution method is chosen, as well as specific implementation(s).
	\item To calculate the gain $\Xi_{\tilde{\beta}}$ the following are assumed to be known: problem size $\nsdof$ and number of analysis sets $a$ as well as the index sets $\sethdof\ii$ and $\setqdof\ii$, indices of the DOFs of interest $\setmdof\ii$, prescribed state $\abf[]{\U}{p}{}{\i[i]}$ and applied load $\abf[]{\F}{f}{}{\i[i]}$ for each analysis set $i \in \setxdof$.
\end{itemize}

The accuracy of $\Xi_{\tilde{\beta}}$  with respect to $\Xi_\text{t}$ under the given assumptions, as well as the range of validity will be shown by numerical experiments in \cref{sec:prob1}.

\subsubsection{Theoretical approximated gain}
Based on the above assumptions we can approximate the expected gain of using the condensation approach compared to the elementary approach for a MPOP with a single response function by
\begin{equation}
	\Xi_{\tilde{\beta}}\left[\nsdof,\nmdof,a,l,b\right] \approx \frac{\sum_i^{a} \tilde{\beta}_{\text{s}}\left[\nsdof,\nrhs\ii + b\ii\right]}{\tilde{\beta}_\text{s}\left[\nsdof -\nmdof, \nmdof\right] + \sum_i^{a} \tilde{\beta}_{\text{d}}\left[\nmdof, \nrhs\ii + b\ii\right]},
	\label{eq:gainnma}
\end{equation}
where $\tilde{\beta}_\text{s}\left[i,j\right]$ and  $\tilde{\beta}_\text{d}\left[i,j\right]$ are the cost (in number of FLOPs) of solving respectively a sparse and dense system of equations of size $i$ with $j$ right-hand-sides. 
The parameter $b\ii$ denotes the number of non-zero right-hand-sides for the adjoint solve of analysis set $i$. The gain is based upon the response evaluation \emph{and} sensitivity analysis, where we assume the preprocessing is reused where possible.
In addition, we assumed that for a sufficiently large problems the ratio of prescribed DOFs to free DOFs becomes so small, that it is valid to assume the size of the analysis equals the system size, that is $f\ii \approx \nsdof$, $\abf[\m]{f}{}{}{\i[i]} \approx \nmdof$ for all $i \in \setxdof$.

In order to show the general applicability of the estimation, the estimation is validated on numerical examples using two different approaches to solve the system of equations:
\begin{enumerate}
	\item Solving both sparse and dense system of equations using a direct solution method, and
	\item Solving sparse system of equations using an iterative solution method and dense systems using a direct solution method.
\end{enumerate}

\subsubsection{Theoretical complexity of solution methods}
The Cholesky factorization (CHOL) with subsequent back-substitutions is the most commonly applied direct solution method (both for sparse and dense matrices) to solve symmetric positive definite system matrices of small to moderate size.
The FLOP count of a CHOL (without reordering) and subsequent back-substitutions of a symmetric sparse banded system matrix depends on dimensionality $\nsdof$, bandwidth $k$ and number of right-hand-sides $l$. Note that the bandwidth for discretized PDEs depends on the number of dimensions, \textit{e.g.} in 2D $k = \sqrt{n}$, whereas in 3D $k = n^{\frac{2}{3}}$ \citep{Yano2012}. In this work we will---without loss of generality---study 2D examples, and hence approximate
\begin{equation}
	\tilde{\beta}_\text{s}\left[n,l\right] = n^2 + 2ln^\frac{3}{2},
\end{equation}
for a sparse system, and
\begin{equation}
	\tilde{\beta}_\text{d}\left[n,l\right] = \frac{1}{3}n^3 + 2ln
\end{equation}
for a dense system \citep{boyd2004convex}.

One of the most common iterative solution methods to solve symmetric positive definite system of equations is to first construct an incomplete Cholesky factorization (ICHOL) preconditioner and then find the solution using a Conjugate Gradient (CG) algorithm. 
The computational cost of the preconditioning (for simple preconditioners) is assumed negligible compared to the cost of the iterative solve. Therefore, in what follows, this contribution is neglected. 
This assumption also emphasizes the usability of the method for a wide range of solution methods.
Each CG iteration involves multiple operations, of which one matrix-vector product is dominating. For a $k$-banded matrix, the total FLOP count per iteration is thus $2nk$ \citep{saad2003iterative}.
The required number of CG iterations is approximately proportional to $\sqrt{n}$.
The FLOP count of a CG solve on a discretized 2D PDE can thus be estimated as
\begin{equation}
	\label{eq:iterative}
	\tilde{\beta}_\text{s}\left[n,l\right] = 2ln^2.
\end{equation}
Note that both functions $\tilde{\beta}_\text{s}\left[n,l\right]$ and $\tilde{\beta}_\text{d}\left[n,l\right]$ and coefficients therein depend on both type of solver, implementation and computer efficiency, and are thus subjected to change depending on the solver of choice.

\section{Numerical examples}
\label{sec:examples}

This section demonstrates the computational efficiency of the condensation approach for two different example problems, each representative for a specific type of MPOP. 

We consider the following distinct examples:
\begin{enumerate}
	\item A self-adjoint thermal conductivity MPOP where, by definition of the problem, the number of analysis sets equals the number of DOFs of interest, that is $a = \nmdof$, and each additional DOF of interest introduces an additional analysis set based on the reduced system matrix.
	\item A non-self-adjoint displacement-based compliant mechanism design MPOP. The problem emphasizes the effect of reduction of computational effort of the sensitivity analysis.
\end{enumerate}
These examples are defined to easy allow variations of parameters to emphasize the validation of the efficiency improvements. As a result, the examples are conceptual and lack direct usefulness in practice.
Both examples are topology optimization problems, parametrized in a 2D structured grid of $N$
bilinear quadrilaterals with a single design variable per element. The constrained nonlinear optimization problem is defined as
\begin{equation}
	\begin{aligned}
		\underset{\x}{\text{minimize}} && \quad & g_0\left[\x\right]\\
		\text{subject to}&&& g_j\left[\x\right] \leq 0, &j &=1,...,m\\
		&&& \mathbf{x} \in \mathbb{X}^N
	\end{aligned}
\end{equation}
where $\x$ is the field of design variables, $\mathbb{X} := \left\{x_i \in \mathbb{R} ~ |~ 0  <  x_i \leq 1 \right\}$.
The design variable field is blurred using a filter \citep{Bruns2001} with a radius of $r = 2.0$ finite elements. The element conductivity/stiffness is related to the filtered design variable via the modified SIMP\footnote{We omit further explanation as these terms, arguments and implications are considered common knowledge within the field.}  interpolation function \citep{Sigmund2007} with a constant penalty value of $p=3.0$.

\subsection{Problem 1}
\label{sec:prob1}
Consider the well known heat conduction topology optimization problem in a two-dimensional domain \citep{Bendsoe2004}. We investigate a variation of this problem with multiple analysis sets, by considering efficient heat conduction between $\nmdof$ randomly distributed DOFs.  In each of the $a$ analysis sets, a single DOF acts as a heat sink (temperature prescribed to zero), and a heat load of random magnitude (between \SI{0}{\joule\per\second} and \SI{1}{\joule\per\second}) is applied to the $m-1$ remaining primary DOFs. 

\begin{figure*}
	\centering
	\includegraphics[width=\textwidth]{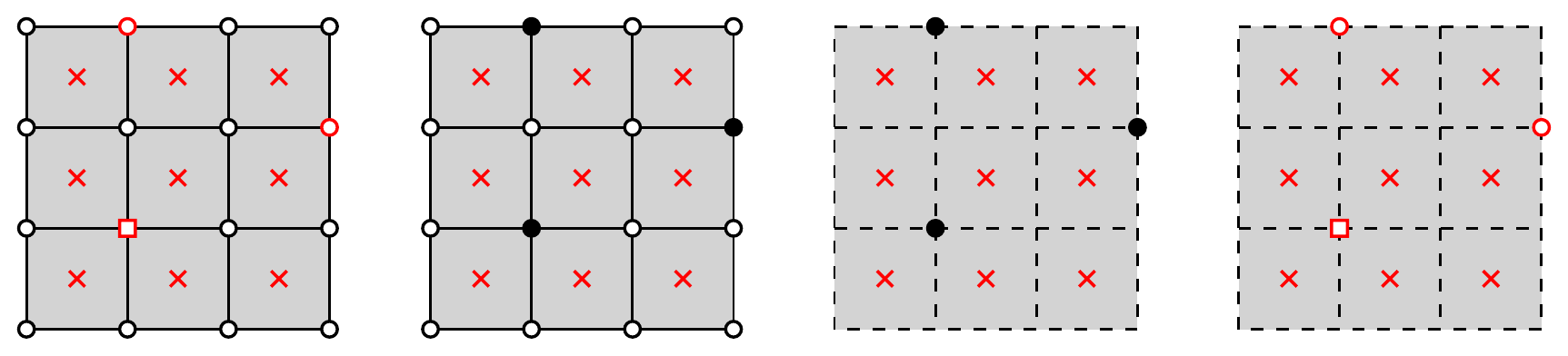}
	\caption{Schematic representation of Problem 1 for $N = 9$ and $\nmdof = 3$. The DOFs of interest are chosen arbitrarily. From left to right: one of the three analyses as solved for in the elementary approach, subdivision into primary and secondary DOFs, resulting system after static condensation and, the equivalent analysis as solved for in the condensation approach. The DOFs with prescribed states are denoted by a square (\mybox), the DOFs with free states by a circle (\myocircle) and primary DOFs by a black dot (\mydot) (analogously to \cref{fig:condd,fig:staticc}). DOFs of interest are coloured in red (\textcolor{red}{\myocircle}, \textcolor{red}{\mybox}).}
	\label{fig:heatschematic}
\end{figure*}

\begin{figure}
	\centering
	\includegraphics[width=0.45\textwidth]{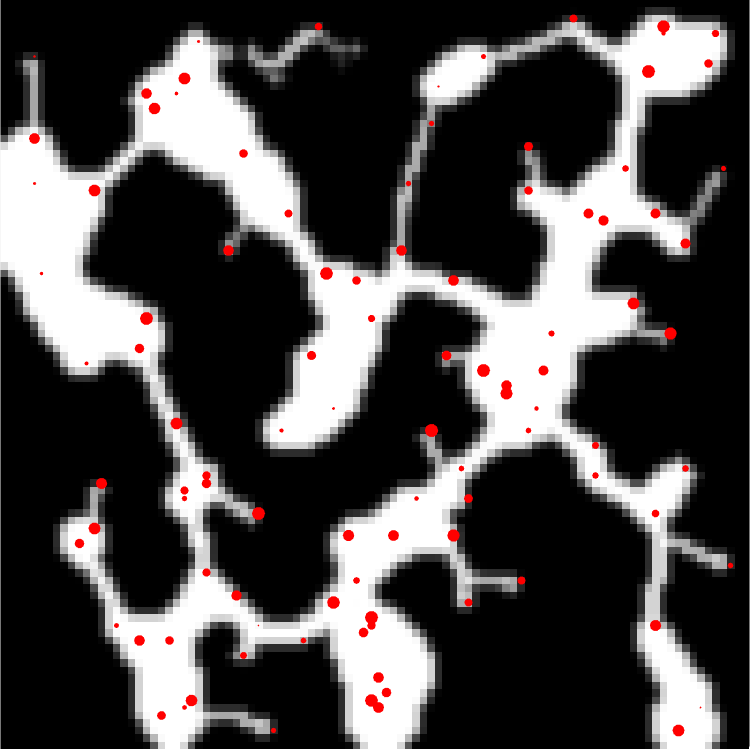}
	\caption{Resulting topology of Problem 1 as schematically sketched in \cref{fig:heatschematic}. The white material is highly conductive, whereas the black material has low conductivity. For this MPOP the number of variables $N = 100 \times 100$, the number of structural DOFs $\nsdof = 10000$, and the number of primary DOFs and analyses $\nmdof = a = 100$. The maximum amount of material usage $\overline{v} = 0.2$. The location and magnitude (size of red dot) of the primary DOFs and applied loads are indicated in red.}
	\label{fig:design}
\end{figure}

Following the drawing conventions as previously used, a representation of this problem is sketched in \cref{fig:heatschematic}. For this MPOP, the number of primary DOFs is equal to the number of distinct analysis sets, \textit{i.e.} $\nmdof = a$. 
The aim is to find a design that efficiently conducts heat between all primary DOFs, for each considered scenario. Thereto, we minimize the overall conductive resistance for all considered analyses, that is
\begin{equation}
	g_0\big[\abf[]{\U}{}{}{\i[1]}\left[\x\right], ..., \abf[]{\U}{}{}{\i[a]}\left[\x\right]\big]= \sum_{i\in\setxdof} \sum_{j}^{\nrhs\ii} \abf[]{\u}{}{j}{\i[i]} \cdot \abf[]{\K}{}{}{}\left[\x\right] \abf[]{\u}{}{j}{\i[i]}
\end{equation}
for the elementary approach and
\begin{equation}
	g_0\big[\abf[\m]{\U}{}{}{\i[1]}\left[\x\right], ..., \abf[\m]{\U}{}{}{\i[a]}\left[\x\right]\big]= \sum_{i\in\setxdof} \sum_{j}^{\nrhs\ii} \abf[\m]{\u}{}{j}{\i[i]} \cdot \abf[]{\tilde{\K}}{}{}{}\left[\x\right] \abf[\m]{\u}{}{j}{\i[i]}
\end{equation}
for the condensation approach. The material usage is constrained as
\begin{equation}
	g_1 = \sum_{i}^{N} \frac{\tilde{x}_i}{N\overline{v}} -1,
\end{equation} 
where $\overline{v}$ is the maximum ratio of solid to void material based upon the filtered field $\tilde{\mathbf{x}}$.
An example of a resulting topology for $\nmdof = 100$ is shown in \cref{fig:design}, which can be reproduced using the attached MATLAB code.

\subsubsection{Computational efficiency}
Although the choice of approach, \textit{i.e.} elementary or condensation-based, has no influence on the resulting topology or design performance, it does influence the computational effort. 
For this specific MPOP $\nrhs\ii = \nmdof - 1$ for all $i \in \setxdof$ and $\der[g_0]{\abf[]{\U}{}{}{\i[i]}} = \abf[]{\F}{}{}{\i[i]}$. 
As a result, we can simply write $\abf[]{\bm{\Lambda}}{}{}{\i[i]} = \abf[]{\U}{}{}{\i[i]}$ and fully equivalent for the condensation approach $\abf[\m]{\bm{\Lambda}}{}{}{\i[i]} = \abf[\m]{\U}{}{}{\i[i]}$. 
Hence, the problem is self-adjoint and there is no need to solve the additional system of equations of \cref{eq:yjadjoint2}. Therefore, $b\ii = 0$ for all $i \in \setxdof$. 
For the proposed optimization problem, \cref{eq:gainnma} can thereto be simplified to
\begin{equation}
	\Xi_{\tilde{\beta}}\left[\nsdof,\nmdof\right] \approx \frac{\sum_i^{\nmdof} \tilde{\beta}_{\text{s}}\left[\nsdof,\nmdof-1\right]}{\tilde{\beta}_{\text{s}}\left[\nsdof -\nmdof, \nmdof\right] + \sum_i^{\nmdof} \tilde{\beta}_{\text{d}}\left[\nmdof, \nmdof-1\right]}.
	\label{eq:gainp1}
\end{equation}

\begin{figure*}
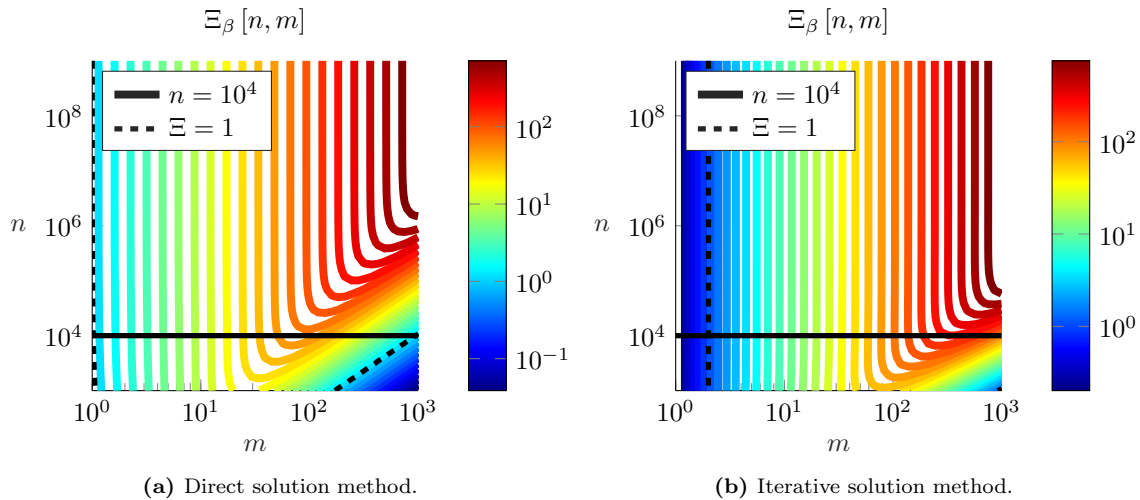

	\begin{center}
		\begin{subfigure}[t]{0.475\textwidth}
			\setlength{\figW}{0.65\textwidth}
			\setlength{\figH}{0.2\textheight} 
			\centering
			\input{figures/gainp1d.tikz}
			\caption{Direct solution method.}
			\label{fig:prob1directcontour}
		\end{subfigure}
		~
		\begin{subfigure}[t]{0.475\textwidth}
			\setlength{\figW}{0.65\textwidth}
			\setlength{\figH}{0.2\textheight} 
			\centering
			\input{figures/gainp1i.tikz}
			\caption{Iterative solution method.}
			\label{fig:prob1iterativecontour}
		\end{subfigure}
		\caption{Theoretical estimated gain of the condensation approach compared to the elementary approach for Problem 1. \cref{fig:prob1directcontour,fig:prob1iterativecontour}  show a triple-log contour plot of the gain $\Xi_\beta$ as a function of system size $n$ and number of DOFs of interest $m$ for the direct (a) and iterative (b) solution methods respectively.}
		\label{fig:prob1direct}
	\end{center}
\end{figure*}

\begin{figure*}
	\begin{center}
		\begin{subfigure}[t]{0.475\textwidth}
			\setlength{\figW}{0.8\textwidth}
			\setlength{\figH}{0.3\textheight} 
			\centering
%
%
\definecolor{mycolor1}{rgb}{0.00000,1.00000,1.00000}%
\begin{tikzpicture}
\begin{axis}[%
width=\figW,
height=\figH,
at={(0\figW,0\figH)},
scale only axis,
xmode=log,
xmin=1,
xmax=1000,
xminorticks=true,
xlabel style={font=\color{white!15!black}},
xlabel={$\nmdof$},
ymode=log,
ymin=0.1,
ymax=1000,
yminorticks=true,
ylabel style={font=\color{white!15!black}, rotate=-90},
ylabel={$\Xi$},
axis background/.style={fill=white},
title style={font=\bfseries},
axis x line*=bottom,
axis y line*=left,
legend style={at={(0.04,1.04)}, anchor=north west, legend cell align=left, align=left, draw=none, fill=none}
]
\addplot [color=blue, line width=2.0pt]
  table[row sep=crcr]{%
1	0.942372026333918\\
1.15139539932645	1.08597355441223\\
1.32571136559011	1.25160153146123\\
1.52641796717523	1.44267494014113\\
1.75751062485479	1.66315322368477\\
2.02358964772516	1.91762297799548\\
2.32995181051537	2.21139773817125\\
2.68269579527973	2.55063165212795\\
3.08884359647748	2.9424465919163\\
3.55648030622313	3.39506952317107\\
4.09491506238043	3.91797133071169\\
4.71486636345739	4.52198714509127\\
5.42867543932386	5.21937665677643\\
6.25055192527397	6.02374215662605\\
7.19685673001152	6.94964687650061\\
8.28642772854684	8.0116419910447\\
9.54095476349994	9.22218282125386\\
10.9854114198756	10.5875613278338\\
12.648552168553	12.1005294065354\\
14.5634847750124	13.7279903344985\\
16.7683293681101	15.3928590126582\\
19.3069772888325	16.9529152206827\\
22.2299648252619	18.1887789165112\\
25.5954792269954	18.826309154712\\
29.4705170255181	18.6190198484858\\
33.9322177189533	17.4745479790157\\
39.0693993705462	15.5384798080328\\
44.9843266896945	13.1469529139385\\
51.7947467923121	10.6753368213625\\
59.6362331659464	8.40384915273702\\
68.66488450043	6.47431334015468\\
79.060432109077	4.91801860726006\\
91.0298177991522	3.70370751921783\\
104.811313415469	2.77552123016972\\
120.679264063933	2.07471768367117\\
138.949549437314	1.54928857680553\\
159.985871960606	1.15679404378457\\
184.206996932672	0.864079963791956\\
212.095088792019	0.645868642447544\\
244.205309454865	0.483147883831115\\
281.176869797423	0.361722760028715\\
323.745754281764	0.271033604020025\\
372.759372031494	0.203236329362728\\
429.193426012878	0.152504892924574\\
494.171336132383	0.114509518350293\\
568.98660290183	0.0860292179369178\\
655.128556859551	0.0646650274653472\\
754.312006335462	0.0486280183572683\\
868.511373751353	0.0365825571428814\\
1000	0.027530300679224\\
};
\addlegendentry{$\Xi_\beta\left[\nsdof=10^3\right]$}

\addplot [color=green, line width=2.0pt]
  table[row sep=crcr]{%
1	0.980139768625012\\
1.15139539932645	1.12863383144494\\
1.32571136559011	1.29964286892214\\
1.52641796717523	1.49658615536185\\
1.75751062485479	1.72340389984803\\
2.02358964772516	1.98463708160042\\
2.32995181051537	2.28551963329922\\
2.68269579527973	2.63208489445424\\
3.08884359647748	3.03128854315943\\
3.55648030622313	3.49115052584242\\
4.09491506238043	4.02091882441837\\
4.71486636345739	4.63125819016307\\
5.42867543932386	5.33446715182995\\
6.25055192527397	6.14472650569312\\
7.19685673001152	7.07838178779875\\
8.28642772854684	8.15426027231122\\
9.54095476349994	9.39401860204379\\
10.9854114198756	10.8225078974679\\
12.648552168553	12.4681247158414\\
14.5634847750124	14.3630804118524\\
16.7683293681101	16.5434535092782\\
19.3069772888325	19.0487634855573\\
22.2299648252619	21.9205755909235\\
25.5954792269954	25.199245486746\\
29.4705170255181	28.9172472695284\\
33.9322177189533	33.086527837391\\
39.0693993705462	37.6761197922793\\
44.9843266896945	42.575643014036\\
51.7947467923121	47.5429011403975\\
59.6362331659464	52.1450821824102\\
68.66488450043	55.7296775302663\\
79.060432109077	57.4966596233045\\
91.0298177991522	56.7399850407411\\
104.811313415469	53.2074265547412\\
120.679264063933	47.3344222514908\\
138.949549437314	40.1072362222538\\
159.985871960606	32.6311594557989\\
184.206996932672	25.7410174648915\\
212.095088792019	19.8684582846332\\
244.205309454865	15.1166861235702\\
281.176869797423	11.3988643200167\\
323.745754281764	8.55071345180774\\
372.759372031494	6.39653142561653\\
429.193426012878	4.77927244114322\\
494.171336132383	3.56997120286787\\
568.98660290183	2.66742835930146\\
655.128556859551	1.99423595026714\\
754.312006335462	1.49203433287467\\
868.511373751353	1.11717460790483\\
1000	0.837144318197726\\
};
\addlegendentry{$\Xi_\beta\left[\nsdof=10^4\right]$}

\addplot [color=mycolor1, line width=2.0pt]
  table[row sep=crcr]{%
1	0.998141321416658\\
1.15139539932645	1.14925623044416\\
1.32571136559011	1.32324953572059\\
1.52641796717523	1.52358501673705\\
1.75751062485479	1.75425088423659\\
2.02358964772516	2.01983918670192\\
2.32995181051537	2.32563724183805\\
2.68269579527973	2.67773291460461\\
3.08884359647748	3.08313583944958\\
3.55648030622313	3.54991700239105\\
4.09491506238043	4.08736946485705\\
4.71486636345739	4.7061934330686\\
5.42867543932386	5.41870936269599\\
6.25055192527397	6.23910334828086\\
7.19685673001152	7.18370969171737\\
8.28642772854684	8.27133628683317\\
9.54095476349994	9.52363931262833\\
10.9854114198756	10.9655547129866\\
12.648552168553	12.6257950749868\\
14.5634847750124	14.5374218230989\\
16.7683293681101	16.7385041466744\\
19.3069772888325	19.2728777991162\\
22.2299648252619	22.1910188748738\\
25.5954792269954	25.551049907245\\
29.4705170255181	29.4198981464759\\
33.9322177189533	33.874628656512\\
39.0693993705462	39.0039778321845\\
44.9843266896945	44.9101158843682\\
51.7947467923121	51.7106693173007\\
59.6362331659464	59.5410354884583\\
68.66488450043	68.5570191046329\\
79.060432109077	78.9378112484002\\
91.0298177991522	90.8893080501709\\
104.811313415469	104.647714672966\\
120.679264063933	120.483274893915\\
138.949549437314	138.703758802048\\
159.985871960606	159.656941884586\\
184.206996932672	183.730557248186\\
212.095088792019	211.346816700614\\
244.205309454865	242.946102385158\\
281.176869797423	278.950096469342\\
323.745754281764	319.687494400473\\
372.759372031494	365.254882775496\\
429.193426012878	415.272918172237\\
494.171336132383	468.492729023702\\
568.98660290183	522.237173670523\\
655.128556859551	571.784562271736\\
754.312006335462	610.084800572064\\
868.511373751353	628.563836218167\\
1000	619.716160592936\\
};
\addlegendentry{$\Xi_\beta\left[\nsdof=10^6\right]$}

\addplot [color=red, line width=2.0pt]
  table[row sep=crcr]{%
1	0.999936760446386\\
1.15139539932645	1.15132258664113\\
1.32571136559011	1.32562753078558\\
1.52641796717523	1.52632144200488\\
1.75751062485479	1.75739948865419\\
2.02358964772516	2.02346168924504\\
2.32995181051537	2.32980448399195\\
2.68269579527973	2.68252616987481\\
3.08884359647748	3.0886482980916\\
3.55648030622313	3.55625545053562\\
4.09491506238043	4.0946561778003\\
4.71486636345739	4.71456830247311\\
5.42867543932386	5.42833227651639\\
6.25055192527397	6.2501568400016\\
7.19685673001152	7.19640187148216\\
8.28642772854684	8.28590406065977\\
9.54095476349994	9.54035188645804\\
10.9854114198756	10.9847173651361\\
12.648552168553	12.6477531631927\\
14.5634847750124	14.5625649710253\\
16.7683293681101	16.7672705315252\\
19.3069772888325	19.3057584428306\\
22.2299648252619	22.2285618406703\\
25.5954792269954	25.5938643526515\\
29.4705170255181	29.468658350003\\
33.9322177189533	33.9300785541019\\
39.0693993705462	39.066937545947\\
44.9843266896945	44.9814937460961\\
51.7947467923121	51.7914870604663\\
59.6362331659464	59.632482715959\\
68.66488450043	68.6605699451871\\
79.060432109077	79.055469243802\\
91.0298177991522	91.0241100576797\\
104.811313415469	104.804750122451\\
120.679264063933	120.671718460953\\
138.949549437314	138.940876459736\\
159.985871960606	159.975905740762\\
184.206996932672	184.195548003006\\
212.095088792019	212.081940957587\\
244.205309454865	244.190216294642\\
281.176869797423	281.159550730728\\
323.745754281764	323.725890084058\\
372.759372031494	372.736599570476\\
429.193426012878	429.167331735166\\
494.171336132383	494.141447329458\\
568.98660290183	568.952374752674\\
655.128556859551	655.089351148408\\
754.312006335462	754.267053503127\\
868.511373751353	868.459701424139\\
1000	999.940294125923\\
};
\addlegendentry{$\Xi_\beta\left[\nsdof=10^9\right]$}

\addplot [color=black, line width=2.0pt, mark=o, mark options={solid, black}]
table[row sep=crcr]{%
	2	1.35017366428154\\
	4	2.09926175315216\\
	7	2.82972446592265\\
	13	4.62231938180007\\
	24	10.0124420251116\\
	46	15.8168496616955\\
	88	29.082806063249\\
	167	36.0850968101496\\
	316	24.6474293951563\\
};
\addlegendentry{$\Xi_\text{t}\left[\nsdof=10^4\right] $}

\addplot [color=black, dashed, line width=2.0pt]
  table[row sep=crcr]{%
1	1\\
1000	1\\
};
\addlegendentry{$\Xi = 1$}

\end{axis}

\begin{axis}[%
width=1.227\figW,
height=1.227\figH,
at={(-0.16\figW,-0.135\figH)},
scale only axis,
xmin=0,
xmax=1,
ymin=0,
ymax=1,
axis line style={draw=none},
ticks=none,
axis x line*=bottom,
axis y line*=left,
legend style={legend cell align=left, align=left, draw=white!15!black}
]
\end{axis}
\end{tikzpicture}%
			\caption{Direct solution method.}
			\label{fig:prob1directlines}
		\end{subfigure}
		~
		\begin{subfigure}[t]{0.475\textwidth}
			\setlength{\figW}{0.8\textwidth}
			\setlength{\figH}{0.3\textheight} 
			\centering
%
%
\definecolor{mycolor1}{rgb}{0.00000,1.00000,1.00000}%
\begin{tikzpicture}

\begin{axis}[%
width=\figW,
height=\figH,
at={(0\figW,0\figH)},
scale only axis,
xmode=log,
xmin=1,
xmax=1000,
xminorticks=true,
xlabel style={font=\color{white!15!black}},
xlabel={$\nmdof$},
ymode=log,
ymin=0.1,
ymax=1000,
yminorticks=true,
ylabel style={font=\color{white!15!black}, rotate=-90},
ylabel={$\Xi$},
axis background/.style={fill=white},
title style={font=\bfseries},
axis x line*=bottom,
axis y line*=left,
legend style={at={(0.04,1.04)}, anchor=north west, legend cell align=left, align=left, draw=none, fill=none}
]
\addplot [color=blue, line width=2.0pt]
  table[row sep=crcr]{%
1	0\\
1.15139539932645	0.151744565084768\\
1.32571136559011	0.326576369875592\\
1.52641796717523	0.528027758383316\\
1.75751062485479	0.76017785160477\\
2.02358964772516	1.02773915669397\\
2.32995181051537	1.33615846296016\\
2.68269579527973	1.69173450738612\\
3.08884359647748	2.10175532028371\\
3.55648030622313	2.57465862914369\\
4.09491506238043	3.12021916959438\\
4.71486636345739	3.74976714439456\\
5.42867543932386	4.4764422143705\\
6.25055192527397	5.31548696467329\\
7.19685673001152	6.2845821280093\\
8.28642772854684	7.40422177225853\\
9.54095476349994	8.69811798039477\\
10.9854114198756	10.1936072816916\\
12.648552168553	11.9219980943411\\
14.5634847750124	13.9187371387314\\
16.7683293681101	16.2231616242316\\
19.3069772888325	18.8774076524412\\
22.2299648252619	21.9237103025474\\
25.5954792269954	25.3987898020908\\
29.4705170255181	29.3232259265923\\
33.9322177189533	33.6827870776621\\
39.0693993705462	38.3982106740254\\
44.9843266896945	43.2817427842793\\
51.7947467923121	47.9866489942334\\
59.6362331659464	51.9745600465481\\
68.66488450043	54.552018763263\\
79.060432109077	55.0336659779281\\
91.0298177991522	53.0219984675476\\
104.811313415469	48.6538292377342\\
120.679264063933	42.6045777799953\\
138.949549437314	35.8132599391489\\
159.985871960606	29.1320083569949\\
184.206996932672	23.1208212970784\\
212.095088792019	18.0316446885715\\
244.205309454865	13.8956028094205\\
281.176869797423	10.6237843180861\\
323.745754281764	8.08074561597828\\
372.759372031494	6.12638503434179\\
429.193426012878	4.635192278372\\
494.171336132383	3.50253454132115\\
568.98660290183	2.64462495103751\\
655.128556859551	1.99594088392753\\
754.312006335462	1.50597030207909\\
868.511373751353	1.13611120983981\\
1000	0.857020303116957\\
};
\addlegendentry{$\Xi_\beta\left[\nsdof=10^3\right]$}

\addplot [color=green, line width=2.0pt]
  table[row sep=crcr]{%
1	0\\
1.15139539932645	0.151431945755516\\
1.32571136559011	0.325801895914497\\
1.52641796717523	0.526586435953152\\
1.75751062485479	0.757789752480333\\
2.02358964772516	1.02402391986074\\
2.32995181051537	1.33060147470427\\
2.68269579527973	1.68364218114882\\
3.08884359647748	2.09019619128841\\
3.55648030622313	2.5583861560624\\
4.09491506238043	3.09757125085766\\
4.71486636345739	3.71853655672458\\
5.42867543932386	4.43371179353564\\
6.25055192527397	5.2574240477848\\
7.19685673001152	6.20618988764603\\
8.28642772854684	7.29905312289423\\
9.54095476349994	8.5579754546097\\
10.9854114198756	10.0082883660875\\
12.648552168553	11.6792158068429\\
14.5634847750124	13.6044784472054\\
16.7683293681101	15.8229913799071\\
19.3069772888325	18.3796678112966\\
22.2299648252619	21.3263409302695\\
25.5954792269954	24.7228136690784\\
29.4705170255181	28.6380394617907\\
33.9322177189533	33.1514227148296\\
39.0693993705462	38.3541989919875\\
44.9843266896945	44.3508002822516\\
51.7947467923121	51.2600098048486\\
59.6362331659464	59.2155284207018\\
68.66488450043	68.3652513910055\\
79.060432109077	78.8679930989646\\
91.0298177991522	90.8854507418553\\
104.811313415469	104.565668830347\\
120.679264063933	120.011972956547\\
138.949549437314	137.228373125066\\
159.985871960606	156.029877239956\\
184.206996932672	175.907693582838\\
212.095088792019	195.853761598868\\
244.205309454865	214.191143101382\\
281.176869797423	228.535751966963\\
323.745754281764	236.094132105875\\
372.759372031494	234.44807832319\\
429.193426012878	222.635379130861\\
494.171336132383	201.865102454618\\
568.98660290183	175.21894599792\\
655.128556859551	146.458022868821\\
754.312006335462	118.771034870737\\
868.511373751353	94.1486488850008\\
1000	73.4197625486887\\
};
\addlegendentry{$\Xi_\beta\left[\nsdof=10^4\right]$}

\addplot [color=mycolor1, line width=2.0pt]
  table[row sep=crcr]{%
1	0\\
1.15139539932645	0.151395702117648\\
1.32571136559011	0.325712115636304\\
1.52641796717523	0.526419362933851\\
1.75751062485479	0.757512937413889\\
2.02358964772516	1.0235932456697\\
2.32995181051537	1.32995719307578\\
2.68269579527973	1.6827036364879\\
3.08884359647748	2.08885480394731\\
3.55648030622313	2.55649609935513\\
4.09491506238043	3.09493707635893\\
4.71486636345739	3.71489678748836\\
5.42867543932386	4.42871720031985\\
6.25055192527397	5.25060893184907\\
7.19685673001152	6.19693419649662\\
8.28642772854684	7.28653260516073\\
9.54095476349994	8.54109630732638\\
10.9854114198756	9.98560195257874\\
12.648552168553	11.6488080817072\\
14.5634847750124	13.5638278637109\\
16.7683293681101	15.7687885997072\\
19.3069772888325	18.3075911483188\\
22.2299648252619	21.2307844245588\\
25.5954792269954	24.5965724259122\\
29.4705170255181	28.471973891825\\
33.9322177189533	32.9341577599162\\
39.0693993705462	38.0719811059671\\
44.9843266896945	43.987760316628\\
51.7947467923121	50.7993109261758\\
59.6362331659464	58.6422969461942\\
68.66488450043	67.6729367380598\\
79.060432109077	78.0711196458718\\
91.0298177991522	90.0439958607968\\
104.811313415469	103.830111480491\\
120.679264063933	119.704171624787\\
138.949549437314	137.982526940241\\
159.985871960606	159.029493026643\\
184.206996932672	183.264628355802\\
212.095088792019	211.171114122599\\
244.205309454865	243.305398937021\\
281.176869797423	280.308291658252\\
323.745754281764	322.917705488701\\
372.759372031494	371.983272685006\\
429.193426012878	428.483056194763\\
494.171336132383	493.542571648032\\
568.98660290183	568.456281300434\\
655.128556859551	654.711596480654\\
754.312006335462	754.015165993794\\
868.511373751353	868.320727501778\\
1000	999.856869411665\\
};
\addlegendentry{$\Xi_\beta\left[\nsdof=10^6\right]$}

\addplot [color=red, line width=2.0pt]
  table[row sep=crcr]{%
1	0\\
1.15139539932645	0.151395399675079\\
1.32571136559011	0.325711366453708\\
1.52641796717523	0.526417968782301\\
1.75751062485479	0.757510627517458\\
2.02358964772516	1.02358965186781\\
2.32995181051537	1.32995181671282\\
2.68269579527973	1.68269580430805\\
3.08884359647748	2.0888436093817\\
3.55648030622313	2.55648032440727\\
4.09491506238043	3.09491508772725\\
4.71486636345739	3.71486639848759\\
5.42867543932386	4.42867548740754\\
6.25055192527397	5.25055199091167\\
7.19685673001152	6.1968568192073\\
8.28642772854684	7.28642784930376\\
9.54095476349994	8.54095492647766\\
10.9854114198756	9.98541163926328\\
12.648552168553	11.6485524632276\\
14.5634847750124	13.5634851700756\\
16.7683293681101	15.7683298969271\\
19.3069772888325	18.3069779957372\\
22.2299648252619	21.2299657691445\\
25.5954792269954	24.5954804860611\\
29.4705170255181	28.4705187035991\\
33.9322177189533	32.9322199538783\\
39.0693993705462	38.0694023452409\\
44.9843266896945	43.9843306469008\\
51.7947467923121	50.7947520541065\\
59.6362331659464	58.6362401596209\\
68.66488450043	67.6648937928087\\
79.060432109077	78.060444452017\\
91.0298177991522	90.0298341898718\\
104.811313415469	103.811335176534\\
120.679264063933	119.679292949305\\
138.949549437314	137.949587772948\\
159.985871960606	158.985922830849\\
184.206996932672	183.207064427382\\
212.095088792019	211.095178334171\\
244.205309454865	243.205428234847\\
281.176869797423	280.177027348756\\
323.745754281764	322.745963244258\\
372.759372031494	371.759649162817\\
429.193426012878	428.193793529305\\
494.171336132383	493.171823485706\\
568.98660290183	567.987249134035\\
655.128556859551	654.129413722688\\
754.312006335462	753.313142424554\\
868.511373751353	867.512879977905\\
1000	999.001996838491\\
};
\addlegendentry{$\Xi_\beta\left[\nsdof=10^9\right]$}

\addplot [color=black, line width=2.0pt, mark=o, mark options={solid, black}]
table[row sep=crcr]{%
	2	0.469241639869064\\
	4	1.53086461956568\\
	7	4.52363614087883\\
	13	12.7738003352005\\
	24	25.6435194150682\\
	46	84.3631517163212\\
	88	183.968446811175\\
	167	330.449816863438\\
	316	216.96483088757\\
};
\addlegendentry{$\Xi_\text{t}\left[\nsdof=10^4\right]$}

\addplot [color=black, dashed, line width=2.0pt]
  table[row sep=crcr]{%
1	1\\
1000	1\\
};
\addlegendentry{$\Xi = 1$}

\end{axis}

\begin{axis}[%
width=1.227\figW,
height=1.227\figH,
at={(-0.16\figW,-0.135\figH)},
scale only axis,
xmin=0,
xmax=1,
ymin=0,
ymax=1,
axis line style={draw=none},
ticks=none,
axis x line*=bottom,
axis y line*=left,
legend style={legend cell align=left, align=left, draw=none, fill=none}
]
\end{axis}
\end{tikzpicture}%
			\caption{Iterative solution method.}
			\label{fig:prob1iterativelines}
		\end{subfigure}
		\caption{Theoretical estimated and experimental gain of the condensation approach compared to the elementary approach for Problem 1.  \cref{fig:prob1directlines,fig:prob1iterativelines} show  double-log graphs of the gain as a function of the number of primary DOFs $\nmdof$ for a variety of system sizes. The gain based on FLOP count is verified by calculating the gain based on CPU time as shown in \cref{fig:prob1directlines,fig:prob1iterativelines}.}
		\label{fig:prob1exp}
	\end{center}
\end{figure*}
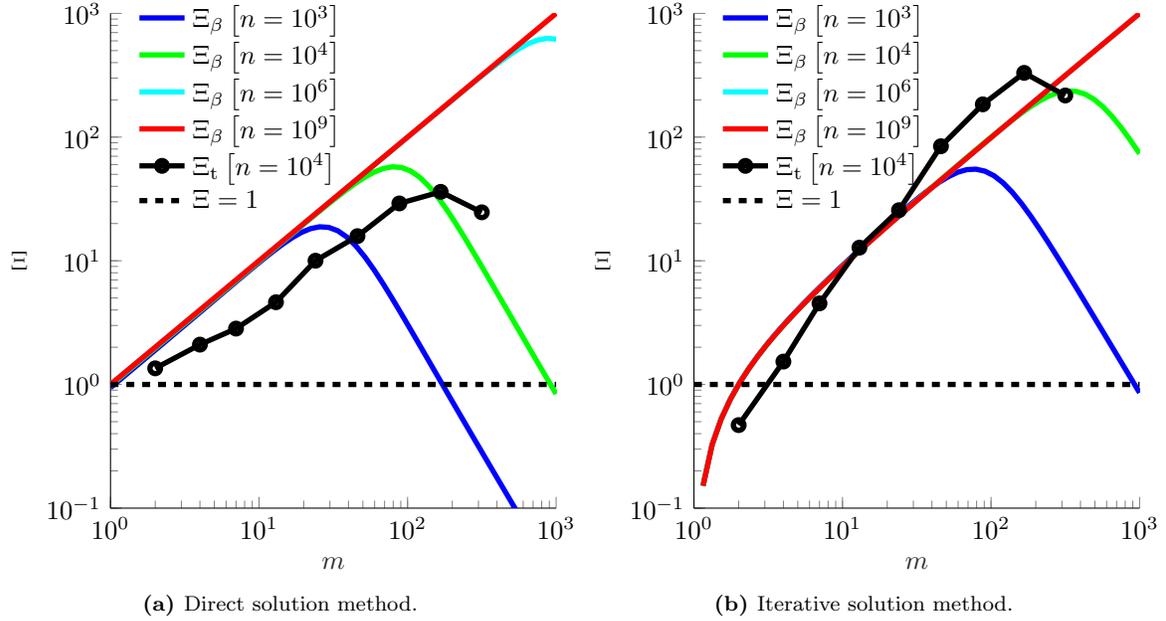

The computational gain $\Xi_\beta$ as a function of $\nsdof$ and $\nmdof$ is plotted in \cref{fig:prob1directcontour,fig:prob1iterativecontour} for direct and iterative solution methods respectively.  
The contour graphs show the condensation approach clearly outperforms the elementary approach, especially for large systems and small to moderate $m$. 
For this type of MPOP, a problem with many DOFs and a low ratio of $\nmdof$ to $\nsdof$ can be solved up to \textit{1000 times faster} using the condensation approach independent of the chosen solution method. \Cref{fig:prob1directlines,fig:prob1iterativelines} give a clear view of the gain as a function of $\nmdof$ for different system sizes using direct and iterative solution methods, respectively. 
In addition to the gain as a function of FLOP count $\Xi_{\tilde{\beta}}$, the gain obtained via a time measurement\footnote{The gain based on time measurements, equivalently to the gain based on FLOP count, only considers the time of solving the systems of equations.} for $n=10^4$ is plotted. 
Although $\Xi_\text{t}$ deviates from $\Xi_{\tilde{\beta}}$, the trend is captured, which confirms the validity of the assumptions on FLOP count. 
Therefore, \cref{eq:gainnma} can be used to roughly \emph{estimate} the expected gain of using the condensation approach.

\subsection{Problem 2}
\label{sec:prob2}

\begin{figure*}
	\centering
	\includegraphics[width=\textwidth]{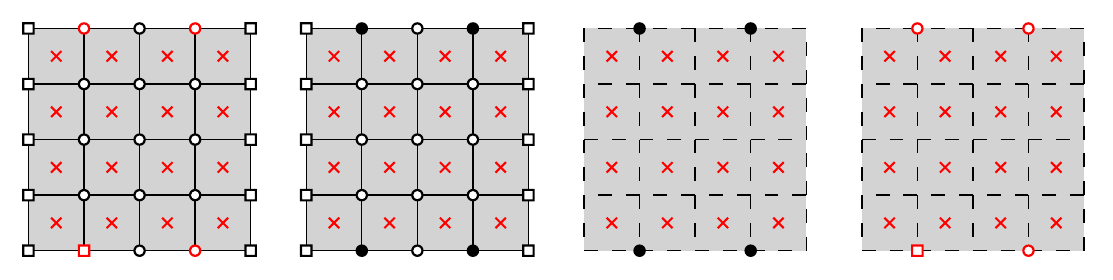}
	\caption{Schematic representation of Problem 2; a multi-input-multi-output compliant mechanism with $N = 16$, the number of inputs $x=2$ and $\nmdof = 2x= 4$. The number of DOFs per node is 2, however, for simplicity of visualization, only a single DOF is shown. From left to right (analogous to \cref{fig:heatschematic}): one of the two analyses as solved for in the elementary approach, subdivision into primary and secondary DOFs, resulting system after static condensation and, the equivalent analysis as solved for in the condensation approach. Note that the displacements of both sides are fixed. }
	\label{fig:mimogrid}
\end{figure*}

The second problem aims to maximize the amount of material required to achieve a displacement-based $x$-input $x$-output compliant mechanism satisfying prescribed input-output transmission ratios, see \textit{e.g.} \citet{Larsen1997}. The problem is sketched in \cref{fig:mimogrid}.
This second problem emphasizes the effect of reduction of computational effort of the sensitivity analysis.

\begin{figure}
	\centering
	\includegraphics[width=0.45\textwidth]{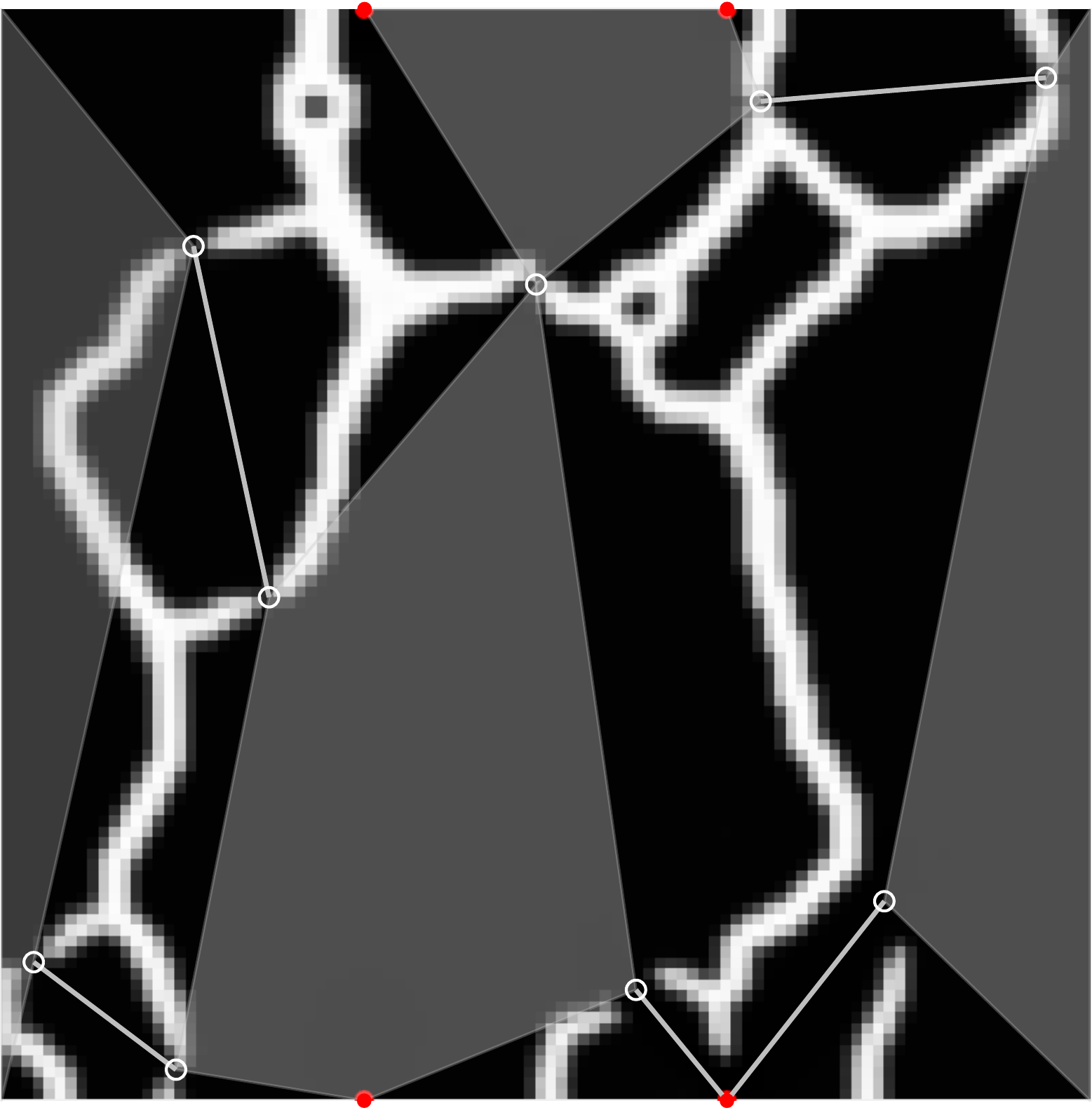}
	\caption{Resulting topology of the MPOP posed in \cref{sec:prob2} as schematically sketched in \cref{fig:mimogrid}. The white material is void, whereas the black material is solid. For this MPOP the number of variables $N = 100 \times 100$, the number of structural DOFs $\nsdof = 10000$, the number of analysis sets $a=2$ and the number of primary DOFs $\nmdof = 2x = 4$. The location of primary DOFs and applied loads are indicated in red. This multi-input-multi-output compliant mechanism is optimized for a Jacobian matrix $\overline{\mathbf{J}} = \left[0.5 ~2.0 ; ~1.0 ~-1.0\right]$. An interpretation of the resulting mechanism is illustrated as a rigid-body mechanism with rigid links and bodies as well as rotation points.}
	\label{fig:design2}
\end{figure}

The objective function for this MPOP is
\begin{equation}
	g_0 = \sum_{i}^{N} \tilde{x}_i.
\end{equation}
One can write the mechanism output displacements as a function of the input displacements as
\begin{equation}
	\mathbf{u}_{i} = \mathbf{J}_{ij} \mathbf{u}_{j}, \quad \forall~ i,j = 1,...,x
\end{equation}
where $\mathbf{J}$ is the so-called Jacobian matrix, $i$ denotes the output and $j$ the input. For example, $\mathbf{J}_{12}$ is the displacement of Output 1 due to an input displacement on Input 2. One requires $a=x$ analyses with prescribed unit input displacements to obtain $\mathbf{J}$. In order to control all entries in the Jacobian matrix, we introduce $2x$ transmission constraints, \textit{i.e.}
\begin{equation}
	g_{ij} = \frac{\mathbf{J}_{ij}}{\mathbf{\overline{J}}_{ij}} + 1,
\end{equation}
with $\mathbf{\overline{J}}_{ij} \not= 0$ the minimum transmission between input $i$ and output $j$. Note this response can be used for both positive and negative ratios. If one assigns $\mathbf{\overline{J}}_{ij} = u^*$, then the required $\mathbf{J}_{ij} \leq -u^*$ and if $\mathbf{\overline{J}}_{ij} = -u^*$, then $\mathbf{J}_{ij} \geq u^*$, for any target displacement $u^* \in \mathbb{R}^+$.

An example of the resulting topology for $x=2$ and
\begin{equation*}
	\overline{\mathbf{J}} = \begin{bmatrix}
		0.5 & 2.0 \\
		1.0 & -1.0
	\end{bmatrix}
\end{equation*}
is shown in \cref{fig:design2}. Due to the conflicting behaviour between volume maximization and minimum transmission ratio the resulting topology almost exactly satisfies the target Jacobian with a minimum amount of holes.

\subsubsection{Computational efficiency}
\begin{figure*}
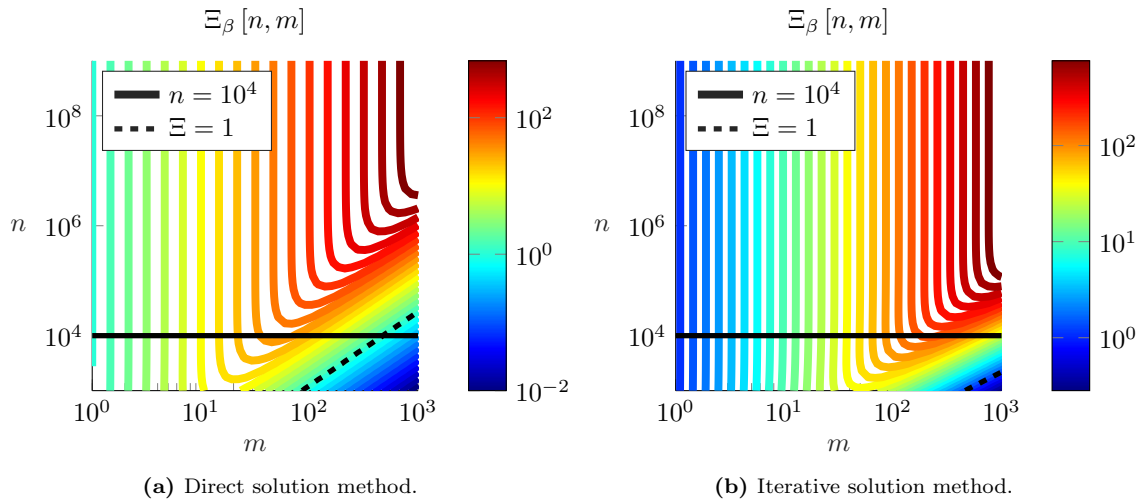

	\begin{center}
		\begin{subfigure}[t]{0.475\textwidth}
			\setlength{\figW}{0.65\textwidth}
			\setlength{\figH}{0.2\textheight} 
			\centering
			\input{figures/gainp2direct.tex}
			\caption{Direct solution method.}
			\label{fig:gainp2direct}
		\end{subfigure}
		~
		\begin{subfigure}[t]{0.475\textwidth}
			\setlength{\figW}{0.65\textwidth}
			\setlength{\figH}{0.2\textheight} 
			\centering
			\input{figures/gainp2iterative.tex}
			\caption{Iterative solution method.}
			\label{fig:gainp2iterative}
		\end{subfigure}
		\caption{Theoretical estimated gain of the condensation approach compared to the elementary approach for Problem 2. \cref{fig:gainp2direct,fig:gainp2iterative} show a triple-log contour plot of the gain $\Xi_\beta$ as a function of system size $n$ and number of DOFs of interest $m$ for direct (a) and iterative (b) solution methods respectively.}
		\label{fig:prob2direct}
	\end{center}
\end{figure*}

For this specific MPOP $\nrhs\ii = 1$ for all $i \in \setxdof$. Since the problem is non-self-adjoint an additional $x$ right-hand-sides per analysis has to be solved for, that is $b\ii = x$ for all $i \in \setxdof$. Note that the number of DOFs of interest equals twice the number of inputs\footnote{Assuming the number of inputs and outputs are equal}, \textit{i.e.} $\nmdof = 2x$. For the proposed optimization formulation, \cref{eq:gainnma} can therefore be simplified to
\begin{equation}
	\Xi_{\tilde{\beta}}\left[\nsdof,\nmdof\right] \approx \frac{\sum_i^{\frac{\nmdof}{2}} \tilde{\beta}_{\text{s}}\left[\nsdof,\nmdof\right]}{\tilde{\beta}_{\text{s}}\left[\nsdof -\nmdof, \nmdof\right] + \sum_i^{\frac{\nmdof}{2}} \tilde{\beta}_{\text{d}}\left[\nmdof, \nmdof\right]},
	\label{eq:gainp2}
\end{equation}
with $\nmdof = 2x$ and $x$ the number of inputs and outputs. 
Note the computational gain for this type of problem is twofold. 
One requires a single analysis set with $2x$ analyses, as opposed to $x$ analysis sets to obtain the responses plus an additional $x$ analysis sets to obtain the sensitivities. 
Simple reasoning as well as the FLOP count analysis learns that for this type of problem the gain approximately equals the number of inputs (and thus the number of analysis sets) for low to moderate number of inputs, see \cref{fig:prob2direct}.

\section{Extensions}
\label{sec:discussion}
The proposed method is applicable to any linear(ized) PDE as well as multi-physical coupled problems and can be extended or adapted to a range of other problems.
The method is still applicable to asymmetric system matrices. The advantageous property of inexpensive sensitivities vanishes for asymmetric matrices since the self-adjointness property no longer holds, hence the method is expected to reduce efficiency. The method is applicable to reduce the dimensionality of eigen-problems, with tremendous reduction in computational cost. However, in the reduced mass matrix, combinations of stiffness and mass elements appear. The result is that the eigenvalue-eigenvector problem is closely but not exactly preserved, with scenario-dependent accuracy issues \citep{Guyan1965}.

A response function may depend on secondary DOFs, in which case an additional large system of linear equations has to be solved to obtain the sensitivities, see \cref{sec:sensubc}.
Although the additional effort is marginal since the preconditioner can be reused, it makes the use of responses dependent on secondary DOFs uneconomical. As opposed to the general guidelines presented in this work, one may want to put secondary DOFs  with varying freedom per analysis set subjected to loading into the set of primary DOFs. 
This will increase the number of primary DOFs, however circumvents the necessity to calculate and use the reduced load. This will reduce the computational effort if the number of loaded secondary DOFs are small.


For scenarios with multiple DOFs following the same freedom \emph{and} magnitude throughout each analysis set (for example the DOFs contained in a non-design domain), pre-processing of the system matrix by multi-point constraints\footnote{Primary-secondary, penalty augmentation or Lagrange multiplier adjunction methods.} is recommended to limit the dimensionality of the reduced system matrix. This has no further consequences apart from implementation, possibly reduction in global dimensionality and change of sparsity pattern / band density. 


\section{Conclusion}
\label{sec:conclusion}
This study aims to find a solution to efficiently solve for large-scale structural optimization problems involving multiple analyses with varying boundary conditions.
With an increasing number of analyses, the computational effort to solve such problems quickly increases. 
Using static condensation to eliminate the DOFs that do not change freedom between analyses \underline{and} are not of interest to any of the optimization responses, the effort can be highly reduced.
In contrast to solving for multiple systems of equations with different sparse system matrices of high dimensionality, this allows for solving multiple systems of equations with \underline{the same} sparse system matrix of high dimensionality \underline{and} multiple systems of equations with different dense system matrices of low dimensionality.
A profound finding is that the sensitivities of the static condensation procedure can be calculated without solving additional systems of equations, even if the response functions are not self-adjoint. Consequently, the dimensionality of the adjoint equations is highly reduced.

Numerical examples indicate an algorithmic complexity analysis can be used to roughly estimate the expected computational effort of a broad range of problems, even if solely the major contributors are taken into account.
In line with the prognosis the numerical examples demonstrate that substantial computational gains can be achieved. 
However, the presented examples only cover a small part of the total applicability. 

It is expected that future research will reveal more applications that benefit from this method. 
We expect at least applications in structural and multidisciplinary optimization with many conflicting requirements based on a variety of load cases.
Given the limited additional implementation effort, future work on---or including---multi-partition structural optimization problem(s) should consider a back-of-the-envelop estimation of computational gain by using the condensation approach.

\section*{Declaration of interest}
The authors declare that there is no conflict of interest.

\section*{Acknowledgements}
The authors would like to thank Freek Broeren (set theory), Can Ayas (Einstein notation), Arnoud Delissen (dyadic product) and Max van der Kolk (algorithmic complexity analysis) for their help.
This work was supported by the Dutch Research Council
(NWO) Applied and Engineering Sciences (AES) project
16191 entitled Stable and Adjustable Mechanisms for Optical Instruments and Implants (SAMOII).

\bibliography{ref.bib}

\begin{thebibliography}{51}
\expandafter\ifx\csname natexlab\endcsname\relax\def\natexlab#1{#1}\fi
\providecommand{\url}[1]{\texttt{#1}}
\providecommand{\href}[2]{#2}
\providecommand{\path}[1]{#1}
\providecommand{\DOIprefix}{doi:}
\providecommand{\ArXivprefix}{arXiv:}
\providecommand{\URLprefix}{URL: }
\providecommand{\Pubmedprefix}{pmid:}
\providecommand{\doi}[1]{\href{http://dx.doi.org/#1}{\path{#1}}}
\providecommand{\Pubmed}[1]{\href{pmid:#1}{\path{#1}}}
\providecommand{\bibinfo}[2]{#2}
\ifx\xfnm\relax \def\xfnm[#1]{\unskip,\space#1}\fi
\bibitem[{Amir et~al.(2009)Amir, Bends{\o}e, and Sigmund}]{Amir2009}
\bibinfo{author}{O.~Amir}, \bibinfo{author}{M.~Bends{\o}e},
  \bibinfo{author}{O.~Sigmund}, \bibinfo{journal}{Int. J. Numer. Methods Eng.}
  \bibinfo{volume}{78} (\bibinfo{year}{2009}) \bibinfo{pages}{1474--1491}.
  \DOIprefix\doi{10.1002/nme.2536}. \href{http://arxiv.org/abs/1010.1724}{\tt
  arXiv:1010.1724}.
\bibitem[{Rozvany et~al.(1993)Rozvany, Sigmund, Lewi{\'{n}}ski, Gerdes, and
  Birker}]{Rozvany1993}
\bibinfo{author}{G.~Rozvany}, \bibinfo{author}{O.~Sigmund},
  \bibinfo{author}{T.~Lewi{\'{n}}ski}, \bibinfo{author}{D.~Gerdes},
  \bibinfo{author}{T.~Birker}, \bibinfo{journal}{Struct. Optim.}
  \bibinfo{volume}{5} (\bibinfo{year}{1993}) \bibinfo{pages}{204--206}.
  \DOIprefix\doi{10.1007/bf01743359}.
\bibitem[{Cavazzuti et~al.(2011)Cavazzuti, Baldini, Bertocchi, Costi,
  Torricelli, and Moruzzi}]{Cavazzuti2010}
\bibinfo{author}{M.~Cavazzuti}, \bibinfo{author}{A.~Baldini},
  \bibinfo{author}{E.~Bertocchi}, \bibinfo{author}{D.~Costi},
  \bibinfo{author}{E.~Torricelli}, \bibinfo{author}{P.~Moruzzi},
  \bibinfo{journal}{Struct. Multidiscip. Optim.} \bibinfo{volume}{44}
  (\bibinfo{year}{2011}) \bibinfo{pages}{45--56}.
  \DOIprefix\doi{10.1007/s00158-010-0578-7}.
\bibitem[{Diaz and Bends{\o}e(1992)}]{Diaz1992}
\bibinfo{author}{A.~R. Diaz}, \bibinfo{author}{M.~P. Bends{\o}e},
  \bibinfo{journal}{Struct. Optim.} \bibinfo{volume}{4} (\bibinfo{year}{1992})
  \bibinfo{pages}{17--22}. \DOIprefix\doi{10.1007/BF01894077}.
\bibitem[{Zhang et~al.(2017)Zhang, de~Sturler, and Paulino}]{Zhang2017}
\bibinfo{author}{X.~S. Zhang}, \bibinfo{author}{E.~de~Sturler},
  \bibinfo{author}{G.~H. Paulino}, \bibinfo{journal}{Comput. Methods Appl.
  Mech. Eng.} \bibinfo{volume}{325} (\bibinfo{year}{2017})
  \bibinfo{pages}{463--487}. \DOIprefix\doi{10.1016/j.cma.2017.06.035}.
\bibitem[{Zhang et~al.(2020)Zhang, de~Sturler, and Shapiro}]{Zhang2020}
\bibinfo{author}{X.~S. Zhang}, \bibinfo{author}{E.~de~Sturler},
  \bibinfo{author}{A.~Shapiro}, \bibinfo{journal}{J. Appl. Mech.}
  \bibinfo{volume}{87} (\bibinfo{year}{2020}).
  \DOIprefix\doi{10.1115/1.4045902}.
\bibitem[{Liu and Korvink(2009)}]{Liu2009}
\bibinfo{author}{Z.~Liu}, \bibinfo{author}{J.~Korvink},
  \bibinfo{journal}{Finite Elem. Anal. Des.} \bibinfo{volume}{45}
  (\bibinfo{year}{2009}) \bibinfo{pages}{555--568}.
  \DOIprefix\doi{10.1016/j.finel.2009.03.005}.
\bibitem[{Jin et~al.(2017)Jin, Zhang, Yang, and Zhu}]{Jin2017}
\bibinfo{author}{M.~Jin}, \bibinfo{author}{X.~Zhang},
  \bibinfo{author}{Z.~Yang}, \bibinfo{author}{B.~Zhu}, \bibinfo{journal}{J.
  Mech. Des.} \bibinfo{volume}{140} (\bibinfo{year}{2017})
  \bibinfo{pages}{011402}. \DOIprefix\doi{10.1115/1.4038332}.
\bibitem[{Evgrafov et~al.(2008)Evgrafov, Rupp, Maute, and Dunn}]{Evgrafov2008}
\bibinfo{author}{A.~Evgrafov}, \bibinfo{author}{C.~Rupp},
  \bibinfo{author}{K.~Maute}, \bibinfo{author}{M.~Dunn},
  \bibinfo{journal}{Struct. Multidiscip. Optim.} \bibinfo{volume}{36}
  (\bibinfo{year}{2008}) \bibinfo{pages}{329--345}.
  \DOIprefix\doi{10.1007/s00158-007-0190-7}.
\bibitem[{Amir et~al.(2014)Amir, Aage, and Lazarov}]{Amir2014}
\bibinfo{author}{O.~Amir}, \bibinfo{author}{N.~Aage},
  \bibinfo{author}{B.~Lazarov}, \bibinfo{journal}{Struct. Multidiscip. Optim.}
  \bibinfo{volume}{49} (\bibinfo{year}{2014}) \bibinfo{pages}{815--829}.
  \DOIprefix\doi{10.1007/s00158-013-1015-5}.
\bibitem[{Aage et~al.(2015)Aage, Andreassen, and Lazarov}]{Aage2015}
\bibinfo{author}{N.~Aage}, \bibinfo{author}{E.~Andreassen},
  \bibinfo{author}{B.~Lazarov}, \bibinfo{journal}{Struct. Multidiscip. Optim.}
  \bibinfo{volume}{51} (\bibinfo{year}{2015}) \bibinfo{pages}{565--572}.
  \DOIprefix\doi{10.1007/s00158-014-1157-0}.
\bibitem[{Wu et~al.(2016)Wu, Dick, and Westermann}]{Wu2016}
\bibinfo{author}{J.~Wu}, \bibinfo{author}{C.~Dick},
  \bibinfo{author}{R.~Westermann}, \bibinfo{journal}{IEEE Trans. Vis. Comput.
  Graph.} \bibinfo{volume}{22} (\bibinfo{year}{2016})
  \bibinfo{pages}{1195--1208}. \DOIprefix\doi{10.1109/TVCG.2015.2502588}.
\bibitem[{Kirsch(1993)}]{Kirsch1993}
\bibinfo{author}{U.~Kirsch}, \bibinfo{journal}{Struct. Optim.}
  \bibinfo{volume}{6} (\bibinfo{year}{1993}) \bibinfo{pages}{143--150}.
  \DOIprefix\doi{10.1007/BF01743505}.
\bibitem[{Amir(2015)}]{Amir2015}
\bibinfo{author}{O.~Amir}, \bibinfo{journal}{Struct. Multidiscip. Optim.}
  \bibinfo{volume}{51} (\bibinfo{year}{2015}) \bibinfo{pages}{41--57}.
  \DOIprefix\doi{10.1007/s00158-014-1098-7}.
\bibitem[{Yoon(2010)}]{Yoon2010a}
\bibinfo{author}{G.-H. Yoon}, \bibinfo{journal}{Comput. Methods Appl. Mech.
  Eng.} \bibinfo{volume}{199} (\bibinfo{year}{2010})
  \bibinfo{pages}{1744--1763}. \DOIprefix\doi{10.1016/j.cma.2010.02.002}.
\bibitem[{Choi et~al.(2019)Choi, Oxberry, White, and Kirchdoerfer}]{Choi2019}
\bibinfo{author}{Y.~Choi}, \bibinfo{author}{G.~Oxberry},
  \bibinfo{author}{D.~White}, \bibinfo{author}{T.~Kirchdoerfer}
  (\bibinfo{year}{2019}). \DOIprefix\doi{10.13140/RG.2.2.16056.08965}.
  \href{http://arxiv.org/abs/1909.11320}{\tt arXiv:1909.11320}.
\bibitem[{Guyan(1965)}]{Guyan1965}
\bibinfo{author}{R.~Guyan}, \bibinfo{journal}{AIAA J.} \bibinfo{volume}{3}
  (\bibinfo{year}{1965}) \bibinfo{pages}{380--380}.
  \DOIprefix\doi{10.2514/3.2874}.
\bibitem[{Irons(1965)}]{Irons1965}
\bibinfo{author}{B.~Irons}, \bibinfo{journal}{AIAA J.} \bibinfo{volume}{3}
  (\bibinfo{year}{1965}) \bibinfo{pages}{961--962}.
  \DOIprefix\doi{10.2514/3.3027}.
\bibitem[{Przemieniecki(1963)}]{Przemieniecki1963}
\bibinfo{author}{J.~Przemieniecki}, \bibinfo{journal}{AIAA J.}
  \bibinfo{volume}{1} (\bibinfo{year}{1963}) \bibinfo{pages}{138--147}.
\bibitem[{Schmit and Hirokazu(1976)}]{Schmit1976}
\bibinfo{author}{L.~Schmit}, \bibinfo{author}{M.~Hirokazu},
  \bibinfo{title}{{Approximation concepts for efficient structural synthesis}},
  \bibinfo{type}{Technical Report}, NASA, \bibinfo{year}{1976}.
\bibitem[{Noor et~al.(1978)Noor, Kamel, and Fulton}]{Noor1978}
\bibinfo{author}{A.~Noor}, \bibinfo{author}{H.~Kamel},
  \bibinfo{author}{R.~Fulton}, \bibinfo{journal}{Comput. Struct.}
  \bibinfo{volume}{8} (\bibinfo{year}{1978}) \bibinfo{pages}{621--632}.
  \DOIprefix\doi{10.1016/0045-7949(78)90100-1}.
\bibitem[{Kron(1955)}]{Kron1955}
\bibinfo{author}{G.~Kron}, \bibinfo{journal}{J. Appl. Mech.}
  \bibinfo{volume}{22} (\bibinfo{year}{1955}) \bibinfo{pages}{235--244}.
\bibitem[{Wilson(1974)}]{Wilson1974}
\bibinfo{author}{E.~Wilson}, \bibinfo{journal}{Int. J. Numer. Methods Eng.}
  \bibinfo{volume}{8} (\bibinfo{year}{1974}) \bibinfo{pages}{198--203}.
  \DOIprefix\doi{10.1002/nme.1620080115}.
\bibitem[{Benscoter(1948)}]{Benscoter1948}
\bibinfo{author}{S.~Benscoter}, \bibinfo{journal}{J. Appl. Mech.}
  \bibinfo{volume}{15} (\bibinfo{year}{1948}) \bibinfo{pages}{303--307}.
\bibitem[{Haynsworth(1968)}]{Haynsworth1968}
\bibinfo{author}{E.~Haynsworth}, \bibinfo{title}{{On the Schur Complement}},
  \bibinfo{year}{1968}.
\bibitem[{Schur(1917)}]{Schur1917}
\bibinfo{author}{I.~Schur}, \bibinfo{journal}{J. f{\"{u}}r die Reine und Angew.
  Math.} \bibinfo{volume}{147} (\bibinfo{year}{1917})
  \bibinfo{pages}{205--232}.
\bibitem[{Botkin and Yang(1989)}]{Botkin1989}
\bibinfo{author}{M.~Botkin}, \bibinfo{author}{R.~Yang}, \bibinfo{journal}{AIAA
  J.} \bibinfo{volume}{29} (\bibinfo{year}{1989}) \bibinfo{pages}{486--488}.
  \DOIprefix\doi{10.2514/3.59924}.
\bibitem[{Gangadharan et~al.(1990)Gangadharan, Haftka, and
  Nikolaidis}]{Gangadharan1990}
\bibinfo{author}{S.~Gangadharan}, \bibinfo{author}{R.~Haftka},
  \bibinfo{author}{E.~Nikolaidis}, \bibinfo{journal}{Commun. Appl. Numer.
  Methods} \bibinfo{volume}{6} (\bibinfo{year}{1990})
  \bibinfo{pages}{161--171}. \DOIprefix\doi{10.1002/cnm.1630060302}.
\bibitem[{Yang and Lu(1996)}]{Yang1996}
\bibinfo{author}{R.~Yang}, \bibinfo{author}{C.~Lu}, \bibinfo{journal}{AIAA J.}
  \bibinfo{volume}{34} (\bibinfo{year}{1996}) \bibinfo{pages}{1533--1535}.
  \DOIprefix\doi{10.2514/3.60028}.
\bibitem[{Yoon et~al.(2007)Yoon, Joung, and Kim}]{Yoon2007}
\bibinfo{author}{G.-H. Yoon}, \bibinfo{author}{Y.-S. Joung},
  \bibinfo{author}{Y.-Y. Kim}, \bibinfo{journal}{Int. J. Numer. Methods Eng.}
  \bibinfo{volume}{69} (\bibinfo{year}{2007}) \bibinfo{pages}{1278--1304}.
  \DOIprefix\doi{10.1002/nme.1808}.
\bibitem[{Xia et~al.(2012)Xia, Zhu, and Zhang}]{Xia2012}
\bibinfo{author}{L.~Xia}, \bibinfo{author}{J.-H. Zhu}, \bibinfo{author}{W.-H.
  Zhang}, \bibinfo{journal}{Struct. Multidiscip. Optim.} \bibinfo{volume}{45}
  (\bibinfo{year}{2012}) \bibinfo{pages}{643--655}.
  \DOIprefix\doi{10.1007/s00158-011-0720-1}.
\bibitem[{Groen et~al.(2017)Groen, Langelaar, Sigmund, and Ruess}]{Groen2017}
\bibinfo{author}{J.~Groen}, \bibinfo{author}{M.~Langelaar},
  \bibinfo{author}{O.~Sigmund}, \bibinfo{author}{M.~Ruess},
  \bibinfo{journal}{Int. J. Numer. Methods Eng.} \bibinfo{volume}{110}
  (\bibinfo{year}{2017}) \bibinfo{pages}{903--920}.
  \DOIprefix\doi{10.1002/nme.5432}.
\bibitem[{Amir and Amir(2019)}]{Amir2019}
\bibinfo{author}{E.~Amir}, \bibinfo{author}{O.~Amir}, \bibinfo{journal}{Struct.
  Multidiscip. Optim.} \bibinfo{volume}{59} (\bibinfo{year}{2019})
  \bibinfo{pages}{165--184}. \DOIprefix\doi{10.1007/s00158-018-2058-4}.
\bibitem[{Zhang and Sun(2006)}]{Zhang2006}
\bibinfo{author}{W.-H. Zhang}, \bibinfo{author}{S.~Sun}, \bibinfo{journal}{Int.
  J. Numer. Methods Eng.} \bibinfo{volume}{68} (\bibinfo{year}{2006})
  \bibinfo{pages}{993--1011}. \DOIprefix\doi{10.1002/nme.1743}.
\bibitem[{Wu et~al.(2019)Wu, Xia, Wang, and Shi}]{Wu2019}
\bibinfo{author}{Z.~Wu}, \bibinfo{author}{L.~Xia}, \bibinfo{author}{S.~Wang},
  \bibinfo{author}{T.~Shi}, \bibinfo{journal}{Comput. Methods Appl. Mech. Eng.}
  \bibinfo{volume}{345} (\bibinfo{year}{2019}) \bibinfo{pages}{602--617}.
  \DOIprefix\doi{10.1016/j.cma.2018.11.003}.
\bibitem[{Wang(2009)}]{Wang2009a}
\bibinfo{author}{M.~Y. Wang}, \bibinfo{journal}{J. Mech. Robot.}
  \bibinfo{volume}{1} (\bibinfo{year}{2009}) \bibinfo{pages}{021011}.
  \DOIprefix\doi{10.1115/1.3056476}.
\bibitem[{Hasse and Campanile(2009)}]{Hasse2009}
\bibinfo{author}{A.~Hasse}, \bibinfo{author}{L.~Campanile},
  \bibinfo{title}{{Compliant mechanisms with selective compliance}},
  \bibinfo{year}{2009}. \DOIprefix\doi{10.1088/0964-1726/18/11/115016}.
\bibitem[{Hasse et~al.(2017)Hasse, Franz, and Mauser}]{Hasse2017}
\bibinfo{author}{A.~Hasse}, \bibinfo{author}{M.~Franz},
  \bibinfo{author}{K.~Mauser}, in: \bibinfo{booktitle}{Mech. Mach. Sci.},
  volume~\bibinfo{volume}{45}, \bibinfo{publisher}{Springer, Cham},
  \bibinfo{year}{2017}, pp. \bibinfo{pages}{227--238}.
  \DOIprefix\doi{10.1007/978-3-319-45387-3_20}.
\bibitem[{Li and Zhu(2019)}]{Li2019a}
\bibinfo{author}{L.~Li}, \bibinfo{author}{X.~Zhu}, \bibinfo{journal}{Struct.
  Multidiscip. Optim.} \bibinfo{volume}{60} (\bibinfo{year}{2019})
  \bibinfo{pages}{1475--1489}. \DOIprefix\doi{10.1007/s00158-019-02278-8}.
\bibitem[{Cook et~al.(2001)Cook, Malkus, Plesha, and Witt}]{Cook2001}
\bibinfo{author}{R.~Cook}, \bibinfo{author}{D.~Malkus},
  \bibinfo{author}{M.~Plesha}, \bibinfo{author}{R.~Witt},
  \bibinfo{title}{{Concepts and Applications of Finite Element Analysis}},
  \bibinfo{edition}{4th editio} ed., \bibinfo{publisher}{John Wiley {\&} Sons,
  Ltd.}, \bibinfo{year}{2001}.
\bibitem[{Arora and Haug(1979)}]{Arora1979}
\bibinfo{author}{J.~Arora}, \bibinfo{author}{E.~Haug}, \bibinfo{journal}{AIAA
  J.} \bibinfo{volume}{17} (\bibinfo{year}{1979}) \bibinfo{pages}{970--974}.
\bibitem[{Belegundu(1986)}]{Belegundu1986}
\bibinfo{author}{A.~Belegundu}, \bibinfo{journal}{J. Struct. Eng. (United
  States)} \bibinfo{volume}{112} (\bibinfo{year}{1986})
  \bibinfo{pages}{1971--1976}.
  \DOIprefix\doi{10.1061/(ASCE)0733-9445(1986)112:8(1971)}.
\bibitem[{Vanderplaats(1980)}]{Vanderplaats1980}
\bibinfo{author}{G.~Vanderplaats}, \bibinfo{journal}{AIAA J.}
  \bibinfo{volume}{18} (\bibinfo{year}{1980}) \bibinfo{pages}{1406--1407}.
\bibitem[{Bends{\o}e and Kikuchi(1988)}]{MartinPhilipBendsoe1988}
\bibinfo{author}{M.~P. Bends{\o}e}, \bibinfo{author}{N.~Kikuchi},
  \bibinfo{journal}{Computer Methods in Applied Mechanics and Engineering}
  \bibinfo{volume}{71} (\bibinfo{year}{1988}) \bibinfo{pages}{197--224}.
  \DOIprefix\doi{10.1016/0045-7825(88)90086-2}.
\bibitem[{Yano et~al.(2012)Yano, Penn, Konidaris, and Patera}]{Yano2012}
\bibinfo{author}{M.~Yano}, \bibinfo{author}{J.~D. Penn},
  \bibinfo{author}{G.~Konidaris}, \bibinfo{author}{A.~T. Patera},
  \bibinfo{number}{September}, \bibinfo{publisher}{MIT}, \bibinfo{year}{2012}.
\bibitem[{Boyd et~al.(2004)Boyd, Boyd, and Vandenberghe}]{boyd2004convex}
\bibinfo{author}{S.~Boyd}, \bibinfo{author}{S.~P. Boyd},
  \bibinfo{author}{L.~Vandenberghe}, \bibinfo{title}{Convex optimization},
  \bibinfo{publisher}{Cambridge university press}, \bibinfo{year}{2004}.
\bibitem[{Saad(2003)}]{saad2003iterative}
\bibinfo{author}{Y.~Saad}, \bibinfo{title}{Iterative methods for sparse linear
  systems}, \bibinfo{publisher}{SIAM}, \bibinfo{year}{2003}.
\bibitem[{Bruns and Tortorelli(2001)}]{Bruns2001}
\bibinfo{author}{T.~Bruns}, \bibinfo{author}{D.~Tortorelli},
  \bibinfo{journal}{Comput. Methods Appl. Mech. Eng.} \bibinfo{volume}{190}
  (\bibinfo{year}{2001}) \bibinfo{pages}{3443--3459}.
  \DOIprefix\doi{10.1016/S0045-7825(00)00278-4}.
\bibitem[{Sigmund and Clausen(2007)}]{Sigmund2007}
\bibinfo{author}{O.~Sigmund}, \bibinfo{author}{P.~Clausen},
  \bibinfo{journal}{Comput. Methods Appl. Mech. Eng.} \bibinfo{volume}{196}
  (\bibinfo{year}{2007}) \bibinfo{pages}{1874--1889}.
  \DOIprefix\doi{10.1016/j.cma.2006.09.021}.
\bibitem[{Bends{\o}e and Sigmund(2004)}]{Bendsoe2004}
\bibinfo{author}{M.~P. Bends{\o}e}, \bibinfo{author}{O.~Sigmund},
  \bibinfo{title}{{Topology Optimization}}, \bibinfo{publisher}{Springer Berlin
  Heidelberg}, \bibinfo{year}{2004}. \DOIprefix\doi{10.1007/978-3-662-05086-6}.
\bibitem[{Larsen et~al.(1997)Larsen, Sigmund, and Bouwstra}]{Larsen1997}
\bibinfo{author}{U.~Larsen}, \bibinfo{author}{O.~Sigmund},
  \bibinfo{author}{S.~Bouwstra}, \bibinfo{journal}{J. Microelectromechanical
  Syst.} \bibinfo{volume}{6} (\bibinfo{year}{1997}) \bibinfo{pages}{99--106}.
  \DOIprefix\doi{10.1109/84.585787}.

\end{thebibliography}

\appendix
\section{Sensitivity analysis}
\label{sec:sensitivityfull}
Dependencies will, after first mention, be omitted for brevity without further notice.

\subsection{Sensitivities of the reduced system matrix}
\label{sec:sensS}
Consider a response function $g \big[\abf{\tilde{\K}}{}{}{}\left[\abf{\K}{}{}{}\right]\big]$, with
\begin{equation*}
	\abf{\tilde{\K}}{}{}{} := \abf[\mm]{\K}{}{}{} - \abf[\ms]{\K}{f}{}{}\Zi. \tag{\ref{eq:kbarfbar} revisited}
\end{equation*}
Here the system matrix is a function of the design variables, \textit{i.e.} $\abf{\K}{}{}{}\left[\mathbf{x}\right]$. For now we are solely interested in the sensitivities of the reduced system matrix with respect to the system matrix and therefore omit this dependency from here onward.
The full derivative can be written as
\begin{equation*}
	\text{d}g \big[\abf{\tilde{\K}}{}{}{}\left[\abf{\K}{}{}{}\right]\big]  = 
	\der[g]{\abf{\tilde{\K}}{}{}{}} : 
	\der[\abf{\tilde{\K}}{}{}{}]{\abf{\K}{}{}{}}:
	\text{d}\abf{\K}{}{}{}, 
	\tag{\ref{eq:dA2} revisited}
\end{equation*}
where $\der[g]{\abf{\tilde{\K}}{}{}{}}$ is the sensitivity of the response function with respect to $\abf{\tilde{\K}}{}{}{}$ and $\text{d}\abf{\K}{}{}{}$ the sensitivity of the system matrix. In the following we assume both are known.
The aim is to find the fourth order tensor $\der[\abf{\tilde{\K}}{}{}{}]{\abf{\K}{}{}{}}$, \textit{i.e.} the sensitivities of the reduced system matrix with respect to the system matrix.

\subsubsection{Direct method}\label{sec:senskbarkbdirect}
Direct differentiation of \cref{eq:kbarfbar} yields
\begin{equation*}
	\text{d}\abf{\tilde{\K}}{}{}{} = \text{d}\abf[\mm]{\K}{}{}{} - \text{d}\abf[\ms]{\K}{f}{}{} \Zi - \abf[\ms]{\K}{f}{}{} \text{d}\Zi. \tag{\ref{eq:dAk} revisited}
\end{equation*}
Here $\text{d}\Zi$ can be obtained by direct differentiation of \cref{eq:solves}, that is
\begin{equation*}
	\text{d}\Zi =  - \abf[\ss]{\K}{ff}{}{-1} \text{d}\abf[\ss]{\K}{ff}{}{} \Zi + \abf[\ss]{\K}{ff}{}{-1} \text{d}\abf[\sm]{\K}{f}{}{}. \tag{\ref{eq:dZ} revisited}
\end{equation*}
Substitution of \cref{eq:dZ} into \cref{eq:dAk} gives
\begin{equation}
	\text{d}\abf{\tilde{\K}}{}{}{} = \text{d}\abf[\mm]{\K}{}{}{} - \text{d}\abf[\ms]{\K}{f}{}{} \Zi + \Zit \text{d}\abf[\ss]{\K}{ff}{}{} \Zi - \Zit \text{d} \abf[\sm]{\K}{f}{}{}.
	\label{eq:dkbar}
\end{equation}
Substitution of \cref{eq:submatrices} into \cref{eq:dkbar} yields 
\begin{equation}
	\text{d}\abf{\tilde{\K}}{}{}{} = \abf[\m]{\mathbf{S}}{}{}{\tran} \text{d}\abf{\K}{}{}{} \abf[\m]{\mathbf{S}}{}{}{} - \abf[\m]{\mathbf{S}}{}{}{\tran}\text{d}\abf{\K}{}{}{} \abf[\s]{\mathbf{S}}{f}{}{} \Zi + \Zit \abf[\s]{\mathbf{S}}{f}{}{\tran}\text{d}\abf{\K}{}{}{} \abf[\s]{\mathbf{S}}{f}{}{} \Zi - \Zit \abf[\s]{\mathbf{S}}{f}{}{\tran}\text{d} \abf{\K}{}{}{} \abf[\m]{\mathbf{S}}{}{}{}.
\end{equation}
Subsequent substitution of the result into \cref{eq:dA2} gives the expression for the sensitivities of the response function, \textit{i.e.}
\begin{align*}
	\begin{aligned}
		\text{d}g 
		& = \der[g]{\abf{\tilde{\K}}{}{}{}}:\Big( \left(\abf[\m]{\mathbf{S}}{}{}{\tran} - \Zit \abf[\s]{\mathbf{S}}{f}{}{\tran}\right) \text{d}\abf{\K}{}{}{} \left(\abf[\m]{\mathbf{S}}{}{}{} - \abf[\s]{\mathbf{S}}{f}{}{}\Zi \right)\Big)\\
		& = \der[g]{\abf{\tilde{\K}}{}{}{}}: \left(\ses\tran \text{d}\abf{\K}{}{}{} \ses\right),
	\end{aligned} \tag{\ref{eq:kkbar} revisited}
\end{align*}
where
\begin{equation*}
	\ses := \abf[\m]{\mathbf{S}}{}{}{} - \abf[\s]{\mathbf{S}}{f}{}{}\Zi. \tag{\ref{eq:xghz} revisited}
\end{equation*}
We can simplify\footnote{\label{foot:abcd}$\mathbf{A} : \left(\mathbf{B}\mathbf{C}\mathbf{D}\right) = \left(\mathbf{B}\tran \mathbf{A} \otimes \mathbf{D} \right) : \mathbf{C}$} this to obtain
\begin{equation}
	\label{eq:check1}
	\text{d}g = \left(\ses\der[g]{\abf{\tilde{\K}}{}{}{}}  \otimes \ses\right) : \text{d}\abf{\K}{}{}{},
\end{equation}
In addition, rewriting\footref{foot:atbc} of \cref{eq:kkbar} gives
\begin{equation*}
	\der[\abf{\tilde{\K}}{}{}{}]{\abf{\K}{}{}{}} = \ses \otimes \ses. \tag{\ref{eq:xx} revisited}
\end{equation*}

\subsubsection{Adjoint method}\label{sec:senskbarkbadjoint}
Alternatively to the direct method, one can use the adjoint method \citep{Arora1979, Vanderplaats1980}. After augmentation of the response function as defined in \cref{eq:dA2} with \cref{eq:solves}, we get
\begin{equation*}
	\lag\big[\abf{\tilde{\K}}{}{}{}\left[\abf{\K}{}{}{}\right], \abf[\sm]{\bm{\Lambda}}{f}{}{} \big]: = g \big[\abf{\tilde{\K}}{}{}{}\left[\abf{\K}{}{}{}\right]\big] + \abf[\sm]{\bm{\Lambda}}{f}{}{}: \left(\abf[\ss]{\K}{ff}{}{} \Zi - \abf[\sm]{\K}{f}{}{}\right), \tag{\ref{eq:initadj} revisited}
\end{equation*}
where $\lag\big[\abf{\tilde{\K}}{}{}{}\left[\abf{\K}{}{}{}\right], \abf[\sm]{\bm{\Lambda}}{f}{}{}\big]$ is the Lagrangian and $\abf[\sm]{\bm{\Lambda}}{f}{}{} \in \mathbb{R}^{\ncdof \times \nmdof}$ the Lagrange multipliers related to $\Zi$. Full differentiation of both sides gives
\begin{align}
	\label{eq:dL}
	\begin{aligned}
		\text{d}\lag = \der[g]{\abf{\tilde{\K}}{}{}{}} : \text{d}\abf{\tilde{\K}}{}{}{} +\abf[\sm]{\bm{\Lambda}}{f}{}{} : \left(\text{d}\abf[\ss]{\K}{ff}{}{} \Zi + \abf[\ss]{\K}{ff}{}{}\text{d}\Zi - \text{d}\abf[\sm]{\K}{f}{}{} \right).
	\end{aligned}
\end{align}
Substitution of \cref{eq:dAk} into \cref{eq:dL} yields
\begin{align}
	\label{eq:adjoint}
	\begin{aligned}
		\text{d}\lag= \der[g]{\abf{\tilde{\K}}{}{}{}} : \left(\text{d}\abf[\mm]{\K}{}{}{} - \text{d}\abf[\ms]{\K}{f}{}{}\Zi - \abf[\ms]{\K}{f}{}{} \text{d} \Zi\right) + \abf[\sm]{\bm{\Lambda}}{f}{}{}: \left(\text{d}\abf[\ss]{\K}{ff}{}{} \Zi + \abf[\ss]{\K}{ff}{}{} \text{d}\Zi - \text{d}\abf[\sm]{\K}{f}{}{}\right).
	\end{aligned}
\end{align}
Since $\text{d}\Zi$ is computationally expensive to calculate, the aim is to circumvent this by a suitable choice of $\abf[\sm]{\bm{\Lambda}}{f}{}{}$. Collecting terms involving $\text{d}\Zi$ gives
\begin{equation}
	\abf[\sm]{\bm{\Lambda}}{f}{}{} : \left(\abf[\ss]{\K}{ff}{}{} \text{d}\Zi\right) - \der[g]{\abf{\tilde{\K}}{}{}{}} : \left(\abf[\ms]{\K}{f}{}{} \text{d}\Zi\right).
\end{equation}
This can be rewritten\footnote{\label{foot:abc}$\mathbf{A} : \left(\mathbf{B}\mathbf{C}\right) = \left(\mathbf{B}\tran \mathbf{A} \right) : \mathbf{C}$} into
\begin{equation}
	\label{eq:adjointeq}
	\left( \abf[\ss]{\K}{ff}{}{} \abf[\sm]{\bm{\Lambda}}{f}{}{} -\abf[\sm]{\K}{f}{}{}\der[g]{\abf{\tilde{\K}}{}{}{}} \right) : \text{d}\Zi.
\end{equation}
To make the terms involving $\text{d}\Zi$ vanish, the term in parentheses must be zero. This can be achieved by choosing $\abf[\sm]{\bm{\Lambda}}{f}{}{}$ as
\begin{equation*}
	\abf[\sm]{\bm{\Lambda}}{f}{}{} = \abf[\ss]{\K}{ff}{}{-1} \abf[\sm]{\K}{f}{}{} \der[g]{\abf{\tilde{\K}}{}{}{}} = \Zi \der[g]{\abf{\tilde{\K}}{}{}{}}. \tag{\ref{eq:lambdaaa} revisited}
\end{equation*}
Here the self-adjointness property appears; no additional solves are needed to compute $\abf[\sm]{\bm{\Lambda}}{f}{}{}$.
With these multipliers, the terms involving $\text{d}\Zi$ vanish from \cref{eq:adjoint}, that is
\begin{align}
	\label{eq:adjoint3}
	\begin{aligned}
		\text{d}\lag= \der[g]{\abf{\tilde{\K}}{}{}{}} : \left(\text{d}\abf[\mm]{\K}{}{}{} - \text{d}\abf[\ms]{\K}{f}{}{}\Zi\right) + \abf[\sm]{\bm{\Lambda}}{f}{}{}: \left(\text{d}\abf[\ss]{\K}{ff}{}{} \Zi  - \text{d}\abf[\sm]{\K}{f}{}{}\right),
	\end{aligned}
\end{align}
with $\abf[\sm]{\bm{\Lambda}}{f}{}{}$ as defined in \cref{eq:lambdaaa}. Next, substitution of \cref{eq:submatrices} into \cref{eq:adjoint3} yields
\begin{align*}
	\begin{aligned}
		\text{d}\lag& =\der[g]{\abf{\tilde{\K}}{}{}{}} : \left(\abf[\m]{\mathbf{S}}{}{}{\tran}\text{d}\abf{\K}{}{}{}\abf[\m]{\mathbf{S}}{}{}{}
		- \abf[\m]{\mathbf{S}}{}{}{\tran}\text{d}\abf{\K}{}{}{}\abf[\s]{\mathbf{S}}{f}{}{}\Zi\right)+   \abf[\sm]{\bm{\Lambda}}{f}{}{} : \left(\abf[\s]{\mathbf{S}}{f}{}{\tran}\text{d}\abf{\K}{}{}{}\abf[\s]{\mathbf{S}}{f}{}{} \Zi - \abf[\s]{\mathbf{S}}{f}{}{\tran}\text{d}\abf{\K}{}{}{}\abf[\m]{\mathbf{S}}{}{}{}\right)\\
		& = \der[g]{\abf{\tilde{\K}}{}{}{}} : \left(\abf[\m]{\mathbf{S}}{}{}{\tran}\text{d}\abf{\K}{}{}{}\ses\right)-  \abf[\sm]{\bm{\Lambda}}{f}{}{} : \left(\abf[\s]{\mathbf{S}}{f}{}{\tran}\text{d}\abf{\K}{}{}{}\ses\right)\\
		& = \left(\abf[\m]{\mathbf{S}}{}{}{}\der[g]{\abf{\tilde{\K}}{}{}{}} - \abf[\s]{\mathbf{S}}{f}{}{} \abf[\sm]{\bm{\Lambda}}{f}{}{}\right)\tran \text{d}\abf{\K}{}{}{} \ses.
	\end{aligned} \tag{\ref{eq:dLkbar} revisited}
\end{align*}
The reader can verify that substitution of \cref{eq:lambdaaa} into \cref{eq:dLkbar} results in \cref{eq:check1}.

\subsection{Sensitivities of the reduced load}
\label{sec:sensfbar}
Consider a response function $g \left[\tilde{\F}\left[\abf{\K}{}{}{}, \abf[\s]{\F}{f}{}{}, \abf[\s]{\U}{p}{}{}\right]\right]$, with
\begin{equation*}
	\tilde{\F}\left[\abf{\K}{}{}{}, \abf[\s]{\F}{f}{}{}, \abf[\s]{\U}{p}{}{}\right] := \abf[\ms]{\K}{f}{}{} \Zc\left[\abf{\K}{}{}{}, \abf[\s]{\F}{f}{}{}, \abf[\s]{\U}{p}{}{}\right] - \abf[\ms]{\K}{p}{}{}\abf[\s]{\U}{p}{}{}. \tag{\ref{eq:fbar} revisited}
\end{equation*}
By applying the chain-rule, the full derivative of this response function can be written as
\begin{align}
	\begin{aligned}
		\label{eq:dvvv}
		\text{d}g\left[\tilde{\F}\left[\abf{\K}{}{}{}, \abf[\s]{\F}{f}{}{}, \abf[\s]{\U}{p}{}{}\right]\right] ={}&\der[g]{\tilde{\F}} : \text{d}\tilde{\F}\left[\abf{\K}{}{}{}, \abf[\s]{\F}{f}{}{}, \abf[\s]{\U}{p}{}{}\right]\\={}& \der[g]{\tilde{\F}} : \left(\pder[\tilde{\F}]{\abf{\K}{}{}{}}:\text{d}\abf{\K}{}{}{} + \pder[\tilde{\F}]{\abf[\s]{\F}{f}{}{}}:\text{d}\abf[\s]{\F}{f}{}{} + \pder[\tilde{\F}]{\abf[\s]{\U}{p}{}{}}:\text{d}\abf[\s]{\U}{p}{}{}\right).
	\end{aligned}
\end{align}
The aim is to find $\pder[\tilde{\F}]{\abf{\K}{}{}{}}$, $\pder[\tilde{\F}]{\abf[\s]{\F}{f}{}{}}$ and $\pder[\tilde{\F}]{\abf[\s]{\U}{p}{}{}}$.

\subsubsection{Direct method}\label{sec:sensfbarkbdirect}
Direct differentiation of both sides of \cref{eq:fbar} yields
\begin{align}
	\label{eq:dfbar3}
	\begin{aligned}
		\text{d}\tilde{\F} = \text{d}\abf[\ms]{\K}{f}{}{} \Zc  + \abf[\ms]{\K}{f}{}{} \text{d}\Zc - \text{d}\abf[\ms]{\K}{p}{}{} \abf[\s]{\U}{p}{}{} - \abf[\ms]{\K}{p}{}{} \text{d}\abf[\s]{\U}{p}{}{}.
	\end{aligned}
\end{align}
Here $\text{d}\Zc$ can be obtained by direct differentiation of \cref{eq:solvesq}, that is
\begin{equation}
	\label{eq:dzc}
	\text{d} \Zc = \abf[\ss]{\K}{ff}{}{-1} \left(- \text{d}\abf[\ss]{\K}{ff}{}{} \Zc + \text{d}\abf[\ss]{\K}{fp}{}{} \abf[\s]{\U}{p}{}{} -\text{d}\abf[\s]{\F}{f}{}{}+ \abf[\ss]{\K}{fp}{}{} \text{d}\abf[\s]{\U}{p}{}{}  \right).
\end{equation}
Substitution of \cref{eq:dzc} into \cref{eq:dfbar3}, and using the definition of $\Zc$ from \cref{eq:solvesq} yields
\begin{align}
	\label{eq:dfbar4}
	\begin{aligned}
		\text{d}\tilde{\F} = \text{d}\abf[\ms]{\K}{f}{}{} \Zc  + \Zit \left(- \text{d}\abf[\ss]{\K}{ff}{}{} \Zc + \text{d}\abf[\ss]{\K}{fp}{}{} \abf[\s]{\U}{p}{}{} -\text{d}\abf[\s]{\F}{f}{}{}+ \abf[\ss]{\K}{fp}{}{} \text{d}\abf[\s]{\U}{p}{}{}  \right) - \text{d}\abf[\ms]{\K}{p}{}{} \abf[\s]{\U}{p}{}{} - \abf[\ms]{\K}{p}{}{} \text{d}\abf[\s]{\U}{p}{}{}.
	\end{aligned}
\end{align}
For conciseness, we define the variable
\begin{equation}
	\label{eq:defA}
	\mathbf{C}\tran :=  \Zit\abf[\ss]{\K}{fp}{}{} -\abf[\ms]{\K}{p}{}{},
\end{equation}
and substitute this into \cref{eq:dfbar4}, which yields
\begin{align}
	\label{eq:dfbar}
	\begin{aligned}
		\text{d}\tilde{\F} = \text{d}\abf[\ms]{\K}{f}{}{} \Zc - \Zit\text{d}\abf[\ss]{\K}{ff}{}{} \Zc +\Zit \text{d}\abf[\ss]{\K}{fp}{}{} \abf[\s]{\U}{p}{}{} -\text{d}\abf[\ms]{\K}{p}{}{}\abf[\s]{\U}{p}{}{} -\Zit \text{d}\abf[\s]{\F}{f}{}{}+ \mathbf{C}\tran \text{d}\abf[\s]{\U}{p}{}{}.
	\end{aligned}
\end{align}
Subsequent substitution of \cref{eq:submatrices} into \cref{eq:dfbar}, and \cref{eq:dfbar} into \cref{eq:dvvv} gives an expression for the sensitivities of the response function, \textit{i.e.}
\begin{align}
	\label{eq:sensfbar}
	\begin{aligned}
		\text{d}g = \der[g]{\tilde{\F}} :\left( \abf[\m]{\mathbf{S}}{}{}{\tran} \text{d}\abf{\K}{}{}{} \abf[\s]{\mathbf{S}}{f}{}{}\Zc  - \Zit\abf[\s]{\mathbf{S}}{f}{}{\tran} \text{d} \abf{\K}{}{}{} \abf[\s]{\mathbf{S}}{f}{}{} \Zc
		+ \Zit \abf[\s]{\mathbf{S}}{f}{}{\tran} \text{d} \abf{\K}{}{}{}\abf[\s]{\mathbf{S}}{p}{}{}\abf[\s]{\U}{p}{}{}   - \abf[\m]{\mathbf{S}}{}{}{\tran} \text{d}\abf{\K}{}{}{} \abf[\s]{\mathbf{S}}{p}{}{}\abf[\s]{\U}{p}{}{} - \Zit\text{d}\abf[\s]{\F}{f}{}{}+ \mathbf{C}\tran \text{d}\abf[\s]{\U}{p}{}{} \right).
	\end{aligned}
\end{align}
We can simplify\footref{foot:abcd} this to obtain
\begin{equation}
	\label{eq:fbarall}
	\text{d}g = \left(\ses\der[g]{\tilde{\F}} \otimes  \mathbf{B}\right):\text{d}\abf{\K}{}{}{} - \Zi\der[g]{\tilde{\F}} : \text{d}\abf[\s]{\F}{f}{}{}+ \mathbf{C}\der[g]{\tilde{\F}} : \text{d}\abf[\s]{\U}{p}{}{} ,
\end{equation}
with
\begin{equation*}
	\mathbf{B} := \abf[\s]{\mathbf{S}}{f}{}{} \Zc - \abf[\s]{\mathbf{S}}{p}{}{} \abf[\s]{\U}{p}{}{}. \tag{\ref{eq:Y} revisited}
\end{equation*}
And thus
\begin{equation*}
	\label{eq:one}
	\pder[\tilde{\F}]{\abf{\K}{}{}{}} = \ses \otimes  \mathbf{B}. \tag{\ref{eq:dfdkb} revisited}
\end{equation*}
Note that all terms in \cref{eq:fbarall} are calculated in the corresponding forward analysis, and thus no additional solves are required.

\subsubsection{Adjoint method}\label{sec:sensfbarkbadjoint}
After augmentation of the response function with the linear system of equations from \cref{eq:solvesq}, we get
\begin{equation}
	\begin{aligned}
		\lag\left[\tilde{\F}\left[\abf{\K}{}{}{}, \abf[\s]{\F}{f}{}{}, \abf[\s]{\U}{p}{}{}\right], \abf[\s]{\bm{\Lambda}}{f}{}{} \right]: = 
		g \big[\tilde{\F}\left[\abf{\K}{}{}{}, \abf[\s]{\F}{f}{}{}, \abf[\s]{\U}{p}{}{}\right]\big]  - \abf[\s]{\bm{\Lambda}}{f}{}{} : \left(\abf[\ss]{\K}{ff}{}{} \Zc + \abf[\s]{\F}{f}{}{} - \abf[\ss]{\K}{fp}{}{}\abf[\s]{\U}{p}{}{} \right),
	\end{aligned}
\end{equation}
where $\abf[\s]{\bm{\Lambda}}{f}{}{}\in \mathbb{R}^{\ncdof \times l}$ are the Lagrange multipliers related to $\Zc$.
Differentiation of both sides gives
\begin{equation}
	\label{eq:dll}
	\begin{aligned}
		\text{d}\lag =  \der[g]{\tilde{\F}} : \text{d}\tilde{\F} 
		- \abf[\s]{\bm{\Lambda}}{f}{}{} : \big(\text{d}\abf[\ss]{\K}{ff}{}{}\Zc + \abf[\ss]{\K}{ff}{}{}\text{d}\Zc 
		- \text{d}\abf[\ss]{\K}{fp}{}{}\abf[\s]{\U}{p}{}{} + \text{d}\abf[\s]{\F}{f}{}{}- \abf[\ss]{\K}{fp}{}{}\text{d}\abf[\s]{\U}{p}{}{} \big).
	\end{aligned}
\end{equation}
Substitution of \cref{eq:dfbar3} into \cref{eq:dll} gives
\begin{equation}
	\label{eq:adjointfbar}
	\begin{aligned}
		\text{d}\lag =  \der[g]{\tilde{\F}} : \big(\text{d}\abf[\ms]{\K}{f}{}{} \Zc + \abf[\ms]{\K}{f}{}{} \text{d} \Zc -\text{d}\abf[\ms]{\K}{p}{}{}\abf[\s]{\U}{p}{}{} - \abf[\ms]{\K}{p}{}{} \text{d}\abf[\s]{\U}{p}{}{}\big) 
		- \abf[\s]{\bm{\Lambda}}{f}{}{} : \big(\text{d}\abf[\ss]{\K}{ff}{}{}\Zc + \abf[\ss]{\K}{ff}{}{}\text{d}\Zc 
		- \text{d}\abf[\ss]{\K}{fp}{}{}\abf[\s]{\U}{p}{}{} + \text{d}\abf[\s]{\F}{f}{}{}- \abf[\ss]{\K}{fp}{}{}\text{d}\abf[\s]{\U}{p}{}{} \big).
	\end{aligned}
\end{equation}
Since $\text{d}\Zc$ is computationally expensive to calculate, aim is to circumvent this by a suitable choice of $\abf[\s]{\bm{\Lambda}}{f}{}{}$. Collecting terms involving $\text{d}\Zc$ gives
\begin{equation}
	\der[g]{\tilde{\F}}: \left(\abf[\ms]{\K}{f}{}{} \text{d}\Zc\right) - \abf[\s]{\bm{\Lambda}}{f}{}{} :  \left(\abf[\ss]{\K}{ff}{}{}\text{d}\Zc\right).
\end{equation}
This can be rewritten\footref{foot:abc} into
\begin{equation}
	\left(\abf[\sm]{\K}{}{}{}\der[g]{\tilde{\F}}  - \abf[\ss]{\K}{}{}{}\abf[\s]{\bm{\Lambda}}{f}{}{} \right) : \text{d}\Zc.
\end{equation}
To make the terms $\text{d}\Zc$ vanish we choose $\abf[\s]{\bm{\Lambda}}{f}{}{}$ as
\begin{equation}\label{eq:theta}
	\abf[\s]{\bm{\Lambda}}{f}{}{} = \abf[\ss]{\K}{ff}{}{-1}\abf[\sm]{\K}{f}{}{} \der[g]{\tilde{\F}} = \Zi \der[g]{\tilde{\F}}.
\end{equation}
Note again that no adjoint solve is necessary.
Substitution of \cref{eq:submatrices} into \cref{eq:adjointfbar} yields
\begin{equation}
	\label{eq:laggg}
	\text{d}\lag =  \der[g]{\tilde{\F}} : \big(\abf[\m]{\mathbf{S}}{}{}{\tran}\text{d}\abf{\K}{}{}{} \abf[\s]{\mathbf{S}}{f}{}{} \Zc  -\abf[\m]{\mathbf{S}}{}{}{\tran}\text{d}\abf{\K}{}{}{}\abf[\s]{\mathbf{S}}{p}{}{}\abf[\s]{\U}{p}{}{} - \abf[\ms]{\K}{p}{}{} \text{d}\abf[\s]{\U}{p}{}{}\big)
	-  \abf[\s]{\bm{\Lambda}}{f}{}{}: \big(\abf[\s]{\mathbf{S}}{f}{}{\tran}\text{d}\abf{\K}{}{}{}\abf[\s]{\mathbf{S}}{f}{}{}\Zc 
	- \abf[\s]{\mathbf{S}}{f}{}{\tran}\text{d}\abf{\K}{}{}{}\abf[\s]{\mathbf{S}}{p}{}{}\abf[\s]{\U}{p}{}{} + \text{d}\abf[\s]{\F}{f}{}{}- \abf[\ss]{\K}{fp}{}{}\text{d}\abf[\s]{\U}{p}{}{} \big),
\end{equation}
with $\abf[\s]{\bm{\Lambda}}{f}{}{}$ as defined in \cref{eq:theta}. The reader can verify that substitution of \cref{eq:theta} into \cref{eq:laggg} results in \cref{eq:fbarall}.

\subsection{Sensitivities of the state}
\label{sec:sensubf}
Often responses are a function of the state, \textit{i.e.} $g\left[\abf[\m]{\U}{f}{}{}\left[\abf{\tilde{\K}}{}{}{}\left[\abf{\K}{}{}{}\right], \tilde{\F}\left[\abf{\K}{}{}{}, \abf[\s]{\F}{f}{}{}, \abf[\s]{\U}{p}{}{}{}\right], \abf[\m]{\F}{f}{}{}{}, \abf[\m]{\U}{p}{}{}{}\right]\right]$. 
Note that this may be the state of multiple analyses, that is $g \left[\abf[\m]{\U}{f}{}{\i[1]}, \abf[\m]{\U}{f}{}{\i[2]},..., \abf[\m]{\U}{f}{}{\i[a]}\right]$, where $a$ is the total number of analyses based upon $\abf{\tilde{\K}}{}{}{}$. 
Therefore, one is required to determine $\text{d}\abf[\m]{\U}{f}{}{\i[i]}{}\left[\abf{\tilde{\K}}{}{}{}, \tilde{\F}, \abf[\m]{\F}{f}{}{\i[i]}{}, \abf[\m]{\U}{p}{}{\i[i]}{}\right]$ for all $i\in \setxdof$. For clarity, we omit the superscript $i$ from here on (assume a single analysis set). Taking the full derivative of $g\left[\abf[\m]{\U}{f}{}{}\left[\abf{\tilde{\K}}{}{}{}, \tilde{\F}, \abf[\m]{\F}{f}{}{}{}, \abf[\m]{\U}{p}{}{}{}\right]\right]$ yields
\begin{equation*}
	\label{eq:dubfshort}
	\text{d} g = \der[g]{\abf[\m]{\U}{f}{}{}} : \text{d} \abf[\m]{\U}{f}{}{}  = \der[g]{\abf[\m]{\U}{f}{}{}} : \left(
	\frac{\partial \abf[\m]{\U}{f}{}{}}{\partial \abf{\tilde{\K}}{}{}{}} : 
	\text{d}\abf{\tilde{\K}}{}{}{} + \frac{\partial \abf[\m]{\U}{f}{}{}}{\partial \tilde{\F}}:\text{d}\tilde{\F} + \frac{\partial \abf[\m]{\U}{f}{}{}}{\partial \abf[\m]{\F}{f}{}{}{}}:\text{d}\abf[\m]{\F}{f}{}{}{} + \frac{\partial \abf[\m]{\U}{f}{}{}}{\partial \abf[\m]{\U}{p}{}{}{}}:\text{d}\abf[\m]{\U}{p}{}{}{}\right) , 
	\tag{\ref{eq:dguf} revisited}
\end{equation*}
where we assume $\der[g]{\abf[\m]{\U}{f}{}{}}$ to be known. The terms $\frac{\partial \abf[\m]{\U}{f}{}{}}{\partial \abf{\tilde{\K}}{}{}{}}$, $\frac{\partial \abf[\m]{\U}{f}{}{}}{\partial \tilde{\F}}$, $\frac{\partial \abf[\m]{\U}{f}{}{}}{\partial \abf[\m]{\F}{f}{}{}{}}$ and $\frac{\partial \abf[\m]{\U}{f}{}{}}{\partial \abf[\m]{\U}{p}{}{}{}}$ are to be determined.
To this end, we augment the response function with the linear system of equations from \cref{eq:y}, which yields
\begin{equation}
	\lag\left[\abf[\m]{\U}{f}{}{}, \abf[\m]{\bm{\Lambda}}{f}{}{} \right]:= g \left[\abf[\m]{\U}{f}{}{}\right] - \abf[\m]{\bm{\Lambda}}{f}{}{} : \left(\abf[]{\tilde{\K}}{ff}{}{} \abf[\m]{\U}{f}{}{} - \abf[\m]{\F}{f}{}{}{} + \abf[]{\tilde{\K}}{fp}{}{} \abf[\m]{\U}{p}{}{}{} - \tilde{\F}_\text{f}\right)
\end{equation}
where $\lag\big[\abf[\m]{\U}{f}{}{}, \abf[\m]{\bm{\Lambda}}{f}{}{}\big]$ is the Lagrangian and $\abf[\m]{\bm{\Lambda}}{f}{}{} \in \mathbb{R}^{\nfdof \times l}$ the Lagrange multipliers related to $\abf[\m]{\U}{f}{}{}$. Full differentiation of both sides gives
\begin{equation}
	\label{eq:dlagman}
	\text{d} \lag = \der[g]{\abf[\m]{\U}{f}{}{}} : \text{d} \abf[\m]{\U}{f}{}{} - \abf[\m]{\bm{\Lambda}}{f}{}{} : \left(\text{d}\abf[]{\tilde{\K}}{ff}{}{} \abf[\m]{\U}{f}{}{} + \abf[]{\tilde{\K}}{ff}{}{} \text{d}\abf[\m]{\U}{f}{}{} - \text{d}\abf[\m]{\F}{f}{}{}{} + \text{d}\abf[]{\tilde{\K}}{fp}{}{} \abf[\m]{\U}{p}{}{}{} + \abf[]{\tilde{\K}}{fp}{}{}\text{d}\abf[\m]{\U}{p}{}{}{} - \text{d}\tilde{\F}_\text{f}\right).
\end{equation}
Since $\text{d}\abf[\m]{\U}{f}{}{}$ is computationally expensive to calculate, aim is to circumvent this by a suitable choice of $\abf[\m]{\bm{\Lambda}}{f}{}{}$. Collecting terms involving $\text{d}\abf[\m]{\U}{f}{}{}$ gives
\begin{equation}
	\der[g]{\abf[\m]{\U}{f}{}{}} : \text{d} \abf[\m]{\U}{f}{}{}- \abf[\m]{\bm{\Lambda}}{f}{}{} : \abf[]{\tilde{\K}}{ff}{}{} \text{d}\abf[\m]{\U}{f}{}{}.
\end{equation}
This can be rewritten\footref{foot:abc} into
\begin{equation}
	\left(\der[g]{\abf[\m]{\U}{f}{}{}} - \abf[]{\tilde{\K}}{ff}{}{}\abf[\m]{\bm{\Lambda}}{f}{}{}\right) : \text{d}\abf[\m]{\U}{f}{}{}.
\end{equation}
To make the terms $\text{d}\abf[\m]{\U}{f}{}{}$ vanish we choose $\abf[\m]{\bm{\Lambda}}{f}{}{}$ as
\begin{equation*}
	\abf{\tilde{\K}}{ff}{}{} \abf[\m]{\bm{\Lambda}}{f}{}{} = \der[g]{\abf[\m]{\U}{f}{}{}}. \tag{\ref{eq:yjadjoint2} revisited}
\end{equation*}
Note that to obtain $\abf[\m]{\bm{\Lambda}}{f}{}{}$ one has to solve an additional system of equations. Subsequent substitution of \cref{eq:submatrices} into \cref{eq:dlagman} gives
\begin{equation}
	\label{eq:dfdkbar}
	\text{d}\lag = \abf[\m]{\bm{\Lambda}}{f}{}{} : \left(-\abf[]{\tilde{\mathbf{S}}}{f}{}{\tran} \text{d}\abf{\tilde{\K}}{}{}{} \abf[\m]{\U}{}{}{} + \abf[]{\tilde{\mathbf{S}}}{f}{}{\tran}\text{d}\tilde{\F} + \text{d}\abf[\m]{\F}{f}{}{}{} - \abf[]{\tilde{\K}}{fp}{}{} \text{d}\abf[\m]{\U}{p}{}{}{} \right),
\end{equation}
where we also used
\begin{equation}
	\label{eq:ubmm}
	\abf[\m]{\U}{}{}{} = \abf[]{\tilde{\mathbf{S}}}{f}{}{} \abf[\m]{\U}{f}{}{} + \abf[]{\tilde{\mathbf{S}}}{p}{}{} \abf[\m]{\U}{p}{}{}{}.
\end{equation}
Note that \cref{eq:dfdkbar} can be rewritten as
\begin{equation}
	\label{eq:dfdkbar2}
	\text{d}\lag = \left(-\abf[]{\tilde{\mathbf{S}}}{f}{}{}\abf[\m]{\bm{\Lambda}}{f}{}{} \otimes \abf[\m]{\U}{}{}{} \right) : \text{d}\abf{\tilde{\K}}{}{}{} + \abf[]{\tilde{\mathbf{S}}}{f}{}{}\abf[\m]{\bm{\Lambda}}{f}{}{} : \text{d}\tilde{\F} + \abf[\m]{\bm{\Lambda}}{f}{}{} : \text{d}\abf[\m]{\F}{f}{}{}{} - \abf[]{\tilde{\K}}{pf}{}{} \abf[\m]{\bm{\Lambda}}{f}{}{} : \text{d}\abf[\m]{\U}{p}{}{}{}.
\end{equation}
Substitution of \cref{eq:xdfdkx,eq:dfbar} into \cref{eq:dfdkbar2} yields
\begin{equation}
	\text{d}\lag = \left(\ses \abf[]{\tilde{\mathbf{S}}}{f}{}{} \abf[\m]{\bm{\Lambda}}{f}{}{} \otimes \mathbf{D}\right) :\text{d}\abf{\K}{}{}{}  - \Zi \abf[]{\tilde{\mathbf{S}}}{f}{}{} \abf[\m]{\bm{\Lambda}}{f}{}{} :\text{d}\abf[\s]{\F}{f}{}{}+ \mathbf{C}\abf[]{\tilde{\mathbf{S}}}{f}{}{} \abf[\m]{\bm{\Lambda}}{f}{}{} :\text{d}\abf[\s]{\U}{p}{}{} + \abf[\m]{\bm{\Lambda}}{f}{}{} :\text{d}\abf[\m]{\F}{f}{}{}{}- \abf[]{\tilde{\K}}{pf}{}{} \abf[\m]{\bm{\Lambda}}{f}{}{} : \text{d}\abf[\m]{\U}{p}{}{}{} ,
\end{equation}
with
\begin{equation}
	\label{eq:defW}
	\mathbf{D}:= \mathbf{B} - \ses\abf[\m]{\U}{}{}{}.
\end{equation}
Thus to obtain $\text{d}\abf[\m]{\U}{f}{}{}$ one requires an extra solve or matrix-vector product (depending on the self-adjointness), \cref{eq:yjadjoint2}, per analysis based on the reduced system.

\subsection{Sensitivities of the reaction load}
\label{sec:sensfbp}
From the solution of \cref{eq:y} the reaction loads can be calculated via
\begin{equation*}
	\abf[\m]{\F}{p}{}{} = \abf[]{\tilde{\K}}{pf}{}{}\abf[\m]{\U}{f}{}{} + \abf[]{\tilde{\K}}{pp}{}{} \abf[\m]{\U}{p}{}{}{} - \tilde{\F}_\text{p}, \tag{\ref{eq:fb} revisited}
\end{equation*}
Note that here we have again omitted the subscript $i$. Sensitivities of a response function $g\left[\abf[\m]{\F}{p}{}{}\left[\abf{\tilde{\K}}{}{}{}, \tilde{\F}, \abf[\m]{\U}{f}{}{}, \abf[\m]{\U}{p}{}{}{}\right]\right]$ are defined as
\begin{equation}
	\label{eq:dfbpshort}
	\text{d} g = \der[g]{\abf[\m]{\F}{p}{}{}} : \text{d} \abf[\m]{\F}{p}{}{}.
\end{equation}
To this end, we augment the response function with the linear system of equations from \cref{eq:y}, which yields
\begin{equation}
	\lag\left[\abf[\m]{\F}{p}{}{}, \abf[\m]{\bm{\Lambda}}{f}{}{}\right] := g \left[\abf[\m]{\F}{p}{}{}\right] - \abf[\m]{\bm{\Lambda}}{f}{}{} : \left(\abf[]{\tilde{\K}}{ff}{}{} \abf[\m]{\U}{f}{}{} - \abf[\m]{\F}{f}{}{}{} + \abf[]{\tilde{\K}}{fp}{}{} \abf[\m]{\U}{p}{}{}{} - \tilde{\F}_\text{f}\right)
\end{equation}
where $\lag\left[\abf[\m]{\F}{p}{}{}, \abf[\m]{\bm{\Lambda}}{f}{}{}\right]$ is the Lagrangian and $\abf[\m]{\bm{\Lambda}}{f}{}{} \in \mathbb{R}^{\nfdof \times l}$ the Lagrange multipliers related to $\abf[\m]{\U}{f}{}{}$. Full differentiation of both sides gives
\begin{equation}
	\label{eq:dlagman2}
	\text{d} \lag = \der[g]{\abf[\m]{\F}{p}{}{}} : \text{d} \abf[\m]{\F}{p}{}{} - \abf[\m]{\bm{\Lambda}}{f}{}{} : \left(\text{d}\abf[]{\tilde{\K}}{ff}{}{} \abf[\m]{\U}{f}{}{} + \abf[]{\tilde{\K}}{ff}{}{} \text{d}\abf[\m]{\U}{f}{}{} - \text{d}\abf[\m]{\F}{f}{}{}{} + \text{d}\abf[]{\tilde{\K}}{fp}{}{} \abf[\m]{\U}{p}{}{}{} + \abf[]{\tilde{\K}}{fp}{}{}\text{d}\abf[\m]{\U}{p}{}{}{} - \text{d}\tilde{\F}_\text{f}\right),
\end{equation}
with
\begin{equation}
	\label{eq:dfbp}
	\text{d} \abf[\m]{\F}{p}{}{} = \text{d} \abf[]{\tilde{\K}}{pf}{}{} \abf[\m]{\U}{f}{}{} + \abf[]{\tilde{\K}}{pf}{}{} \text{d}\abf[\m]{\U}{f}{}{} + \text{d}\abf[]{\tilde{\K}}{pp}{}{} \abf[\m]{\U}{p}{}{}{} + \abf[]{\tilde{\K}}{pp}{}{} \text{d}\abf[\m]{\U}{p}{}{}{} - \text{d}\tilde{\F}_\text{p}.
\end{equation}
Since $\text{d}\abf[\m]{\U}{f}{}{}$ is computationally expensive to calculate, aim is to circumvent this by a suitable choice of $\abf[\m]{\bm{\Lambda}}{f}{}{}$. Collecting terms involving $\text{d}\abf[\m]{\U}{f}{}{}$ gives
\begin{equation}
	\der[g]{\abf[\m]{\F}{p}{}{}} : \abf[]{\tilde{\K}}{pf}{}{} \text{d} \abf[\m]{\U}{f}{}{}- \abf[\m]{\bm{\Lambda}}{f}{}{} : \abf[]{\tilde{\K}}{ff}{}{} \text{d}\abf[\m]{\U}{f}{}{}.
\end{equation}
This can be rewritten\footref{foot:abc} into
\begin{equation}
	\left(\abf[]{\tilde{\K}}{fp}{}{}\der[g]{\abf[\m]{\F}{p}{}{}} - \abf[]{\tilde{\K}}{ff}{}{} \abf[\m]{\bm{\Lambda}}{f}{}{}\right) : \text{d}\abf[\m]{\U}{f}{}{}.
\end{equation}
To make the terms $\text{d}\abf[\m]{\U}{f}{}{}$ vanish, the term in parentheses must be zero. This can be achieved by choosing $\abf[\m]{\bm{\Lambda}}{f}{}{}$ as
\begin{equation}
	\label{eq:lambdafbp}
	\abf[]{\tilde{\K}}{ff}{}{} \abf[\m]{\bm{\Lambda}}{f}{}{} = \abf[]{\tilde{\K}}{fp}{}{} \der[g]{\abf[\m]{\F}{p}{}{}}.
\end{equation}
Note that to obtain $\abf[\m]{\bm{\Lambda}}{f}{}{}$ one has to solve a system of equations. 
Substitution of \cref{eq:dfbp,eq:submatrices,eq:ubmm} into \cref{eq:dlagman2} and rewriting yields
\begin{equation}
	\text{d} \lag = \der[g]{\abf[\m]{\F}{p}{}{}} : \left(\abf[]{\tilde{\mathbf{S}}}{p}{}{\tran} \text{d}\abf{\tilde{\K}}{}{}{} \abf[\m]{\U}{}{}{} + \abf[]{\tilde{\K}}{pp}{}{} \text{d}\abf[\m]{\U}{p}{}{}{} - \abf[]{\tilde{\mathbf{S}}}{p}{}{\tran} \text{d}\tilde{\F}\right) -\abf[\m]{\bm{\Lambda}}{f}{}{} : \left(\abf[]{\tilde{\mathbf{S}}}{f}{}{\tran} \text{d}\abf{\tilde{\K}}{}{}{} \abf[\m]{\U}{}{}{}  - \text{d}\abf[\m]{\F}{f}{}{}{}  + \abf[]{\tilde{\K}}{fp}{}{}\text{d}\abf[\m]{\U}{p}{}{}{} - \abf[]{\tilde{\mathbf{S}}}{f}{}{\tran}\text{d}\tilde{\F}\right)
\end{equation}
Subsequent substitution of \cref{eq:xdfdkx,eq:dfbar} and simplifying using \cref{eq:defA,eq:defW} yields
\begin{equation}
	\text{d}g = \left(\ses\mathbf{E}  \otimes \mathbf{D}\right) : \text{d}\abf{\K}{}{}{}  - \Zi\mathbf{E} :\text{d}\abf[\s]{\F}{f}{}{} + \mathbf{C}\mathbf{E}:\text{d}\abf[\s]{\U}{p}{}{} + \abf[\m]{\bm{\Lambda}}{f}{}{} :\text{d}\abf[\m]{\F}{f}{}{}{} + \left(\abf[]{\tilde{\K}}{pp}{}{} \der[g]{\abf[\m]{\F}{p}{}{}}-\abf[]{\tilde{\K}}{pf}{}{} \abf[\m]{\bm{\Lambda}}{f}{}{}\right):\text{d}\abf[\m]{\U}{p}{}{}{} ,
\end{equation}
with temporary\footnote{Only used within this section.} variable
\begin{equation}
	\mathbf{E} := \abf[]{\tilde{\mathbf{S}}}{f}{}{} \abf[\m]{\bm{\Lambda}}{f}{}{} - \abf[]{\tilde{\mathbf{S}}}{p}{}{} \der[g]{\abf[\m]{\F}{p}{}{}}.
\end{equation}
Thus to obtain $\text{d}\abf[\m]{\F}{p}{}{}$ one requires an extra solve or matrix-vector product (depending on the self-adjointness), \cref{eq:lambdafbp}, per analysis based on using the reduced system.

\subsection{Sensitivities of the state on the secondary DOFs}
\label{sec:sensubc}
After obtaining the solution to the state of the primary DOFs one can obtain the solution of the free secondary DOFs via
\begin{equation*}
	\abf[\s]{\U}{f}{}{} = - \left(\Zi \abf[\m]{\U}{}{}{} + \Zc\right). \tag{\ref{eq:retubc} revisited}
\end{equation*}
We augment the response function with the linear systems of equations from \cref{eq:solves,eq:solvesq,eq:y}, that is
\begin{align}
	\begin{aligned}
		\lag\left[\abf[\s]{\U}{f}{}{},\abf[\sm]{\bm{\Lambda}}{f}{}{},\abf[\s]{\bm{\Lambda}}{f}{}{}, \abf[\m]{\bm{\Lambda}}{f}{}{}\right] := 
		g \left[\abf[\s]{\U}{f}{}{}\right]
		&-\abf[\sm]{\bm{\Lambda}}{f}{}{}: \left(\abf[\ss]{\K}{ff}{}{}\Zi - \abf[\sm]{\K}{f}{}{}\right)\\
		& -\abf[\s]{\bm{\Lambda}}{f}{}{} : \left(\abf[\ss]{\K}{ff}{}{} \Zc - \abf[\ss]{\K}{fp}{}{}\abf[\s]{\U}{p}{}{} + \abf[\s]{\F}{f}{}{}\right)\\
		& - \abf[\m]{\bm{\Lambda}}{f}{}{} : \left(\abf[]{\tilde{\K}}{ff}{}{}\abf[\m]{\U}{f}{}{} -\abf[\m]{\F}{f}{}{}{} + \abf[]{\tilde{\K}}{fp}{}{}\abf[\m]{\U}{p}{}{}{} - \tilde{\F}_\text{f}\right).
	\end{aligned}
\end{align}
Full differentiation of both sides gives
\begin{equation}
	\label{eq:lagh}
	\begin{aligned}
		\text{d}\lag = &-\der[g]{\abf[\s]{\U}{f}{}{}} : \left(\text{d}\Zi \abf[\m]{\U}{}{}{} + \Zi \text{d}\abf[\m]{\U}{}{}{} + \text{d}\Zc\right)\\ 
		&- \abf[\sm]{\bm{\Lambda}}{f}{}{} : \left(\text{d}\abf[\ss]{\K}{ff}{}{} \Zi + \abf[\ss]{\K}{ff}{}{} \text{d}\Zi - \text{d}\abf[\sm]{\K}{f}{}{}\right)\\
		& -\abf[\s]{\bm{\Lambda}}{f}{}{} : \left(\text{d}\abf[\ss]{\K}{ff}{}{} \Zc + \abf[\ss]{\K}{ff}{}{}\text{d}\Zc - \text{d}\abf[\ss]{\K}{fp}{}{}\abf[\s]{\U}{p}{}{} - \abf[\ss]{\K}{fp}{}{} \text{d}\abf[\s]{\U}{p}{}{} + \text{d}\abf[\s]{\F}{f}{}{}\right)\\
		& - \abf[\m]{\bm{\Lambda}}{f}{}{}: \left(\text{d}\abf[]{\tilde{\K}}{ff}{}{}\abf[\m]{\U}{f}{}{}  + \abf[]{\tilde{\K}}{ff}{}{} \text{d}\abf[\m]{\U}{f}{}{} - \text{d}\abf[\m]{\F}{f}{}{}{} + \text{d}\abf[]{\tilde{\K}}{fp}{}{}\abf[\m]{\U}{p}{}{}{} + \abf[]{\tilde{\K}}{fp}{}{}\text{d}\abf[\m]{\U}{p}{}{}{} - \text{d}\tilde{\F}_\text{f}\right).
	\end{aligned}
\end{equation}
Since $\text{d}\Zi$, $\text{d}\Zc$ and $\text{d}\abf[\m]{\U}{f}{}{}$ are computationally expensive to calculate, aim is to circumvent this by a suitable choices of $\abf[\sm]{\bm{\Lambda}}{f}{}{}$, $\abf[\s]{\bm{\Lambda}}{f}{}{}$  and $\abf[\m]{\bm{\Lambda}}{f}{}{}$. 
Collecting terms involving $\text{d}\abf[\m]{\U}{f}{}{}$ gives
\begin{equation}
	\left(-\abf[]{\tilde{\mathbf{S}}}{f}{}{\tran} \Zit \der[g]{\abf[\s]{\U}{f}{}{}} - \abf[]{\tilde{\K}}{ff}{}{} \abf[\m]{\bm{\Lambda}}{f}{}{}\right) : \text{d}\abf[\m]{\U}{f}{}{},
\end{equation}
where we used
\begin{equation}
	\label{eq:dum}
	\text{d}\abf[\m]{\U}{}{}{} = \abf[]{\tilde{\mathbf{S}}}{f}{}{} \text{d}\abf[\m]{\U}{f}{}{} + \abf[]{\tilde{\mathbf{S}}}{p}{}{}\text{d}\abf[\m]{\U}{p}{}{}{}.
\end{equation}
To make the terms $\text{d}\abf[\m]{\U}{f}{}{}$ vanish we define $\abf[\m]{\bm{\Lambda}}{f}{}{}$ as
\begin{equation}
	\abf[]{\tilde{\K}}{ff}{}{} \abf[\m]{\bm{\Lambda}}{f}{}{} = -\abf[]{\tilde{\mathbf{S}}}{f}{}{\tran} \Zit \der[g]{\abf[\s]{\U}{f}{}{}},
\end{equation}
which requires an additional solve using the condensed system.
Collecting terms involving $\text{d}\Zc$ gives
\begin{equation}
	\left(-\der[g]{\abf[\s]{\U}{f}{}{}} - \abf[\ss]{\K}{ff}{}{} \abf[\s]{\bm{\Lambda}}{f}{}{}\right) : \text{d}\Zc.
\end{equation}
To make the terms $\text{d}\Zc$ vanish we define $\abf[\s]{\bm{\Lambda}}{f}{}{}$ as
\begin{equation}
	\label{eq:Lambdaadj}
	\abf[\ss]{\K}{ff}{}{} \abf[\s]{\bm{\Lambda}}{f}{}{} = -\der[g]{\abf[\s]{\U}{f}{}{}},
\end{equation}
for which an additional full system solve is needed.
Finally, collecting terms involving $\text{d}\Zi$ gives
\begin{equation}
	\left(-\der[g]{\abf[\s]{\U}{f}{}{}} \abf[\m]{\U}{}{}{\tran}- \abf[\ss]{\K}{ff}{}{} \abf[\sm]{\bm{\Lambda}}{f}{}{}\right) : \text{d}\Zi.
\end{equation}
To make the terms $\text{d}\Zi$ vanish we define $\abf[\sm]{\bm{\Lambda}}{f}{}{}$ as
\begin{equation}
	\abf[\ss]{\K}{}{}{} \abf[\sm]{\bm{\Lambda}}{f}{}{} = -\der[g]{\abf[\s]{\U}{f}{}{}} \abf[\m]{\U}{}{}{\tran}.
\end{equation}
Thus, with \cref{eq:Lambdaadj}
\begin{equation}
	\label{eq:lambdarel}
	\abf[\sm]{\bm{\Lambda}}{f}{}{}\left[\abf[\s]{\bm{\Lambda}}{f}{}{}\right] = \abf[\s]{\bm{\Lambda}}{f}{}{} \abf[\m]{\U}{}{}{\tran}.
\end{equation}
With all multipliers defined, substitution of \cref{eq:submatrices,eq:lambdarel,eq:dum} into \cref{eq:lagh} and simplifying using \cref{eq:defW} gives
\begin{equation}
	\label{eq:ducpre}
	\begin{aligned}
		\text{d}\lag = & -\left(\abf[\s]{\mathbf{S}}{f}{}{}\abf[\s]{\bm{\Lambda}}{f}{}{} \otimes \mathbf{D}\right) : \text{d}\abf{\K}{}{}{} - \left(\abf[]{\tilde{\mathbf{S}}}{f}{}{}\abf[\m]{\bm{\Lambda}}{f}{}{} \otimes \abf[\m]{\U}{}{}{}\right) : \text{d}\abf{\tilde{\K}}{}{}{} + \abf[]{\tilde{\mathbf{S}}}{f}{}{} \abf[\m]{\bm{\Lambda}}{f}{}{}:\text{d}\tilde{\F}\\
		&- \abf[\s]{\bm{\Lambda}}{f}{}{} : \text{d}\abf[\s]{\F}{f}{}{}  + \abf[\ss]{\K}{pf}{}{} \abf[\s]{\bm{\Lambda}}{f}{}{} : \text{d}\abf[\s]{\U}{p}{}{}  + \abf[\m]{\bm{\Lambda}}{f}{}{} :\text{d}\abf[\m]{\F}{f}{}{}{} - \left(\abf[]{\tilde{\K}}{pf}{}{} \abf[\m]{\bm{\Lambda}}{f}{}{} +  \abf[]{\tilde{\mathbf{S}}}{p}{}{\tran}  \Zit\der[g]{\abf[\s]{\U}{f}{}{}}\right) : \text{d}\abf[\m]{\U}{p}{}{}{} .
	\end{aligned}
\end{equation}
Substitution of \cref{eq:xdfdkx,eq:dfbar} into \cref{eq:ducpre} yields
\begin{equation}
	\label{eq:ducpost}
	\begin{aligned}
		\text{d}\lag = & \left[\left(\ses \abf[]{\tilde{\mathbf{S}}}{f}{}{} \abf[\m]{\bm{\Lambda}}{f}{}{}-\abf[\s]{\mathbf{S}}{f}{}{}\abf[\s]{\bm{\Lambda}}{f}{}{} \right) \otimes \mathbf{D}\right] : \text{d}\abf{\K}{}{}{}  \\
		&- \left(\Zi\abf[]{\tilde{\mathbf{S}}}{f}{}{}\abf[\m]{\bm{\Lambda}}{f}{}{} + \abf[\s]{\bm{\Lambda}}{f}{}{}\right) : \text{d}\abf[\s]{\F}{f}{}{}  +\left(\mathbf{A} \abf[]{\tilde{\mathbf{S}}}{f}{}{} \abf[\m]{\bm{\Lambda}}{f}{}{} + \abf[\ss]{\K}{pf}{}{} \abf[\s]{\bm{\Lambda}}{f}{}{}\right) : \text{d}\abf[\s]{\U}{p}{}{} + \abf[\m]{\bm{\Lambda}}{f}{}{} :\text{d}\abf[\m]{\F}{f}{}{}{} - \left(\abf[]{\tilde{\K}}{pf}{}{} \abf[\m]{\bm{\Lambda}}{f}{}{} +  \abf[]{\tilde{\mathbf{S}}}{p}{}{\tran} \Zit \der[g]{\abf[\s]{\U}{f}{}{}}\right) : \text{d}\abf[\m]{\U}{p}{}{}{} .
	\end{aligned}
\end{equation}

\subsection{Sensitivities of the reaction load on the secondary DOFs}
\label{sec:sensfbd}
Based on the solution of \cref{eq:retubc} one can also analyse the reaction loads of the secondary DOFs via
\begin{equation*}
	\abf[\s]{\F}{p}{}{} = \abf[\sm]{\K}{p}{}{} \abf[\m]{\U}{}{}{} + \abf[\ss]{\K}{pf}{}{}\abf[\s]{\U}{f}{}{} + \abf[\ss]{\K}{pp}{}{} \abf[\s]{\U}{p}{}{}.  \tag{\ref{eq:retfbd} revisited}
\end{equation*}
We augment the response function with the linear systems of equations from \cref{eq:solves,eq:solvesq,eq:y}, that is
\begin{align}
	\begin{aligned}
		\lag\left[\abf[\s]{\F}{p}{}{},\abf[\sm]{\bm{\Lambda}}{f}{}{},\abf[\s]{\bm{\Lambda}}{f}{}{}, \abf[\m]{\bm{\Lambda}}{f}{}{}\right] := g \left[\abf[\s]{\F}{p}{}{}\right] &-\abf[\sm]{\bm{\Lambda}}{f}{}{}: \left(\abf[\ss]{\K}{ff}{}{}\Zi - \abf[\sm]{\K}{f}{}{}\right)\\
		& - \abf[\s]{\bm{\Lambda}}{f}{}{} : \left(\abf[\ss]{\K}{ff}{}{} \Zc - \abf[\ss]{\K}{fp}{}{}\abf[\s]{\U}{p}{}{} + \abf[\s]{\F}{f}{}{}\right)\\
		& - \abf[\m]{\bm{\Lambda}}{f}{}{} : \left(\abf[]{\tilde{\K}}{ff}{}{}\abf[\m]{\U}{f}{}{} -\abf[\m]{\F}{f}{}{}{} + \abf[]{\tilde{\K}}{fp}{}{}\abf[\m]{\U}{p}{}{}{} - \tilde{\F}_\text{f}\right).
	\end{aligned}
\end{align}
Full differentiation of both sides gives
\begin{equation}
	\label{eq:lagh2}
	\begin{aligned}
		\text{d}\lag = &\der[g]{\abf[\s]{\F}{p}{}{}} : \left(\text{d}\abf[\sm]{\K}{p}{}{} \abf[\m]{\U}{}{}{} + \text{d}\abf[\ss]{\K}{pf}{}{}\abf[\s]{\U}{f}{}{} + \text{d}\abf[\ss]{\K}{pp}{}{}\abf[\s]{\U}{p}{}{}\right)\\ 
		&+ \der[g]{\abf[\s]{\F}{p}{}{}} : \left(\abf[\sm]{\K}{p}{}{} \text{d}\abf[\m]{\U}{}{}{} - \abf[\ss]{\K}{pf}{}{} \left(\text{d}\Zi \abf[\m]{\U}{}{}{} + \Zi \text{d}\abf[\m]{\U}{}{}{} + \text{d}\Zc\right) + \abf[\ss]{\K}{pp}{}{} \text{d}\abf[\s]{\U}{p}{}{}\right)\\
		&- \abf[\sm]{\bm{\Lambda}}{f}{}{} : \left(\text{d}\abf[\ss]{\K}{ff}{}{} \Zi + \abf[\ss]{\K}{ff}{}{} \text{d}\Zi - \text{d}\abf[\sm]{\K}{f}{}{}\right)\\
		& - \abf[\s]{\bm{\Lambda}}{f}{}{} : \left(\text{d}\abf[\ss]{\K}{ff}{}{} \Zc + \abf[\ss]{\K}{ff}{}{}\text{d}\Zc - \text{d}\abf[\ss]{\K}{fp}{}{}\abf[\s]{\U}{p}{}{} - \abf[\ss]{\K}{fp}{}{} \text{d}\abf[\s]{\U}{p}{}{} + \text{d}\abf[\s]{\F}{f}{}{}\right)\\
		& - \abf[\m]{\bm{\Lambda}}{f}{}{} : \left(\text{d}\abf[]{\tilde{\K}}{ff}{}{}\abf[\m]{\U}{f}{}{}  + \abf[]{\tilde{\K}}{ff}{}{} \text{d}\abf[\m]{\U}{f}{}{} - \text{d}\abf[\m]{\F}{f}{}{}{} + \text{d}\abf[]{\tilde{\K}}{fp}{}{}\abf[\m]{\U}{p}{}{}{} + \abf[]{\tilde{\K}}{fp}{}{}\text{d}\abf[\m]{\U}{p}{}{}{} - \text{d}\tilde{\F}_\text{f}\right).
	\end{aligned}
\end{equation}
Since $\text{d}\Zi$, $\text{d}\Zc$ and $\text{d}\abf[\m]{\U}{f}{}{}$ are expensive to calculate, aim is to circumvent this by a suitable choices of $\abf[\sm]{\bm{\Lambda}}{f}{}{}$, $\abf[\s]{\bm{\Lambda}}{f}{}{}$ and $\abf[\m]{\bm{\Lambda}}{f}{}{}$. 
Using \cref{eq:dum} and subsequently collecting terms involving $\text{d}\abf[\m]{\U}{f}{}{}$ gives
\begin{equation}
	\left(-\abf[]{\tilde{\mathbf{S}}}{f}{}{\tran} \mathbf{C}\tran \der[g]{\abf[\s]{\F}{p}{}{}}- \abf[]{\tilde{\K}}{ff}{}{} \abf[\m]{\bm{\Lambda}}{f}{}{}\right) : \text{d}\abf[\m]{\U}{f}{}{}.
\end{equation}
To make the terms $\text{d}\abf[\m]{\U}{f}{}{}$ vanish we define $\abf[\m]{\bm{\Lambda}}{f}{}{}$ as
\begin{equation}
	\abf[]{\tilde{\K}}{ff}{}{}\abf[\m]{\bm{\Lambda}}{f}{}{}= -\abf[]{\tilde{\mathbf{S}}}{f}{}{\tran} \mathbf{C}\tran \der[g]{\abf[\s]{\F}{p}{}{}},
\end{equation}
which requires an additional solve using the condensed system.
Collecting terms involving $\text{d}\Zc$ gives
\begin{equation}
	\left(-\abf[\ss]{\K}{fp}{}{}\der[g]{\abf[\s]{\F}{p}{}{}} - \abf[\ss]{\K}{ff}{}{} \abf[\s]{\bm{\Lambda}}{f}{}{}\right) : \text{d}\Zc.
\end{equation}
To make the terms $\text{d}\Zc$ vanish we define $\abf[\s]{\bm{\Lambda}}{f}{}{}$ as
\begin{equation}
	\abf[\ss]{\K}{ff}{}{} \abf[\s]{\bm{\Lambda}}{f}{}{}= -\abf[\ss]{\K}{fp}{}{}\der[g]{\abf[\s]{\F}{p}{}{}},
\end{equation}
for which an additional full system solve is needed.
Finally, collecting terms involving $\text{d}\Zi$ gives
\begin{equation}
	\left(-\abf[\ss]{\K}{fp}{}{}\der[g]{\abf[\s]{\F}{p}{}{}} \abf[\m]{\U}{}{}{\tran} - \abf[\ss]{\K}{ff}{}{} \abf[\sm]{\bm{\Lambda}}{f}{}{}\right) : \text{d}\Zi.
\end{equation}
To make the terms $\text{d}\Zi$ vanish we define $\abf[\sm]{\bm{\Lambda}}{f}{}{}$ as
\begin{equation}
	\abf[\ss]{\K}{ff}{}{} \abf[\sm]{\bm{\Lambda}}{f}{}{} = -\abf[\ss]{\K}{fp}{}{}\der[g]{\abf[\s]{\F}{p}{}{}} \abf[\m]{\U}{}{}{\tran}.
\end{equation}
Thus, again
\begin{equation*}
	\abf[\sm]{\bm{\Lambda}}{f}{}{}\left[\abf[\s]{\bm{\Lambda}}{f}{}{}\right] = \abf[\s]{\bm{\Lambda}}{f}{}{} \abf[\m]{\U}{}{}{\tran}. \tag{\ref{eq:lambdarel} revisited}
\end{equation*}
Costs are comparable to those for the case in \cref{sec:sensubc}.
With all multipliers defined, substitution of \cref{eq:submatrices,eq:lambdarel,eq:dum} into \cref{eq:lagh2} and simplifying using \cref{eq:defW} gives
\begin{equation}
	\label{eq:ducpre2}
	\begin{aligned}
		\text{d}\lag = & -\left(\abf[\s]{\mathbf{S}}{f}{}{}\abf[\s]{\bm{\Lambda}}{f}{}{} \otimes \mathbf{D}\right) : \text{d}\abf{\K}{}{}{}- \left(\abf[]{\tilde{\mathbf{S}}}{f}{}{}\abf[\m]{\bm{\Lambda}}{f}{}{} \otimes \abf[\m]{\U}{}{}{}\right) : \text{d}\abf{\tilde{\K}}{}{}{} + \abf[]{\tilde{\mathbf{S}}}{f}{}{} \abf[\m]{\bm{\Lambda}}{f}{}{}:\text{d}\tilde{\F}\\
		&- \abf[\s]{\bm{\Lambda}}{f}{}{} : \text{d}\abf[\s]{\F}{f}{}{}  + \left(\abf[\ss]{\K}{pp}{}{}\der[g]{\abf[\s]{\F}{p}{}{}} + \abf[\ss]{\K}{pf}{}{}\abf[\s]{\bm{\Lambda}}{f}{}{}\right) : \text{d}\abf[\s]{\U}{p}{}{}  + \abf[\m]{\bm{\Lambda}}{f}{}{} :\text{d}\abf[\m]{\F}{f}{}{}{} - \left(\abf[]{\tilde{\K}}{pf}{}{}\abf[\m]{\bm{\Lambda}}{f}{}{} +  \abf[]{\tilde{\mathbf{S}}}{p}{}{\tran} \mathbf{C}\tran\der[g]{\abf[\s]{\F}{p}{}{}}\right) : \text{d}\abf[\m]{\U}{p}{}{}{} .
	\end{aligned}
\end{equation}
Substitution of \cref{eq:xdfdkx,eq:dfbar} into \cref{eq:ducpre2} yields
\begin{equation}
	\label{eq:ducpost2}
	\begin{aligned}
		\text{d}\lag = & \left[\left(\abf[\s]{\mathbf{S}}{f}{}{}\abf[\s]{\bm{\Lambda}}{f}{}{}- \ses\abf[]{\tilde{\mathbf{S}}}{f}{}{}\abf[\m]{\bm{\Lambda}}{f}{}{}\right) \otimes \mathbf{D}\right] : \text{d}\abf{\K}{}{}{} \\
		&- \left(\Zi\abf[]{\tilde{\mathbf{S}}}{f}{}{}\abf[\m]{\bm{\Lambda}}{f}{}{} + \abf[\s]{\bm{\Lambda}}{f}{}{}\right) : \text{d}\abf[\s]{\F}{f}{}{}  +\left(\abf[\ss]{\K}{pp}{}{}\der[g]{\abf[\s]{\F}{p}{}{}} + \mathbf{C} \abf[]{\tilde{\mathbf{S}}}{f}{}{} \abf[\m]{\bm{\Lambda}}{f}{}{} + \abf[\ss]{\K}{pf}{}{} \abf[\s]{\bm{\Lambda}}{f}{}{}\right) : \text{d}\abf[\s]{\U}{p}{}{}  + \abf[\m]{\bm{\Lambda}}{f}{}{} :\text{d}\abf[\m]{\F}{f}{}{}{} - \left(\abf[]{\tilde{\K}}{pf}{}{} \abf[\m]{\bm{\Lambda}}{f}{}{} +  \abf[]{\tilde{\mathbf{S}}}{p}{}{\tran} \mathbf{C}\tran\der[g]{\abf[\s]{\F}{p}{}{}}\right) : \text{d}\abf[\m]{\U}{p}{}{}{} .
	\end{aligned}
\end{equation}

\subsection{Summary}
\label{sec:senssum}
For clarity the resulting sensitivities as obtained in \cref{sec:sensS,sec:sensfbar,sec:sensubf,sec:sensfbd,sec:sensubc,sec:sensfbp} are summarized in \cref{tab:sensitivities}. It also lists the additional solves, along with their dimensionality and number of right-hand-sides, required to calculate the sensitivities. Here we assumed a worst-case scenario, where a response $g$ depends on \textit{all} the $l\ii$ states $\abf[\m]{\U}{f}{}{\i[i]}$, $\abf[\m]{\F}{p}{}{\i[i]}$, $\abf[\s]{\U}{f}{}{\i[i]}$ and/or $\abf[\s]{\F}{p}{}{\i[i]}$ of analysis $i$. In practice the number of right-hand-sides for a single response depends on the dependency of the response with respect to each state. \Cref{tab:sensitivities} shows expensive additional solves are only required for responses depending on the state(s) or reaction loads of secondary DOFs.

\begin{table}[h!]
	\centering
	\caption{Sensitivities of common response dependencies with respect to the system matrix, applied loads and prescribed states. Format is $\text{d}g_i\left[x_i\right] = \sum_j a_{ij} : \text{d}b_j$, with $i$ the row and $j$ the column. Note the dimensionality of $\abf[\s]{\bm{\Lambda}}{f}{}{} \in \mathbb{R}^{\abf[\s]{f}{}{}{\i[i]} \times \nrhs\ii}$ and $\abf[\m]{\bm{\Lambda}}{f}{}{} \in \mathbb{R}^{\abf[\m]{f}{}{}{\i[i]} \times \nrhs\ii}$.}
	\begin{tabular}{c|ccccc|cc}
		\toprule
		$x$ & $\text{d}\abf{\K}{}{}{}$ & $\text{d}\abf[\s]{\U}{p}{}{\i[i]}$ & $\text{d}\abf[\s]{\F}{f}{}{\i[i]}$& $\text{d}\abf[\m]{\U}{p}{}{\i[i]}$ & $\text{d}\abf[\m]{\F}{f}{}{\i[i]}$ & $a$ & Solves\\
		\midrule
		$\abf{\tilde{\K}}{}{}{}$ & $\ses a \otimes \ses$ & & && & $\der[g]{\abf{\tilde{\K}}{}{}{}}$ &\\
		\midrule
		$\abf[]{\tilde{\F}}{}{}{\i[i]}$ & $\ses a\otimes \abf[]{\mathbf{B}}{}{}{\i[i]}$ & $\mathbf{C}a$ & $-\Zi a $ &&&$\der[g]{\abf{\tilde{\F}}{}{}{\i[i]}}$ &\\
		\midrule
		$\abf[\m]{\U}{f}{}{\i[i]}$ & $\ses \abf{a}{}{}{} \otimes \abf[]{\mathbf{D}}{}{}{\i[i]}$ & $\mathbf{C} a$ & $-\Zi a$ & $-\abf[]{\tilde{\K}}{pf}{}{\i[i]} \abf[\m]{\bm{\Lambda}}{f}{}{\i[i]}$& $\abf[\m]{\bm{\Lambda}}{f}{}{\i[i]}$ &$ \abf[]{\tilde{\mathbf{S}}}{f}{}{\i[i]} \abf[\m]{\bm{\Lambda}}{f}{}{\i[i]}$& $\abf[]{\tilde{\K}}{ff}{}{\i[i]} \abf[\m]{\bm{\Lambda}}{f}{}{\i[i]} = \der[g]{\abf[\m]{\U}{f}{}{\i[i]}}$ \\
		\midrule
		$\abf[\m]{\F}{p}{}{\i[i]}$ & $\ses a\otimes \abf[]{\mathbf{D}}{}{}{\i[i]}$ & $\mathbf{C}a$ & $-\Zi a$ & $\begin{aligned}-\abf[]{\tilde{\K}}{pf}{}{\i[i]} \abf[\m]{\bm{\Lambda}}{f}{}{\i[i]}\\ +\abf[]{\tilde{\K}}{pp}{}{\i[i]}\der[g]{\abf[\m]{\F}{p}{}{\i[i]}}\end{aligned}$& $\abf[\m]{\bm{\Lambda}}{f}{}{\i[i]}$ & $\begin{aligned}\abf[]{\tilde{\mathbf{S}}}{f}{}{\i[i]}\abf[\m]{\bm{\Lambda}}{f}{}{\i[i]} -\\ \abf[]{\tilde{\mathbf{S}}}{p}{}{\i[i]}\der[g]{\abf[\m]{\F}{p}{}{\i[i]}} \end{aligned}$ &$\abf[]{\tilde{\K}}{ff}{}{\i[i]} \abf[\m]{\bm{\Lambda}}{f}{}{\i[i]} = \abf[]{\tilde{\K}}{fp}{}{\i[i]} \der[g]{\abf[\m]{\F}{p}{}{\i[i]}}$\\
		\midrule
		$\abf[\s]{\U}{f}{}{\i[i]}$ & $ \left( \ses a - \abf[\s]{\mathbf{S}}{f}{}{} \abf[\s]{\bm{\Lambda}}{f}{}{\i[i]}\right)\otimes \abf[]{\mathbf{D}}{}{}{\i[i]}$ &$\mathbf{C}a+\abf[\ss]{\K}{pf}{}{} \abf[\s]{\bm{\Lambda}}{f}{}{\i[i]}$& $-\Zi a - \abf[\s]{\bm{\Lambda}}{f}{}{\i[i]}$ &$\begin{aligned}-\abf[]{\tilde{\K}}{pf}{}{\i[i]}\abf[\m]{\bm{\Lambda}}{f}{}{\i[i]} - \\ {\abf[]{\tilde{\mathbf{S}}}{p}{}{\i[i]}}\tran \Zit \der[g]{\abf[\s]{\U}{f}{}{\i[i]}}\end{aligned}$& $\abf[\m]{\bm{\Lambda}}{f}{}{\i[i]}$& $\abf[]{\tilde{\mathbf{S}}}{f}{}{\i[i]}\abf[\m]{\bm{\Lambda}}{f}{}{\i[i]}$& $\begin{aligned}\abf[\ss]{\K}{ff}{}{} \abf[\s]{\bm{\Lambda}}{f}{}{\i[i]} &= -\der[g]{\abf[\s]{\U}{f}{}{\i[i]}}\\ \abf[]{\tilde{\K}}{ff}{}{\i[i]} \abf[\m]{\bm{\Lambda}}{f}{}{\i[i]} &= -{\abf[]{\tilde{\mathbf{S}}}{f}{}{\i[i]} }\tran\Zit \der[g]{\abf[\s]{\U}{f}{}{\i[i]}}\end{aligned}$\\
		\midrule
		$\abf[\s]{\F}{p}{}{\i[i]}$ & $\left( \ses a - \abf[\s]{\mathbf{S}}{}{}{} \abf[\s]{\bm{\Lambda}}{f}{}{\i[i]}\right)\otimes \mathbf{D} $
		&$\begin{aligned}\mathbf{C}a+\abf[\ss]{\K}{pf}{}{} \abf[\s]{\bm{\Lambda}}{f}{}{\i[i]}\\+ \abf[\ss]{\K}{pp}{}{} \der[g]{\abf[\s]{\F}{p}{}{\i[i]}}\end{aligned}$& $-\Zi a - \abf[\s]{\bm{\Lambda}}{f}{}{\i[i]}$ &$\begin{aligned}- \abf[]{\tilde{\K}}{pf}{}{\i[i]} \abf[\m]{\bm{\Lambda}}{f}{}{\i[i]} -\\ {\abf[]{\tilde{\mathbf{S}}}{p}{}{\i[i]}}\tran \mathbf{C} \tran  \der[g]{\abf[\s]{\F}{p}{}{\i[i]}}\end{aligned}$& $\abf[\m]{\bm{\Lambda}}{f}{}{\i[i]}$& $\abf[]{\tilde{\mathbf{S}}}{f}{}{\i[i]}\abf[\m]{\bm{\Lambda}}{f}{}{\i[i]}$ &$\begin{aligned}\abf[\ss]{\K}{}{}{} \abf[\s]{\bm{\Lambda}}{f}{}{\i[i]} &= -\abf[\ss]{\K}{fp}{}{} \der[g]{\abf[\s]{\F}{p}{}{\i[i]}}\\ \abf[]{\tilde{\K}}{ff}{}{\i[i]} \abf[\m]{\bm{\Lambda}}{f}{}{\i[i]} &= -{\abf[]{\tilde{\mathbf{S}}}{f}{}{\i[i]}}\tran\mathbf{C}\tran \der[g]{\abf[\s]{\F}{p}{}{\i[i]}}\end{aligned}$\\
		\bottomrule
	\end{tabular}
	\label{tab:sensitivities}
\end{table}

\end{document}